\documentclass[11pt,a4paper]{article}

\pdfoutput=1
\usepackage{jheppub}
\usepackage[normalem]{ulem}	
\usepackage{graphicx}
\usepackage{comment}
\usepackage{graphics}
\usepackage{feynmp}
\usepackage{epsf}
\usepackage{color}
\usepackage{amsmath}
\usepackage{amssymb}
\usepackage{latexsym}
\usepackage{slashed}
\usepackage{float}
\usepackage{cases}
\usepackage{multirow}
\newcommand{\al}[1]{\begin{align}#1\end{align}}

\newcommand{\bp}{\begin{pmatrix}}
\newcommand{\ep}{\end{pmatrix}}
\newcommand{\bb}{\begin{bmatrix}}
\newcommand{\eb}{\end{bmatrix}}

\newcommand{\paren}[1]{\left(#1\right)}

\newcommand{\cblack}[1]{{\color{black}#1}}
\newcommand{\beq}{\begin{equation}}
\newcommand{\eeq}{\end{equation}}
\newcommand{\bea}{\begin{eqnarray}}
\newcommand{\eea}{\end{eqnarray}}

\newcommand{\pal}{\partial}

\newcommand{\fulltoday}{\number\day\space \ifcase\month\or
    January\or February\or March\or April\or May\or June\or
    July\or August\or September\or October\or November\or December\fi
    \space\number\year}
\def\rinv{R^{-1}}
\def\rqp{r_Q^\prime}
\def\rtp{r_T^\prime}
\def\rgp{r_G^\prime}
\def\ryp{r_Y^\prime}
\def\rbp{r_B^\prime}
\def\rwp{r_W^\prime}
\def\rew{r_{\text{EW}}^\prime}
\def\rhp{r_{H}^\prime}
\def\rlp{r_{L}^\prime}

\def\tonel{t^{(1)}_l}
\def\toneh{t^{(1)}_h}
\def\ttwol{t^{(2)}_l}

\def\tonebarl{\bar{t}^{(1)}_l}
\def\tonebarh{\bar{t}^{(1)}_h}
\def\ttwobarl{\bar{t}^{(2)}_l}

\def\bonel{b^{(1)}_l}
\def\boneh{b^{(1)}_h}

\def\bonebarl{\bar{b}^{(1)}_l}

\def\kkglone{g^{(1)}}
\def\kkqkone{q^{(1)}}
\def\kkaone{\gamma^{(1)}}
\def\kkzone{Z^{(1)}}
\def\kkwone{W^{(1)^\pm}}
\def\kkwonep{W^{(1)^+}}
\def\kkwonem{W^{(1)^-}}
\def\kkhonep{H^{(1)^+}}

\def\kkhoneo{H^{(1)^0}}
\def\kkhpone{A^{(1)^0}}
\def\kkgltwo{g^{(2)}}

\def\kkatwo{\gamma^{(2)}}
\def\kkztwo{Z^{(2)}}
\def\kkwtwo{W^{(2)^\pm}}

\def\mkkglone{m_{g^{(1)}}}
\def\mkkqkone{m_{q^{(1)}}}
\def\mkkaone{m_{\gamma^{(1)}}}
\def\mkkzone{m_{Z^{(1)}}}
\def\mkkwone{m_{W^{(1)^\pm}}}
\def\mkkhonepm{m_{H^{(1)^\pm}}}
\def\mkkgltwo{m_{g^{(2)}}}
\def\mkkqktwo{m_{q^{(2)}}}
\def\mkkatwo{m_{\gamma^{(2)}}}
\def\mkkztwo{m_{Z^{(2)}}}
\def\mkkwtwo{m_{W^{(2)^\pm}}}
\def\mtonel{m_{t^{(1)}_l}}
\def\mtoneh{m_{t^{(1)}_h}}
\def\mttwol{m_{t^{(2)}_l}}
\def\mttwoh{m_{t^{(2)}_h}}
\def\mbone{m_{b^{(1)}}}
\def\mbtwo{m_{b^{(2)}}}
\def\mtopin{m_t^{\text{in}}}
\def\mtopphys{m_t^{\text{phys}}}
\def\rinv{R{^{-1}}}

\def\alphau1{\alpha_1}
\def\alphasu2{\alpha_2}

\renewcommand{\multirowsetup}{\centering}
\title{Non-minimal Universal Extra Dimensions with Brane Local Terms: The Top Quark Sector}

\author[a]{AseshKrishna Datta}
\author[b]{Kenji Nishiwaki}
\author[b]{Saurabh Niyogi}

\affiliation[a]{Harish-Chandra Research Institute, Allahabad 211019, India}
\affiliation[b]{Regional Centre for Accelerator-based Particle Physics \\
                Harish-Chandra Research Institute, Allahabad 211019, India} 

\emailAdd{asesh@hri.res.in, nishiwaki@hri.res.in, sourabh@hri.res.in}
\preprint{HRI-P-13-10-001\\ 
\vspace*{-0.8cm}
\begin{flushright}
RECAPP-HRI-2013-021
\end{flushright}
}

\abstract{We study the physics of Kaluza-Klein (KK) top quarks in the 
framework of a non-minimal Universal Extra Dimension (nmUED) with an 
orbifolded ($S^1/Z_2$) flat extra spatial dimension in the presence of 
brane-localized terms (BLTs). 
In general, BLTs affect the masses and the couplings of the KK
excitations in a non-trivial way including those for the KK top quarks.
On top of that, BLTs also influence the mixing of the top quark chiral
states at each KK level and trigger 
mixings among excitations from different levels with identical KK
parity (even or odd).
The latter phenomenon of mixing of KK levels is not present in the
popular UED scenario known as the minimal UED (mUED) {at the tree level}. Of particular 
interest are the mixings among the KK top quarks from level `0'
and level `2' (driven by the mass of the Standard Model (SM) top
quark). These open up new production modes in 
the form of single production of a KK top quark and the possibility of 
its direct decays to SM particles leading to rather 
characteristic signals at the colliders. Experimental constraints and
the restrictions they impose on the nmUED parameter 
space are discussed. The scenario is implemented in {\tt MadGraph} 5 by 
including the quark, lepton, the gauge-boson and the Higgs sectors up to the
second KK level. A few benchmark scenarios are chosen for preliminary 
studies of the decay patterns of the KK top quarks and their 
production rates at the LHC in various different modes.
Recast of existing experimental analyzes in 
scenarios having similar states is found to 
be not so straightforward for the KK top quarks of the nmUED
scenario under consideration.}

\begin{document} 
\maketitle

%
\section{Introduction}
\label{sec:intro}

The top quark is altogether a different kind of a fermion in the realm 
of the Standard Model (SM) sheerly because of its large mass or
equivalently, its large (Yukawa) coupling to the Higgs boson. Even when 
the discovery of the Higgs boson was eagerly awaited, the implications 
of such a large Yukawa coupling was already much appreciated. Many new 
physics scenarios beyond the SM (BSM), which have extended top 
quark sectors offering top quark partners, derive theoretically 
nontrivial and phenomenologically rich attributes from this aspect.
At colliders, they warrant dedicated searches which generically result
in weaker bounds on them when compared to their peers from the first
two generations.

Naturally, ever since the confirmation of the recent discovery of a 
Higgs-like scalar particle came in, the top quark sectors of different 
new physics scenarios have been in the spotlight triggering a spur of 
focussed activities. While popular supersymmetric (SUSY) scenarios are
excellent hunting grounds for such possibilities and have taken the 
center stage during the recent past and at a time of renewed drives, 
there {exist other physics scenarios} that offer interesting {signatures} 
at the colliders with phenomenologically rich top quark sectors. 
Scenarios with Universal Extra Dimensions (UEDs) are also no exceptions
{even though the setups are not necessarily tied to and/or
address the `naturalness' issue of the Higgs sector like many of the
competing scenarios do thus requiring relatively light `top partner'
(${\cal O}(1)$ TeV). However, on a somewhat different track, attempts to
understand the hierarchy
of masses and mixings of the (4D) SM fermions while conforming with
the strong FCNC constraints for the first two generations often adopt
mechanisms that distinguish the third generation from the first two
\cite{Del Aguila:2001pu}. This could also lead to lighter states
for the former.
Thus, in the absence of a robust principle that prohibits them and
until the experiments exclude them specifically, it is important that
these should make a necessary part of the search programme at the colliders.
This is further appropriate while being under the cloak of the so-called
`SUSY-UED' confusion \cite{Cheng:2002ab} which may not allow us understand
immediately the nature of such a newly-discovered state.}

Thus, there has been a reasonable amount of 
activity involving comparatively light KK top quarks of the UED scenarios
in the past
\cite{Petriello:2002uu, Rai:2005vy, 
Maru:2009cu,Nishiwaki:2011vi,Nishiwaki:2011gk,Nishiwaki:2011gm} 
and also from recent times post 
Higgs-discovery \cite{Belanger:2012mc, Kakuda:2013kba, Dey:2013cqa, 
Flacke:2013nta}. The latter set of works have constrained the
respective scenarios discussed to varying degrees by analyzing the
Higgs results.
In this work, we study the structure of the top quark sector of the 
so-called non-minimal universal extra dimensions (nmUED), the 
nontrivial features it is endowed with and their implications for the 
LHC. 

The particular nmUED scenario we deal with in this work is different
from the popular minimal UED (mUED) scenario 
\cite{Appelquist:2000nn, Cheng:2002iz} (an incarnation of the 
so-called generic TeV-scale extra {dimensions} \cite{Antoniadis:1990ew}) 
in the fact that the former takes into consideration the effect of 
brane-local terms (BLTs) which are already non-vanishing at the tree 
level\footnote{Note that BLTs get renormalized and thus cannot be 
set to zero at all scales.} 
\cite{delAguila:2003bh, delAguila:2003gu, del Aguila:2006kj} 
and that develop at the two fixed points\footnote{{A possibility with
multiple fixed points (branes)
are helpful for explaining the fermion flavor structure~\cite{Fujimoto:2012wv,
Fujimoto:2013ki}.}} of $S^1/Z_2$ orbifold on 
which the extra space dimension of such a 5-dimensional scenario is 
compactified.
As is well-known, BLTs affect both properties of the KK 
modes (corresponding to the fields present in the bulk) that crucially 
govern their phenomenology: they modify the masses of these KK modes 
and alter their wavefunctions thus affecting their physical couplings 
in four dimensions.

The phenomenology of such a scenario at the LHC has recently been
discussed in \cite{Datta:2012tv} with reference to strong productions
of the KK gluons and (vector-like) KK quarks from the first excited 
level\footnote{Phenomenology of KK-parity violating BLTs
are discussed in \cite{Datta:2012xy, Datta:2013lja}.}.
It was demonstrated how such processes could closely mimic the 
corresponding SUSY processes. There, such a scenario was also 
contrasted against the popular mUED scenario. Tentative bounds on 
these excitations were derived from recent LHC results. However, for 
the KK quarks, such bounds referred only to the first two generation 
quarks.

The top quark sector of the mUED had earlier been studied at the LHC
in ref.~\cite{Choudhury:2009kz}.
In the present work we take up the case of KK top quarks in the nmUED 
scenario. These are `vector-like' states and can be lighter than the 
KK quarks from the first two generations. This is exactly the
reason behind the current surge in studies on `top-partners' at the LHC
{\cite{AguilarSaavedra:2009es,Cacciapaglia:2010vn,Cacciapaglia:2011fx,Berger:2012ec, DeSimone:2012fs,Kearney:2013oia,Buchkremer:2013bha,Aguilar-Saavedra:2013qpa}}. 
{From phenomenological considerations,} the nmUED scenario under consideration is different from the 
mUED scenario {in {the following} important aspects:}
(i) {the} KK masses for these excitations and {their
couplings} derived form the compactification
of the extra dimension can be very 
different\footnote{{An extreme example of decoupling the mass scale of 
new physics form the compactification scale can be found in ref.~\cite{Del Aguila:2001pu}.}} 
{from their mUED counterparts} for a given value of the
inverse compactification radius $\rinv$ and (ii) the mixing between 
the (chiral) top quark states driven 
by the top quark mass (which is a generic feature of scenarios with extended 
top quark sector) can be essentially different. 
Further, we highlight a rather characteristic feature of such an
nmUED scenario which triggers
mixing of excitations from similar KK levels of similar parities (even
or odd).
Such \emph{level-mixings} are 
triggered by BLTs \cite{delAguila:2003kd, delAguila:2003gv} due to 
non-vanishing overlap integrals and arise from the Yukawa sector. 
Hence, such effects depend on the corresponding brane-local parameter.
These induce tree level couplings among the resulting states
(mixtures of corresponding states from different KK levels). Note that
in mUED, such couplings are only present beyond Born-level and are
thus suppressed. Also, as we will see later in this work, such mixings
{can be interesting only for the} KK fermions from the third generation and
in particular, for the top quark sector thanks to the large top quark
mass. Moreover, in the context of the LHC, the only relevant
mixings are going to be those involving the SM (level `0') and the 
level `2' KK states.

In the nmUED scenario, the general setup for the quark sector involves
BLTs of both kinetic and Yukawa type. This was discussed in
appropriate details in \cite{Datta:2012tv} for the level `1' KK
excitations including the third generation quarks. In this work, we
extend the scheme to include the level `2' excitations as well with
particular emphasis on the top quark sector. It is demonstrated how
presence of level mixing may potentially open up interesting 
phenomenological possibilities at the LHC in the form of new modes of
their production and decay some of which would necessarily involve KK
excitations of the gauge and the Higgs bosons in crucial ways. This 
would no doubt have significant phenomenological implications at the 
LHC and could provide us with an understanding of how the same can be 
contrasted against other scenarios having similar signatures and/or 
can be deciphered from experimental data.

The paper is organized as follows. In section 2 we discuss the
theoretical framework of the top quark sector at higher KK levels
along with those of the gauge and the Higgs sectors which are
intimately connected to the theory and phenomenology of the KK top
quarks. The resulting mass spectra and the form of the relevant 
couplings are discussed in section 3. In section 4 we discuss in 
some details the experimental constraints that potentially restrict 
the parameter space of the scenario under consideration. A few
benchmark points, which satisfy all these constraints, are also 
chosen for further studies.  Section 5 is devoted to the basic 
phenomenology of the KK top quarks at the LHC by outlining their 
production and decay patterns. In section 6 we conclude.
%

\section{Theoretical framework}
\label{sec:model}
We consider the top quark sector of a 5D nmUED scenario 
compactified on $S^1/Z_2$ in the presence of tree-level BLTs that 
develop at the orbifold fixed points. The compactification is 
characterized by the length parameter {$L$ where $L = \pi R/2$, $R$ being the radius of the
orbifolded extra space dimension.}
The two fixed points of the $S^1/Z_2$ geometry are 
taken to be at $y=\pm L$. The derivations broadly follow the 
notations, the conventions and the treatments 
adopted in reference~\cite{Datta:2012tv}.
The phenomenological relevance of the KK gauge and Higgs
sectors prompts us to incorporate them thoroughly in the present 
analysis, including even the level `2' KK excitations in some of these
cases. In the following we outline the necessary theoretical setup 
involving these sectors. We start with the gauge and the Higgs
sectors first since the issue of Higgs vacuum expectation value (VEV)
is relevant for the top quark (Yukawa) sector.
%
\subsection{The gauge boson and the Higgs sectors}
\label{subsec:gauge-higgs}
The gauge boson and the Higgs sectors of the nmUED scenario had been
discussed in some detail in ref.~\cite{Flacke:2008ne} with due stress
on their mutual relationship and the implications thereof for possible
dark matter candidates of such a scenario. We closely follow the approach 
there and summarize the aspects that are relevant for our present 
study.

We consider the following 5D action \cite{Flacke:2008ne} describing 
the gauge and the Higgs sectors of the nmUED scenario under study:
\al{
S &=
	\int d^4x \int_{-L}^{L} dy \Bigg\{
	- \frac{1}{4} G_{MN}^{a} G^{aMN} - \frac{1}{4} W_{MN}^{i} W^{iMN}
	- \frac{1}{4} B_{MN} B^{MN} \notag \\
&\phantom{S = \int d^4x \int_{-L}^{L} dy \,}
	+ (D_M \Phi)^{\dagger} (D^M \Phi) + \hat{\mu}^2 \Phi^{\dagger} \Phi
	- \frac{\hat{\lambda}}{4} (\Phi^{\dagger} \Phi)^2 \notag \\
&\quad + \Big( \delta(y-L) + \delta(y+L) \Big) \Big[
	- \frac{r_G}{4} G_{\mu\nu}^{a} G^{a\mu\nu} - \frac{r_W}{4} W_{\mu\nu}^{i} W^{i\mu\nu}
	- \frac{r_B}{4} B_{\mu\nu} B^{\mu\nu} \notag \\
&\phantom{\quad + \Big( \delta(y-L) + \delta(y+L) \Big) \Big[ \,\,\,}
	+ r_H (D_{\mu} \Phi)^{\dagger} (D^{\mu} \Phi) + {\mu}_b^2 \Phi^{\dagger} \Phi
	- \frac{\lambda_b}{4} (\Phi^{\dagger} \Phi)^2 \Big] \Bigg\},
\label{eqn:action-gauge-higgs}
}
where $y$ represents the compact extra spatial direction, the Lorentz
indices $M$ and $N$ run over $0,1,2,3,y$ while $\mu$ and $\nu$ run 
over $0,1,2,3$. $G_{MN}^{a}$, $W_{MN}^{i}$ and $B_{MN}$ are the 5D 
field-strengths associated with the gauge groups $SU(3)_C$, $SU(2)_W$ 
and $U(1)_Y$ respectively with the corresponding 5D gauge bosons 
$G_M^a$, $W_M^i$ and $B_M$. $a$ and $i$ are the adjoint indices for 
the groups $SU(3)_C$ and $SU(2)_W$, respectively. The 5D Higgs doublet 
is represented by $\Phi$ {with its components given by}
\al{
\Phi = \begin{pmatrix} \phi^{+} \\ \frac{1}{\sqrt{2}} 
\left( \hat{v}(y) + H + i \chi \right) \end{pmatrix}
		\label{Higgsform}
}
where $\phi^+$ is the charged component, $H$ and $\chi$ are the
neutral components and $\hat{v}(y)$ is the 5D bulk Higgs VEV. 
$D_M$ stands for the 5D covariant 
derivatives and $\hat{\mu}$ and $\hat{\lambda}$ represent the 5D 
bulk Higgs mass and the Higgs self-coupling, respectively.

We take $Z_2$ {eigenvalues for the fields} $G_{\mu}^a,\, W_{\mu}^i,
\, B_{\mu},\, H,\ \chi,\, \phi^{+}$ {to be even at both the fixed points to realize}
the zero modes 
(that correspond to the SM degrees of freedom) {have} vanishing
KK-masses from compactification. This
automatically renders the eigenvalues of $G_{y}^a,\, W_{y}^i,\, B_{y}$
to be odd because of 5D gauge symmetry for which there are no 
corresponding zero modes.

As can be seen in equation \ref{eqn:action-gauge-higgs}, the BLTs 
(proportional to the $\delta$-functions) are introduced at the 
orbifold fixed points for both the gauge and the Higgs sectors. The 
bulk mass term and the Higgs self-interaction term are considered
only for the latter for preserving the 4D gauge invariance.
The six coefficients $r_G$, $r_W$, $r_B$, $r_H$, $\mu_b$ and $\lambda_b$ 
influence the masses of the KK excitations and the effective couplings 
involving them. As is well-known, due to the existence of the BLTs, 
{momentum conservation} along the $y$ direction is violated even 
at the tree level (in contrast to the mUED where this could happen 
only beyond the tree level), but a discrete symmetry, called the
KK-parity, under the reflection $y \to -y$ is still preserved.
KK-parity ensures the existence of a stable dark matter candidate
which is the lightest KK particle (LKP) at level `1' obtained on
compactification.

In this work, for simplicity, we focus on the following situation:
\al{
\sqrt{\frac{4 \hat{\mu}^2}{\hat{\lambda}}} 
= \sqrt{\frac{4 {\mu_b}^2}{{\lambda_b}}}
	\quad \text{and} \quad r_W = r_B \,\equiv\, r_{\text{EW}}.
	\label{condition_one_for_minimalcase}
}
The first condition ensures a constant profile of the Higgs VEV over 
the whole space, \emph{i.e.},
\al{
\hat{v}(y) \to \sqrt{\frac{4 \hat{\mu}^2}{\hat{\lambda}}} = 
\sqrt{\frac{4 {\mu_b}^2}{{\lambda_b}}} \,\equiv\, \hat{v},
}
while with the second condition\footnote{For $r_W \not= r_B$, 
obtaining the correct value of the Weinberg angle in the SM sector is nontrivial. 
We, thus, do not consider this possibility in the present
work although the same could have interesting phenomenological
implications both at colliders or otherwise (see ref.~\cite{Flacke:2008ne} that discusses its implication for {possible KK dark matter candidates}).} 
we can continue to relate the 5D $W$, $Z$ and the photon ($\gamma$) states 
(at tree level) via the {usual} Weinberg angle $\theta_W$ at all KK levels, 
\emph{i.e.},
\al{
W_M^{\pm} = \frac{W^{1}_M \mp i W^{2}_{M}}{\sqrt{2}},
\quad
\begin{pmatrix} Z_M \\ \gamma_M \end{pmatrix}
=
\begin{pmatrix} \cos{\theta_W} & \sin{\theta_W} \\
	-\sin{\theta_W} & \cos{\theta_W} \end{pmatrix}
\begin{pmatrix} W^3_M \\ B_M \end{pmatrix}.
}

The gauge-fixing conditions along with their consequences are discussed
briefly in appendix \ref{app:gauge-fixing}. We choose the unitary 
gauge. For the fields 
$G_{\mu}^a,\, W_{\mu}^{+},\, Z_{\mu},\, H,\, \chi,\, \phi^{+}$ and for 
the ones like  $\pal_y W_y^{+},\, \pal_y Z_y$, the mode functions for 
KK decomposition and the conditions that determine their KK-masses are 
summarized below.
\al{
f_{F_{(n)}}(y) &= N_{F_{(n)}} \times
	\begin{cases}
	\displaystyle \frac{\cos(M_{F_{(n)}} y)}{C_{F_{(n)}}} & \text{for even }n, \\ -
	\displaystyle \frac{\sin(M_{F_{(n)}} y)}{S_{F_{(n)}}} & \text{for odd }n, \\
	\end{cases} \\ \nonumber \\
m_{F_{(n)}}^2 &= m_{F}^2 + M_{F_{(n)}}^2, \\ \nonumber \\
\frac{(r_F m_{F_{(n)}}^2 - m_{F,b}^2)}{M_{F_{(n)}}} &= 
	\begin{cases} - T_{F_{(n)}} & \text{for even } n, 
\\
	+ 1/T_{F_{(n)}} & \text{for odd } n
	\end{cases}
	\label{masscondition}
}
with the following short-hand notations:
\al{
C_{F_{(n)}} = \cos\left( \frac{M_{F_{(n)}} \pi R}{2} \right), \quad
S_{F_{(n)}} = \sin\left( \frac{M_{F_{(n)}} \pi R}{2} \right), \quad
T_{F_{(n)}} = \tan\left( \frac{M_{F_{(n)}} \pi R}{2} \right).
}
The normalization factors $N_{F_{(n)}}$ for the mode functions
$f_{F_{(n)}}(y)$ are given by
\al{
N_{F_{(n)}}^{-2} =
	\begin{cases}
	\displaystyle 2 r_{F} + \frac{1}{C_{F_{(n)}}^2} \left[ \frac{\pi R}{2} + \frac{1}{2 M_{F_{(n)}}} \sin(M_{F_{(n)}} \pi R) \right] & \text{for even } n, \\
	\displaystyle 2 r_{F} + \frac{1}{S_{F_{(n)}}^2} \left[ \frac{\pi R}{2} - \frac{1}{2 M_{F_{(n)}}} \sin(M_{F_{(n)}} \pi R) \right] & \text{for odd } n.
	\end{cases}
}
Here $m_{F_{(n)}}$, $m_F$, $M_{F_{(n)}}$, $r_F$ and $m_{F,b}$ {stand}
for
the physical mass, the bulk mass, the KK mass, the coefficient of the
corresponding brane-local kinetic term (BLKT) 
and brane mass term of the field $F$, 
respectively. Inputs for the mass-determining conditions for all 
these fields are presented in appendix \ref{app:gauge-fixing}.
%
{Further, following conditions must hold to ensure the zero-mode (SM) profiles to be flat which help evade severe constraints from electroweak observables like the Z-boson mass, $\sin^2{\theta_W}$ etc.}
\al{
r_{\text{EW}} = r_H  \qquad &\text{for }W_{\mu}^{+},\, Z_{\mu}, \notag \\ 
r_H (2 \hat{\mu}^2) = 2 \mu_b^2 \qquad  &\text{for } H.
\label{eqn:extra-conds}
}
{Non-compliance of the above relations could result in unacceptable modifications in the level-`0' (SM) Lagrangian~\cite{Flacke:2008ne}.}

\begin{figure}[t]
\centering
%
%
\includegraphics[width=0.4\columnwidth]{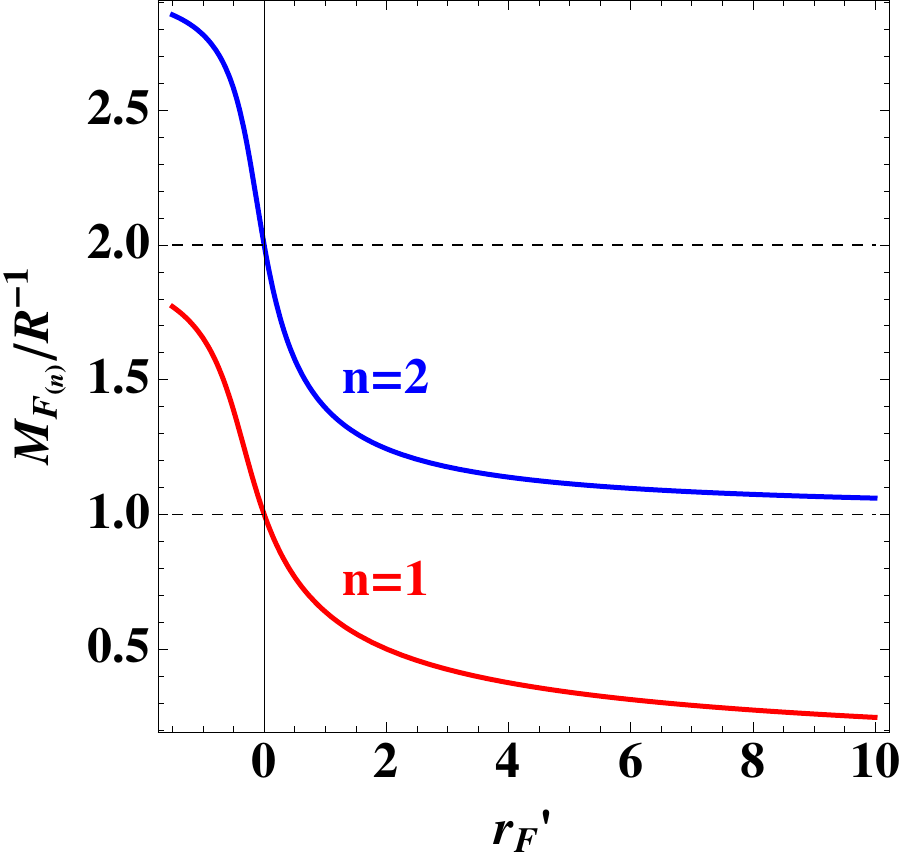}
\caption{
The generic profile of the variation of
$M_{F_{(n)}}/R^{-1}$ as a function of $r'_F \,(= r_F R^{-1})$ for the cases $n=1$ and $n=2$.
}
\label{fig:MKKconfigurations}
\end{figure}

Also, with the above two conditions, 
equation \ref{masscondition} reduces to the following simple form:
\al{
r_F {M_{F_{(n)}}} =
\begin{cases}
- T_{F_{(n)}} & \text{for $n$ even,} \\
1/T_{F_{(n)}} & \text{for $n$ odd}
\end{cases}
\label{masscondition_simplified}
}
where $M_{F_{(0)}}$\,$=$\,$0$ (thus ensuring vanishing KK masses for 
the level `0' (SM) fields).
A theoretical lower bound of $r_F > - \frac{\pi R}{2}$ 
must hold to circumvent tachyonic zero modes.
In figure~\ref{fig:MKKconfigurations}, we illustrate the generic profile of the variation of
$M_{F_{(n)}}/R^{-1}$ as a function of $r'_F \,(= r_F R^{-1})$ for the cases $n=1$ and $n=2$.

On the other hand, vanishing KK masses at level `0' are always 
realized for $\phi^{+}$ and $\chi$ which are eventually 
``eaten up'' by the massless level `0' $W_{\mu}^{+},\, Z_{\mu}$ states 
respectively as they become massive. However, no zero mode appears 
for $W_{y}^{+},\, Z_{y}$ since they are projected out by the $Z_2$-odd 
condition.  
The mode functions for the fields $W_y^{+},\, Z_y$ are given by
\al{
f_{F_{(n)}}(y) &= N_{F_{(n)}} \times
	\begin{cases}
	\displaystyle \frac{\sin(M_{F_{(n)}} y)}{C_{F_{(n)}}} & \text{for even }n, \\ 
	\displaystyle \frac{\cos(M_{F_{(n)}} y)}{S_{F_{(n)}}} & \text{for odd }n \\
	\end{cases}
}
with the mass-determination condition as given in equation 
\ref{masscondition}.
Use of equation \ref{bulk_condition} allows one to 
eliminate $\chi$ in favor of $Z_y$ and $\phi^+$ in favor of 
$W_y^+$. 
Correct normalization of the kinetic terms requires $Z_y$ and $W_y^+$ 
to be renormalized in the following way:
\al{
Z_y^{(n)} \to \left( 1 + \frac{M_{{Z_y}_{(n)}}^2}{M_Z^2}
\right)^{-1/2} Z_y^{(n)}
\, ,
\qquad \qquad
W_y^{{(n)} {+}} \to \left( 1 + \frac{M_{{W_y}_{(n)}}^2}{M_W^2}
\right)^{-1/2} W_y^{{(n)} {+}}.
}
Note that $Z_y^{(n)}$ is the pseudoscalar Higgs state and
$W_y^{{(n)} {+}}$ is the charged {Higgs boson} {from} the $n$-th KK 
level {which has no level `0' counterpart.} In subsequent phenomenological discussions we use the more
transparent notations $A^{(n)^0}$ and $H^{(n)^+}$ for ${Z_y}^{(n)}$
and $W_y^{{(n)} {+}}$, respectively. Thus, up to KK level `1',
the Higgs spectrum consists of the following five Higgs bosons:
the SM (level `0') Higgs boson ($H$) and four Higgs states from level
`1', \emph{i.e.}, the neutral $CP$-even Higgs boson ($\kkhoneo$) which
is the level `1' excitation of the SM Higgs boson, the neutral 
$CP$-odd Higgs boson ($\kkhpone$) and the two charged Higgs bosons
$H^{(1)^\pm}$.
For the rest of the paper, we use a modified convention for the (KK) gluon to be $g^{(n)}$
instead of $G^{(n)}$ for convenience.

%
\subsection{The top quark sector}
\label{subsec:fermion}
We start with the following general framework for the fermion sector 
where, in addition to fermion BLKTs, we incorporate brane-local 
Yukawa terms (BLYTs):
\al{
S_{\text{quark}} &=
\int d^4 x \int_{-L}^{L} dy \sum_{i=1}^{3} \Bigg\{
        + i \overline{U'_i} \Gamma^M \mathcal{D}_M U'_i
        + r_{U_i} \Big( \delta(y-L) + \delta(y+L) \Big) \Big[ i
\overline{U'_i} \Gamma^\mu \mathcal{D}_\mu P_L U'_i \Big] \notag \\
&\phantom{=\int d^4 x \int_{-L}^{L} dy \sum_{i=1}^{3} \Bigg\{ \,\, }
        + i \overline{D'_i} \Gamma^M \mathcal{D}_M D'_i
        + r_{D_i} \Big( \delta(y-L) + \delta(y+L) \Big) \Big[ i
\overline{D'_i} \Gamma^\mu \mathcal{D}_\mu P_L D'_i \Big] \notag \\
&\phantom{=\int d^4 x \int_{-L}^{L} dy \sum_{i=1}^{3} \Bigg\{ \,\, }
        + i \overline{u'_i} \Gamma^M \mathcal{D}_M u'_i
        + r_{u_i} \Big( \delta(y-L) + \delta(y+L) \Big) \Big[ i
\overline{u'_i} \Gamma^\mu \mathcal{D}_\mu P_R u'_i \Big] \notag \\
&\phantom{=\int d^4 x \int_{-L}^{L} dy \sum_{i=1}^{3} \Bigg\{ \,\, }
        + i \overline{d'_i} \Gamma^M \mathcal{D}_M d'_i
        + r_{d_i} \Big( \delta(y-L) + \delta(y+L) \Big) \Big[ i
\overline{d'_i} \Gamma^\mu \mathcal{D}_\mu P_R d'_i \Big] \Bigg\},
\label{eqn:action-kinetic}
}
\al{
S_{\text{Yukawa}} &=
\int d^4 x \int_{-L}^{L} dy \sum_{i,j=1}^{3} \Bigg\{ - \Big( 1 + r_Y (
\delta(y-L) + \delta(y+L)) \Big)
\Big[ \hat{Y}^{u}_{ij} \overline{Q'_i} u'_j \tilde{\Phi} +
\hat{Y}^{d}_{ij} \overline{Q'_i} d'_j \Phi + \text{h.c.} \Big] \Bigg\},
\label{eqn:action-yukawa}
}
where $U'_i, D'_i, u'_i, d'_i$ correspond to the 5D $SU(2)_W$ up-doublet,
down-doublet, up-singlet and down-singlet of the $i$-th generation,
respectively and $Q'_i \,\equiv\,  (U'_i, D'_i)^{\text{T}}$ is the compact
notation used for the $i$-th 5D doublet. $r_{U_i}$ and $r_{u_i}$ are
the coefficients of the corresponding BLKTs. The field $\Phi$ 
represents the 5D Higgs scalar with 
$\tilde{\Phi} \,\equiv\, i \sigma_2 \Phi^\ast$, $\sigma_2$ being the second 
Pauli matrix. $r_Y$ is the {universal} coefficient for the brane-local 
Yukawa term.
We adopt the 5D Minkowski metric to be 
$\eta_{MN} = \text{diag}(1,-1,-1,-1,-1)$ and the representation of the
Clifford algebra is chosen {to be}
$\Gamma^{M} = \{\gamma^{\mu}, i \gamma_5 \}$.
The 4D chiral projectors for (4D) right/left-handed {states} are defined
following the standard convention \emph{i.e.}, 
$P_{R \atop L} = (1 \pm \gamma_5)/2$.
$\mathcal{D}_M$ stands for the 5D covariant derivative. 

In the presence of non-vanishing BLKT in the gauge sector 
(see equation~\ref{Higgsform}), 
the 5D VEV of $\Phi$ is given by
\al{
\langle \Phi \rangle = \begin{pmatrix} 0 \\ \frac{\hat{v}}{\sqrt{2}}
\end{pmatrix} =
\begin{pmatrix} 0 \\ \frac{v}{\sqrt{2}} \frac{1}{\sqrt{\pi R + 2r_{\text{EW}}}}
\end{pmatrix}
}
where $v=246$ GeV is the usual 4D Higgs VEV associated with the 
breaking of electroweak symmetry.
%
The 5D Yukawa couplings $ \hat{Y}^{u}_{ij},  \hat{Y}^{d}_{ij}$ are
related to their 4D counterparts $ {Y}^{u}_{ij},  {Y}^{d}_{ij}$ as
%
\al{
{Y}^{u/d}_{ij} = \frac{\hat{Y}^{u/d}_{ij}}{\sqrt{\pi R + 2r_{\text{EW}}}}.
}

%

The free part of $S_{\text{quark}}$ has already been discussed
in~\cite{Datta:2012tv} and hence we skip the details.
Using that we can KK-expand the mass terms in
$S_{\text{Yukawa}}$ as follows:
\al{
- \int d^4 x \sum_{i,j=1}^{3} \frac{v}{\sqrt{2}} \bigg\{
        Y^{u}_{ij} F^{u,(0,0)}_{ij} \overline{u'^{(0)}_{iL}}
u'^{(0)}_{jR} +
        Y^{d}_{ij} F^{d,(0,0)}_{ij} \overline{d'^{(0)}_{iL}}
d'^{(0)}_{jR} + \text{h.c.}
\bigg\},
		\label{zeromode_Yukawasector}
}
where, for simplicity, we only present the zero-mode part with 
fields redefined (to make them appear more conventional) as
$u'^{(0)}_{iL} \,\equiv\, U'^{(0)}_{iL}$, $d'^{(0)}_{iL} \,\equiv\, D'^{(0)}_{iL}$.
The fermionic mode functions for KK decomposition are described
in an appropriate context in section \ref{sec:massmix}.
The concrete forms of the factors $F^{u/d,(0,0)}_{ij}$ {(which arise from the mode functions of the $L$, $R$ type fields participating in equation~\ref{zeromode_Yukawasector})}
are
\al{
F^{u,(0,0)}_{ij} = \frac{2r_Y + \pi R}{\sqrt{2 r_{U_i} + \pi R}
\sqrt{2 r_{u_i} + \pi R}}, \quad
F^{d,(0,0)}_{ij} = \frac{2r_Y + \pi R}{\sqrt{2 r_{D_i} + \pi R}
\sqrt{2 r_{d_i} + \pi R}}.
\label{eqn:F-values}
}
The $3 \times 3$ matrices $Y^{u}_{ij} F^{u,(0,0)}_{ij}$ and
$Y^{d}_{ij} F^{d,(0,0)}_{ij}$ are diagonalized by the following
bi-unitary transformations
\al{
q'^{(0)}_{iR} = (U_{qR})_{ij} q^{(0)}_{jR}, \quad
q'^{(0)}_{iL} = (U_{qL})_{ij} q^{(0)}_{jL} \qquad (\text{for}\  q=u,
d),
\label{eqn:flavor_rotating_matrices}
}
as follows:
\al{
- \int d^4 x \sum_{i=1}^{3} \frac{v}{\sqrt{2}} \bigg\{
        \mathcal{Y}^{u}_{ii} \overline{u^{(0)}_{iL}} u^{(0)}_{iR} +
        \mathcal{Y}^{d}_{ii} \overline{d^{(0)}_{iL}} d^{(0)}_{iR} +
\text{h.c.} \; \text{(+ KK excitations)}
\bigg\},
}
where $\mathcal{Y}^{u}_{ii}$ and $\mathcal{Y}^{d}_{ii}$ are the
diagonalized Yukawa couplings for up and down quarks, respectively.
We discuss later in this paper that the diagonalized values do not
directly correspond to those in the SM due to level mixing effects.
Also, from now on, we would 
consider universal values of the BLKT
parameters $r_Q$ for the quarks from the first two generations and
$r_T$ for those from the third generation replacing the {many} different ones 
appearing in equation
\ref{eqn:action-kinetic}. We will see later, this provides 
us with a separate handle (modulo some constraints from experiments) on
the top quark sector of the nmUED scenario under consideration.
Further, this simplifies the expressions in equation 
\ref{eqn:F-values}.
%

\section{Mixings, masses and effective couplings}
\label{sec:massmix}

Mixings in the fermion sector, quite generically, could have 
{interesting} implications as these affect both couplings and the 
spectra of the concerned excitations.
Fermions with a certain flavor from a given KK level and belonging to
$SU(2)_W$ doublet and singlet representations always mix once the
electroweak symmetry is broken. 
Presence of BLTs affects such a mixing
at every KK level. On top of this, the dynamics driven by the BLTs 
allows for mixing of fermions from different KK levels that have the 
same KK-parity. Both kinds of mixings are 
proportional to the {Yukawa} mass of the fermion in reference and thus,
are pronounced for the top quark sector.

As pointed out in the introduction, since \emph{level-mixing} 
among the even KK-parity top quarks involves the SM top quark (from 
level `0'), this naturally evokes a reasonable curiosity about its 
consequences and it is indeed found to give rise 
to interesting phenomenological possibilities.
However, {the} phenomenon draws significant
constraints from experiments which we will discuss in some detail. 
We restrict ourselves to the mixing of level `0'-level '2' KK top
quarks ignoring all higher even KK states the
effects of which would be suppressed by their increasing masses.
Also, we do not consider the effects of level-mixings among KK states
from levels with $n>0$, {including say,} those among the excitations from levels
with odd KK-parity. Generally, these could be appreciable.
However, in contrast to the case where SM excitations mix with higher
KK levels, these would only entail details within a sector yet 
to be discovered.

\subsection{Mixing in level `1' top quark sector}
\label{subsec:massmix1}
We first briefly recount \cite{Datta:2012tv} the mixing of the top 
quarks at KK level `1'. In presence of BLTs, the Yukawa part of the
action embodying the mass-matrix is of the form
\al{
- \int d^4 x \Bigg\{
\begin{bmatrix} \overline{T}^{(1)}, \ \overline{t}^{(1)} \end{bmatrix}_L
\begin{bmatrix}
                M_{T_{(1)}} & r'_{T11} \mtopin \\
                -R'_{T11} \mtopin & M_{T_{(1)}}
\end{bmatrix}
\begin{bmatrix} T^{(1)} \\ t^{(1)} \end{bmatrix}_R
+ \text{h.c.} \Bigg\},
\label{1stKKupmasses}
}
with ``input" top mass $\mtopin$
{(which is an additional free parameter in our scenario) and}
\al{
r'_{T11} &= \frac{1}{R_{T00}} \int_{-L}^{L} dy \Big( 1+ r_Y \paren{\delta(y-L) + \delta(y+L)} \Big)
                f_{T_{(1)}}^2 \notag \\
&= \frac{2 r_T + \pi R}{2 r_Y + \pi R} \times
\frac{2r_Y + \frac{1}{S_{T_{(1)}}^2} \left[ \frac{\pi R}{2} - \frac{1}{2M_{T_{(1)}}} \sin(M_{T_{(1)}} \pi R) \right] }
                {2r_T + \frac{1}{S_{T_{(1)}}^2} \left[ \frac{\pi R}{2} - \frac{1}{2M_{T_{(1)}}} \sin(M_{T_{(1)}} \pi R) \right]}, \\
R'_{T11} &= \frac{1}{R_{T00}} \int_{-L}^{L} dy \Big( 1+ r_Y \paren{\delta(y-L) + \delta(y+L)} \Big)
                g_{T_{(1)}}^2 \notag  \\
&= \frac{2 r_T + \pi R}{2 r_Y + \pi R} \times
\frac{2r_Y ( C_{T_{(1)}}/S_{T_{(1)}} )^2 + \frac{1}{S_{T_{(1)}}^2} \left[ \frac{\pi R}{2} + \frac{1}{2M_{T_{(1)}}} \sin(M_{T_{(1)}} \pi R) \right]}
{\frac{1}{S_{T_{(1)}}^2} \left[ \frac{\pi R}{2} + \frac{1}{2M_{T_{(1)}}} \sin(M_{T_{(1)}} \pi R) \right]}
}
{where $R_{T00}$ is given by}
\al{
R_{T00} &= \int_{-L}^{L} dy \Big( 1+ r_Y \paren{\delta(y-L) + \delta(y+L)} \Big)
                f_{T_{(0)}}^2 = \frac{2 r_Y + \pi R}{2 r_T + \pi R}.
		\label{Fromof_RT00}
}
$f_{T_{(n)}}$ and $g_{_{T_{(n)}}}$ 
represent the mode functions for $n$-th KK level and are given by 
\cite{Datta:2012tv}:
\al{
f_{T_{(n)}} \; \equiv \; f_{T_{(n)L}} = f_{t_{(n)R}}  &=
N_{T_{(n)}} \times
                \begin{cases}
                \displaystyle \frac{\cos(M_{T_{(n)}} y)}{C_{T_{(n)}}}
& \text{for $n$ even,} \\
                \displaystyle \frac{{-}\sin(M_{T_{(n)}}
y)}{S_{T_{(n)}}}  & \text{for $n$ odd,}
                \end{cases}  \label{eqn:f-profiles} \\
g_{_{T_{(n)}}} \; \equiv \; f_{T_{(n)R}} = -f_{t_{(n)L}} &=
N_{T_{(n)}} \times
                \begin{cases}
                \displaystyle \frac{\sin(M_{T_{(n)}} y)}{C_{T_{(n)}}}
& \text{for $n$ even,} \\
                \displaystyle \frac{\cos(M_{T_{(n)}} y)}{S_{T_{(n)}}}
& \text{for $n$ odd}
                \end{cases} \label{eqn:g-profiles}
}
with 
\al{
C_{T_{(n)}} =\cos\paren{\frac{M_{T_{(n)}}\pi R}{2}},\quad
S_{T_{(n)}} =\sin\paren{\frac{M_{T_{(n)}}\pi R}{2}}
\label{trigabbrevi}
}
and the normalization factors $N_{T_{(n)}}$ for the mode functions 
{are given by}
\al{
N_{T_{(n)}}^{-2} =
\begin{cases}
\displaystyle 2r_T + \frac{1}{C_{T_{(n)}}^2} \left[ \frac{\pi R}{2} +
\frac{1}{2M_{T_{(n)}}} \sin(M_{T_{(n)}} \pi R) \right] & \text{for $n$
even,} \\
\displaystyle 2r_T + \frac{1}{S_{T_{(n)}}^2} \left[ \frac{\pi R}{2} -
\frac{1}{2M_{T_{(n)}}} \sin(M_{T_{(n)}} \pi R) \right] & \text{for $n$
odd.}
\end{cases}
\label{Psinormalization}
}
The KK {mass} $M_{T_{(n)}}$ for the {`$n$'-th level} top quark {excitation follows
from equation}~\ref{masscondition_simplified} where chiral zero 
modes occur.\footnote{{Here, we consider a situation where the fields $T^{(1)}_{L,R}$ and $t^{(1)}_{L,R}$ are rotated by the same matrices $U_{qR}$ and $U_{qL}$ (of equation~\ref{eqn:flavor_rotating_matrices}) from the basis used in equations~\ref{eqn:action-kinetic} and \ref{eqn:action-yukawa}. We ignore the diagonal and non-diagonal modifications in the boundary conditions.
In our scenario, these modifications are Cabibbo-suppressed (see equation~\ref{eqn:gotomassbases}) and hence such a treatment is justified.}}
%
Note that the off-diagonal terms are asymmetric and pick up nontrivial 
multiplicative factors.
This is because two different mode functions, $f_{T_{(n)}}$ and
$g_{_{T_{(n)}}}$ (associated with the specific states with 
particular chiralities and gauge quantum numbers){,} contribute to
them. On the other hand, the diagonal KK mass terms are now 
solutions of the appropriate transcendental equations. 
{When expanded, the diagonal entries of the mixing matrix
involve} the $L$ and $R$ components of the 
same gauge multiplet ($T$ from $SU(2)_W$ doublet or $t$ from $SU(2)_W$
singlet). In contrast, the off-diagonal entries are of Yukawa-origin
(signalled by the presence of $\mtopin$) and involve both 
$r_T$ and $r_Y$. 
These terms represent the conventional Dirac mass-terms as they 
connect the $L$ and the $R$ components belonging to two 
different multiplets.
It may be noted that even when either $r_T$ or $r_Y$ vanishes,
the mixing remains nontrivial.
Only the case with $r_T$\,$=$\,$r_Y$\,$=$\,$0$ trivially reduces to the (tree-level) mUED.

The mass matrix of equation \ref{1stKKupmasses} can be diagonalized by
bi-unitary transformation with the matrices $V_{tL}^{(1)}$ and
$V_{tR}^{(1)}$ where
\al{
\begin{bmatrix} T^{(1)} \\ t^{(1)} \end{bmatrix}_L
= V^{(1)}_{tL}
\begin{bmatrix} t^{(1)}_{l} \\ t^{(1)}_{h} \end{bmatrix}_L, \quad
\begin{bmatrix} T^{(1)} \\ t^{(1)} \end{bmatrix}_R
= V^{(1)}_{tR}
\begin{bmatrix} t^{(1)}_{l} \\ t^{(1)}_{h} \end{bmatrix}_R.
\label{eqn:Vmats}
}
{Then,} equation \ref{1stKKupmasses} takes the diagonal form 
\al{
- \int d^4 x
\begin{bmatrix} \overline{t}^{(1)}_{l}, \ \overline{t}^{(1)}_{h} \end{bmatrix}
\begin{bmatrix}
                m_{t^{(1)}_{l}} &  \\
                & m_{t^{(1)}_{h}}
\end{bmatrix}
\begin{bmatrix} t^{(1)}_{l} \\ t^{(1)}_{h} \end{bmatrix}
\label{eqn:diogonal}
}
where $t^{(1)}_{l}, \; t^{(1)}_{h}$ are the level
`1' top quark mass eigenstates and $(m_{t^{(1)}_{l}})^2$ and 
$(m_{t^{(1)}_{l}})^2$ are the mass-eigenvalues of the squared 
mass-matrix with $m_{t^{(1)}_{h}} > m_{t^{(1)}_{l}}$. 
Note that, for clarity and convenience, we have modified 
{the} notations and {the} ordering of the states in the presentations above 
{from} what appear in ref.~\cite{Datta:2012tv}.

\subsection{Mixing among level `0' and level `2' top quark states}
\label{subsec:massmix2}
The formulation described above can be extended in a straight-forward 
manner for the level `2' KK top quarks when this sector is augmented 
by the level `0' (SM) top quark. Thus, the mass-matrix for the even 
KK parity top quark sector (keeping only level `0' and level `2' KK 
excitations) takes the following form:
%
\al{
- \int d^4 x \Bigg\{
\begin{bmatrix} \overline{t^{(0)}}, \ \overline{T}^{(2)}, \ \overline{t}^{(2)} \end{bmatrix}_L
\begin{bmatrix}
                \mtopin & 0 & \mtopin R'_{T02} \\
                \mtopin R'_{T02} & M_{T_{(2)}} & \mtopin r'_{T22}  \\
                0 & - \mtopin R'_{T22}  & M_{T_{(2)}}
\end{bmatrix}
\begin{bmatrix} t^{(0)} \\ T^{(2)} \\ t^{(2)} \end{bmatrix}_R
+ \text{h.c.} \Bigg\}
\label{massmatrixwithmixing}
}
where $r'_{T22}$, $R'_{T22}$, $R'_{T02}$ are defined as follows, in a
way similar to the case for level `1' top quarks:
\al{
r'_{T22} &= \frac{1}{R_{T00}} \int_{-L}^{L} dy \Big( 1+ r_Y \paren{\delta(y-L) + \delta(y+L)} \Big)
                f_{T_{(2)}}^2 \notag \\
&= \frac{2 r_T + \pi R}{2 r_Y + \pi R} \times \frac{2r_Y + \frac{1}{C_{T_{(2)}}^2} \left[ \frac{\pi R}{2} + \frac{1}{2M_{T_{(2)}}} \sin(M_{T_{(2)}} \pi R) \right]}
                {2r_T + \frac{1}{C_{T_{(2)}}^2} \left[ \frac{\pi R}{2} + \frac{1}{2M_{T_{(2)}}} \sin(M_{T_{(2)}} \pi R) \right]}, \\
R'_{T22} &= \frac{1}{R_{T00}} \int_{-L}^{L} dy \Big( 1+ r_Y \paren{\delta(y-L) + \delta(y+L)} \Big)
                g_{T_{(2)}}^2 \notag  \\
&= \frac{2 r_T + \pi R}{2 r_Y + \pi R} \times \frac{2r_Y ( S_{T_{(2)}}/C_{T_{(2)}} )^2 + \frac{1}{C_{T_{(2)}}^2} \left[ \frac{\pi R}{2} - \frac{1}{2M_{T_{(2)}}} \sin(M_{T_{(2)}} \pi R) \right]}
{\frac{1}{C_{T_{(2)}}^2} \left[ \frac{\pi R}{2} - \frac{1}{2M_{T_{(2)}}} \sin(M_{T_{(2)}} \pi R) \right]}, \\
R'_{T02} &= \frac{1}{R_{T00}} \int_{-L}^{L} dy \Big( 1+ r_Y \paren{\delta(y-L) + \delta(y+L)} \Big)
                f_{T_{(0)}} f_{T_{(2)}} \notag \\
&= \frac{2 r_T + \pi R}{2 r_Y + \pi R} \times
                \frac{2r_Y + 2(S_{T_{(2)}}/M_{T_{(2)}} C_{T_{(2)}}) }
                {\sqrt{2r_T + \pi R} \sqrt{2r_T + \frac{1}{C_{T_{(2)}}^2} \left[ \frac{\pi R}{2} + \frac{1}{2M_{T_{(2)}}} \sin(M_{T_{(2)}} \pi R) \right]} },
}
{with $R_{T00}$ given by equation~\ref{Fromof_RT00}.}
The lower $2 \times 2$ block of the mass-matrix in equation 
\ref{massmatrixwithmixing} is reminiscent of the level `1' top quark
mass-matrix of equation \ref{1stKKupmasses}. 
{Beyond this,} the
mass-matrix contains as the first diagonal element the `input' top
quark mass, $\mtopin$ and two other non-vanishing off-diagonal elements as the
13 and 21 elements. Obviously, the latter two play direct roles in 
the mixings of the level `0' and level `2' top quarks. Note that all
the off-diagonal terms of the mass-matrix are proportional to $\mtopin$
which is clearly indicative of their origins in the Yukawa sector.
The zeros in turn reflect $SU(2)_W$ invariance.

Diagonalization of this $3 \times 3$ mass-matrix yields the physical 
states (3 of them) along with their mass-eigenvalues. Thus, the level 
`0' top quark (\emph{i.e.}, the SM top quark) ceases to be a physical 
state and mixes with the level `2' top states. Given the rather 
involved structure of the mass-matrix, neither is it possible to 
express the eigenvalues analytically in a compact way nor they would 
be much illuminating theoretically. We, thus, diagonalize the 
mass-matrix numerically.
Similar to the case of the level `1' states, we adopt the 
following conventions:
\al{
\begin{bmatrix}
	t^{(0)} \\ T^{(2)} \\ t^{(2)}
\end{bmatrix}_L
= V_{tL}^{(2)}
\begin{bmatrix}
	t \\ t^{(2)}_{l} \\ t^{(2)}_{h}
\end{bmatrix}_L, \quad
\begin{bmatrix}
	t^{(0)} \\ T^{(2)} \\ t^{(2)}
\end{bmatrix}_R
= V_{tR}^{(2)}
\begin{bmatrix}
	t \\ t^{(2)}_{l} \\ t^{(2)}_{h}
\end{bmatrix}_R
}
with the physical masses $\mtopphys$, $m_{t^{(2)}_{l}}$ and 
$m_{t^{(2)}_{h}}$ and with the ordering
$\mtopphys < m_{t^{(2)}_{l}} < m_{t^{(2)}_{h}}$.
%

%
\subsection{Quantitative estimates
\label{sec:Quantitative_estimates}}
%

As can be seen from the equations above, the free parameters of the 
top-quark sector in the nmUED scenario under consideration are $R$, 
$r_T$ and $r_Y$. For the latter two, we use \cite{Datta:2012tv} the 
dimensionless quantities $\rtp$ and $\ryp$ 
where $\rtp = r_T \rinv$ and $\ryp = r_Y \rinv$. In addition, $\mtopin$ 
serves as an extra free parameter from the SM sector. 
%
%
\subsubsection{Top quark masses}
\label{subsubsec:mass-quant}
%
\begin{figure}[t]
\centering
\includegraphics[width=0.3\columnwidth]{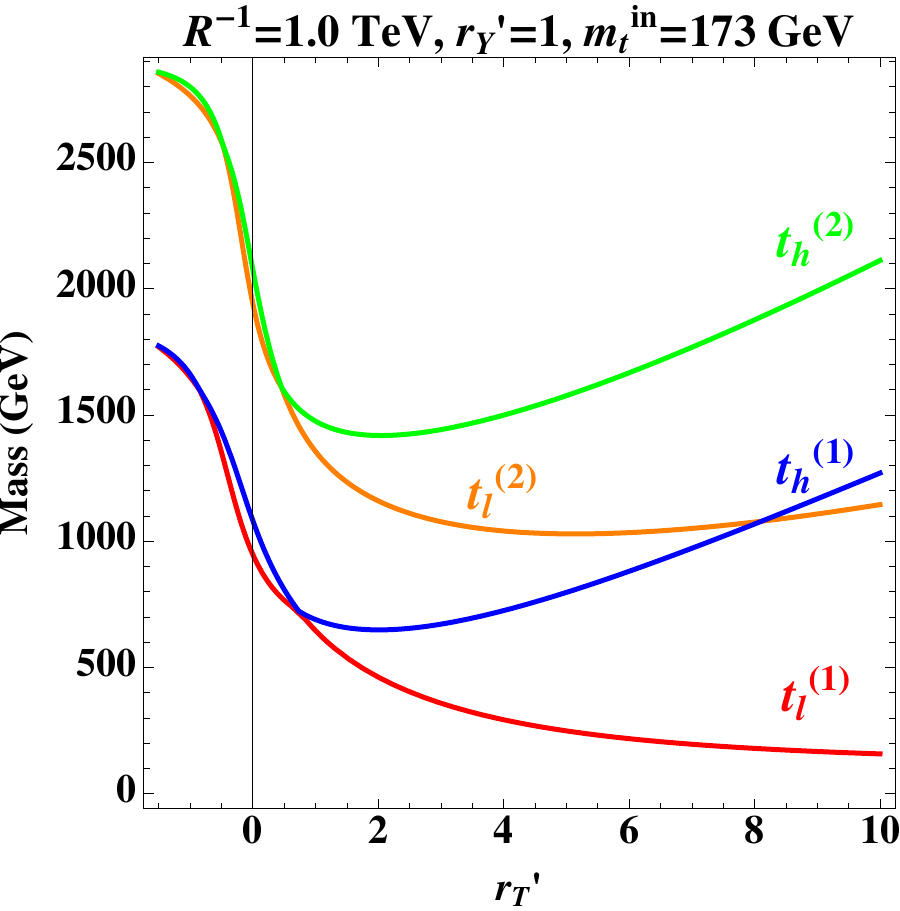}
\hspace*{0.15in}
\includegraphics[width=0.3\columnwidth]{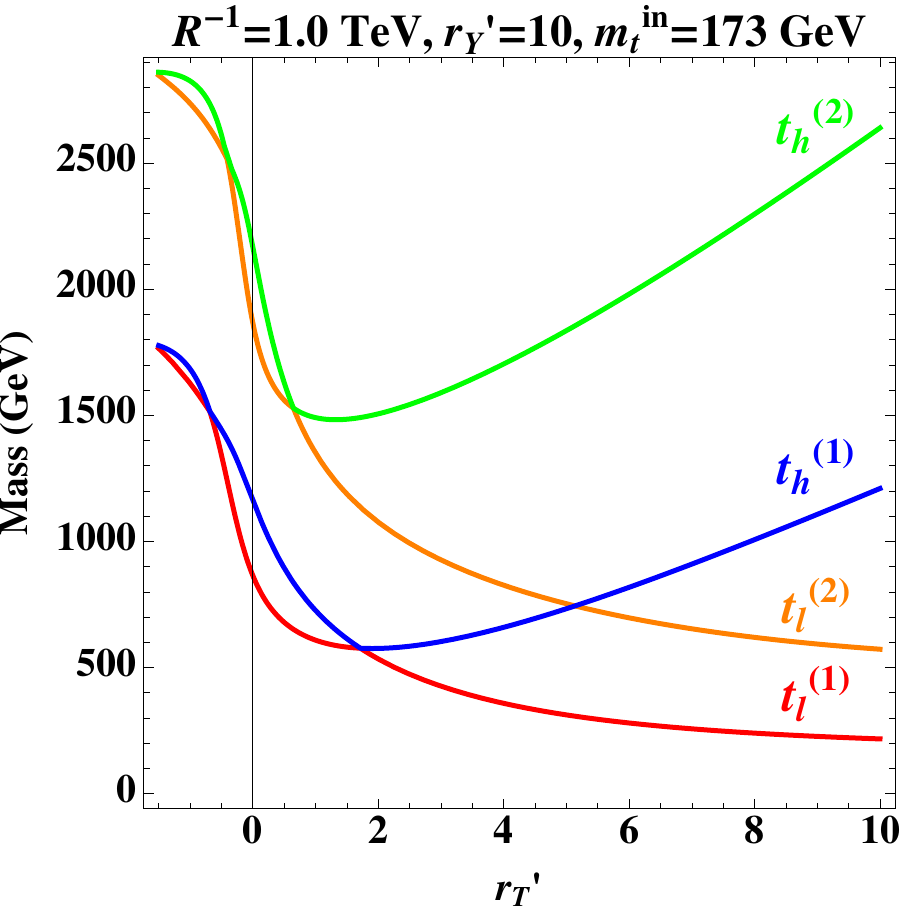}
\hspace*{0.15in}
\includegraphics[width=0.3\columnwidth]{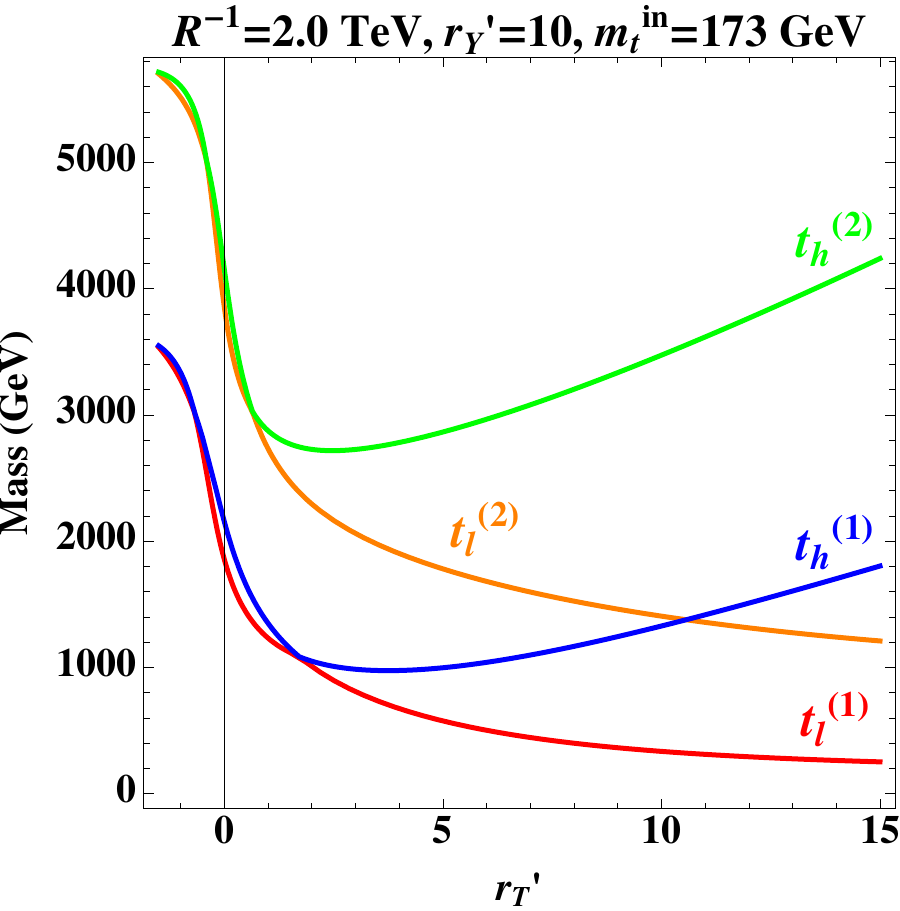}
\caption{Masses of level `1' and {level} `2' KK top quarks as functions of 
$\rtp$ for given $\ryp$ and $\rinv$ with $\mtopin = 173\,\text{GeV}$.}
\label{fig:top-masses}
\end{figure}
%

In figure \ref{fig:top-masses} we illustrate the variations of the masses,
as functions of $\rtp$, of the two KK top quark eigenstates from level 
`1' and the two heavier mass eigenstates that result from the mixing 
of level `0' and level `2'. The plot in the middle, when compared to
the one in the left, demonstrates how the spectrum changes as $\ryp$
varies with $\rinv$ held fixed.
We set the input top mass $\mtopin$ to $173\,\text{GeV}$ in all the plots 
{of} figure~\ref{fig:top-masses}.
In turn, the effect of changing $\rinv$ 
can be seen as one goes from the plot in the middle ($\rinv=1$ TeV) to 
the one on the right ($\rinv=2$ TeV).
An interesting feature common to all these plots is that there is a 
cross-over of the curves for $\mtoneh$ and 
$\mttwol$, \emph{i.e.}, as a function of $\rtp$, at some point, the 
lighter of the mixed level `2' state top quark eigenstates becomes 
less massive compared to the heavier of the level `1' KK top quark 
eigenstate. The cross-overs take place at smaller values of $\rtp$ 
when $\ryp$ is increased for a given $\rinv$ and at larger values of 
$\rtp$ when $\rinv$ is increased with $\ryp$ held fixed. Accordingly, 
the mass-values at those flipping points also go down or up,
respectively. Here, the dominant role is being played by the 
`chiral mixing' while \emph{level-mixing} is unlikely to have much
bearing.
{These plots also
reveal that} achieving a `flipped-spectrum' (in the above sense) is 
difficult if one requires the light level `1' KK top quark to be 
heavier than about 400 GeV.  Nonetheless, the overall trend could 
provide easier reach for a KK top quark from level `2' at the LHC. 
Thus, it may be possible for up to three excited top quark 
states ($\mtonel, \, \mtoneh, \, \mttwol$) to pop up at the LHC.
\subsubsection{Top quark mixings}
\label{subsubsec:mix-quant}
In this subsection we take a quantitative look at the mixings in the
top quark sector from the first KK level discussed earlier in section
\ref{subsec:massmix1}. The mixing is known to be near-maximal in the
case of quarks (fermions) from the lighter generations
\cite{Datta:2012tv}.
Deviations from such maximal mixings occur in the top quark sector
due to its nontrivial structure\footnote{This is in direct contrast
with competing SUSY scenarios where mixings in the light sfermion
sector are always negligible while for top squark sector it could
attain the maximal value.}. Such mixings are expected to follow
similar trends at level `2' (and higher) KK levels and hence 
we do not present {them}
separately. However, some deviations are expected in the presence of
\emph{level-mixings} which can at best be modest for the case of
$t^{(0)}-t^{(2)}$ system that we focus on in this work.
%
\begin{figure}[t]
\centering
\includegraphics[width=0.4\columnwidth]{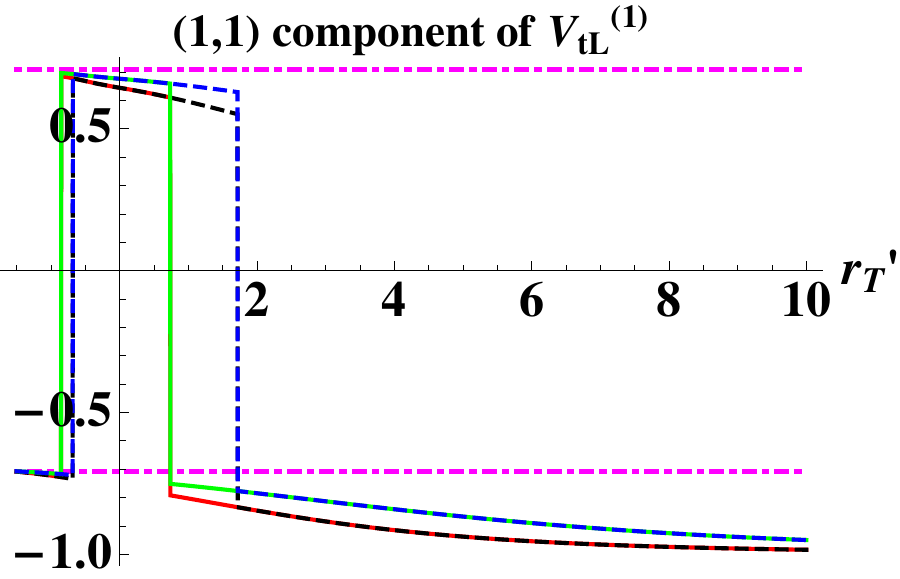}
\hspace*{0.1in}
\includegraphics[width=0.4\columnwidth]{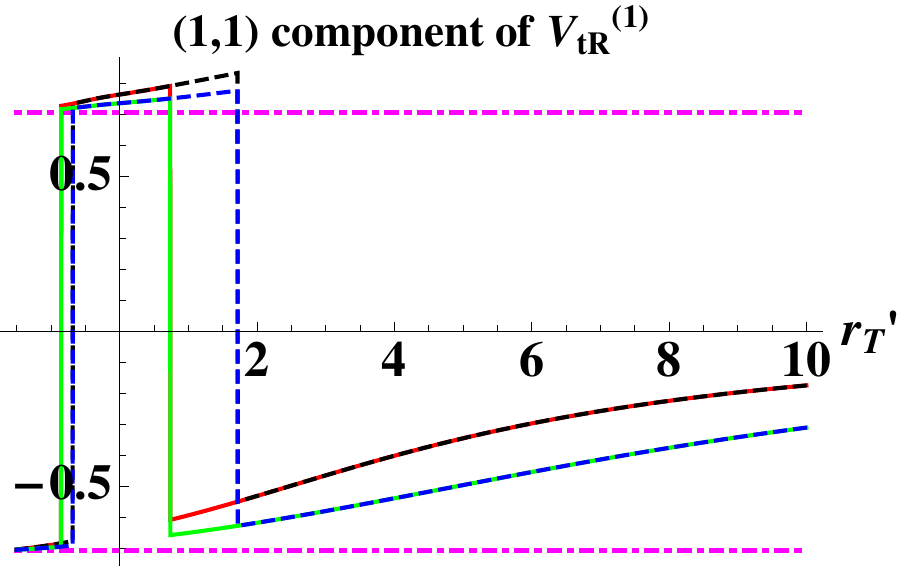}
\caption{Variations of the (1,1) elements of the matrices $V^{(1)}_{tL}$ (left) and 
$V^{(1)}_{tL}$ (right) as functions of $\rtp$ for fixed set of values
of $\rinv$ and $\ryp$. Conventions used  for
different sets of $\rinv$ and $\ryp$ values are:
bold red for $\rinv = 1$ TeV and $\ryp = 1$,
dashed black for  $\rinv = 1$ TeV and $\ryp = 10$,
bold green for $\rinv = 2$ TeV and $\ryp = 1$ and
dashed blue for $\rinv = 2$ TeV and $\ryp = 10$.
}
\label{fig:mix11}
\end{figure}
%
\begin{figure}[t]
\centering
\includegraphics[width=0.4\columnwidth]{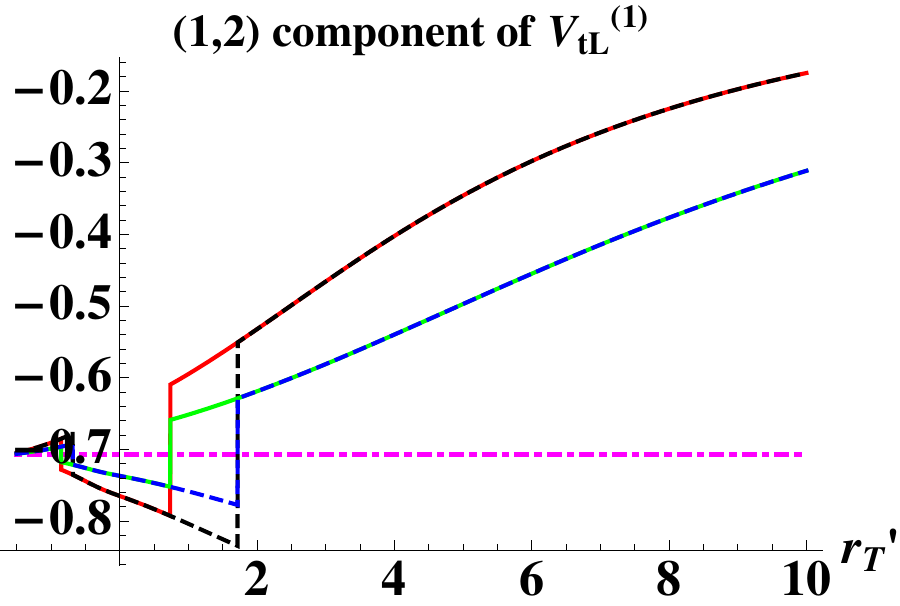}
\hspace*{0.1in}
\includegraphics[width=0.4\columnwidth]{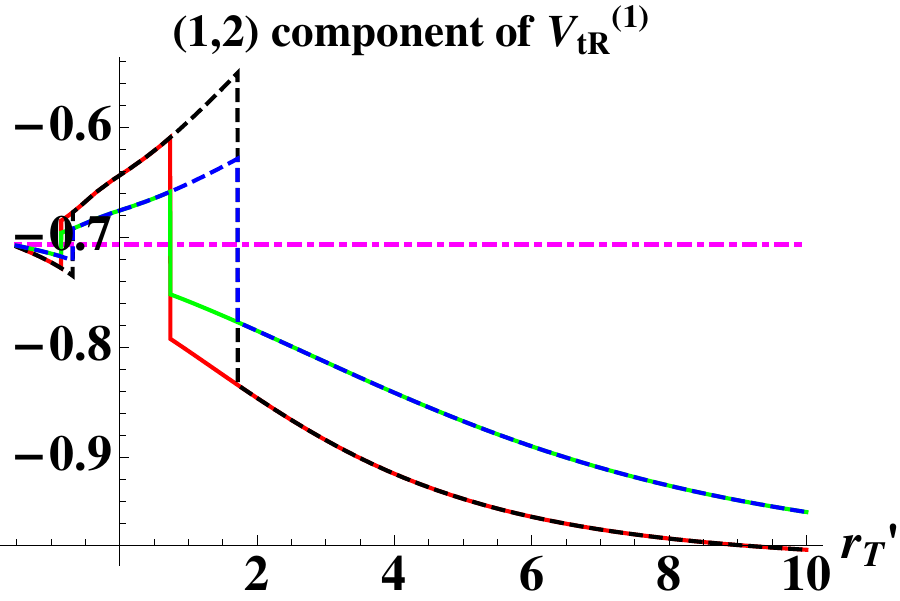}
\caption{Same as in figure \ref{fig:mix11} but for the variations of the 
(1,2) elements of the matrices $V^{(1)}_{tL}$ (left) and 
$V^{(1)}_{tL}$ (right). The respective (2,1) elements can be
obtained from the orthogonality of these matrices.
}
\label{fig:mix12}
\end{figure}
%
%

The elements of the $V$-matrices in equation \ref{eqn:Vmats} give the
admixtures of different participating states in the KK top quark
eigenstates. To be precise, $V^{(1)}_{{tL}_{(1,1)}}$ and
$V^{(1)}_{{tL}_{(2,2)}}$ represent the admixture of $T_L^{(1)}$
in $t^{(1)}_{lL}$ and $t_L^{(1)}$ in
$t^{(1)}_{hL}$ respectively while
$V^{(1)}_{{tL}_{(1,2)}}$ and $V^{(1)}_{{tL}_{(2,1)}}$ indicate
the same for $t_L^{(1)}$ in $t^{(1)}_{lL}$ and
$T_L^{(1)}$ in $t^{(1)}_{hL}$ {in that order}. Similar
descriptions hold for the $V_R^{(1)}$ matrix.
In figures \ref{fig:mix11} and \ref{fig:mix12} we illustrate the
deviations from maximal mixing {in the level `1' top quark sector}
in terms of these components of the $V$ matrices as functions 
of $\rtp$. Each figure contains multiples curves which present
situations for different combinations of $\rinv$ and $\ryp$
(see the captions for details). 
Note that the abrupt changes in sign of the mixings that happen
between $-1< \rtp < 2$ can be understood in terms of
the trends of the red and blue curves in figure 
\ref{fig:top-masses} (the blue curves smoothly evolve {to} the red ones
and vice-versa).

The flat, broken magenta lines indicate maximal mixing 
($|V^{(1)}_{{tL}_{(1,1)}}| = |V^{(1)}_{{tL}_{(1,2)}}| = 1/\sqrt{2}$).
It is clear from these figures
that there can be appreciable deviations from maximal mixing in all
these cases. As can be seen, the effects are bigger for larger values of
$\rtp$ and smaller $\rinv$. Some dependence {on} $\ryp$ is observed
for smaller values of $\rtp$. However, it is to be kept in mind that
the effective deviations arise from the interplay of these elements which
is {again} neither easy to present nor much illuminating.

\subsection{Effective couplings}
\label{subsec:couplings}
As mentioned earlier, not only masses undergo modifications in the
presence of BLTs but also the wavefunctions get {distorted}. The latter 
affects the couplings through the overlap integrals. These are 
integrals over the extra dimension of a product of mode functions of 
the states that appear at a given interaction vertex. 
In this section we briefly discuss the generic properties of some of
these overlap integrals which play roles in the present study. 
Assuming the wavefunctions to be real, the general form of the 
multiplicative factor that scales the corresponding SM
coupling strengths is given by
\al{
g_{f_i^{(l)} f_j^{(m)} f_k^{(n)}} =
\mathcal{N}_{ijk}
\int_{-L}^{L} dy  \Big[ 1 + r_{ijk}^{(l,m,n)} \left( \delta(y-L) +
\delta(y+L) \right) \Big] f_i^{(l)}(y) f_j^{(m)}(y) f_k^{(n)}(y)
\label{eqn:overlap1}
}
where $i,j,k$ represent different interacting fields and $f_i^{(l)},
f_j^{(m)}, f_k^{(n)}$ 
are the corresponding mode functions with the KK
indices $l,m,n$, respectively, as defined in sections 
\ref{subsec:gauge-higgs}, \ref{subsec:massmix1} and \ref{subsec:massmix2}. 
The factor $r_{ijk}^{(l,m,n)}$ stands for relevant BLT parameter(s)
while the
normalization factor $\mathcal{N}_{ijk}$ is suitably chosen to 
recover the SM vertices when $l$=$m$=$n$=0
{(except for the Yukawa sector of the nmUED scenario under
consideration)}.

The key to understand the general structure is the flatness of the
zero-mode ($n=0$) profiles in our minimal configuration. For these,
the factor takes the following form:
\al{
g_{f_i^{(l)} f_j^{(m)} f_k^{(0)}} =
\mathcal{N}_{ijk} f_k^{(0)}
\int_{-L}^{L} dy \Big[ 1 + r_{ijk}^{(l,m,0)} \left( \delta(y-L) +
\delta(y+L) \right) \Big] f_i^{(l)}(y) f_j^{(m)}(y),
\label{eqn:overlap2}
}
where we see the zero-mode field has been taken out of the integral
in equation \ref{eqn:overlap1}. 
For $i=j$, the overlap integral 
reduces to Kronecker's delta function, $\delta_{l,m}$ and the overall 
strength {turns out to be} identically equal to 1.
Orthonormality of the involved states constrains the possibilities.
In table \ref{tab:overlap} we collect some of these interactions
and group them in terms of their effective strengths (given by equation
\ref{eqn:overlap2}). This list, in particular, the set {of couplings} in the third
column, is not exhaustive and presented for demonstrative purposes 
only.

In addition to these, mixings in the top quark sector in the form of
both {chiral mixing and \emph{level-mixing}} play roles in determining
the effective couplings. In this subsection we briefly discuss such 
effects on some of the important interaction-vertices involving 
the top quarks, the gauge and the Higgs bosons from different KK
levels.
{As in section~\ref{sec:Quantitative_estimates}, we further introduce the dimensionless
parameters $\rew\,(=R^{-1} r_{\text{EW}})$, $\rqp\,(=R^{-1} r_Q)$ and $\rgp\,(=R^{-1} r_{G})$
replacing $r_{\text{EW}}\,(=r_H)$, $r_Q$ and $r_G$, the BLKT parameters for the electroweak gauge boson and Higgs sectors, the first two generation quark sector and the gluon sector, respectively.}
In addition, we also introduce a corresponding universal parameter $r_L$
for the lepton sector which
we will use in section~\ref{subsec:precision}.
Later, in section \ref{sec:pheno}, we will refer back to this 
discussion in the context of phenomenological analyses of the 
scenario. 

\begin{table}[t]
\begin{center}
\begin{tabular}{|c|c|c|}
\hline
  & & \\
  &
  & $Q_{R/L}^{(1)} - {V}^{(1)} - Q_{R/L}^{(0)}$ \\  
  & $V^{(2)} - V^{(2)} - V^{(0)}$
  & $q_{R/L}^{(1)} - {V}^{(1)} - q_{R/L}^{(0)}$ \\
  & $V^{(1)} - V^{(1)} - V^{(0)}$
  & $Q_{R/L}^{(0)} - {V}^{(2)} - Q_{R/L}^{(0)}$ \\ 
  & $Q_{R/L}^{(1)} - {V}^{(0)} - Q_{R/L}^{(1)}$ 
  & $Q_{L}^{(1)} - {H}^{(0)} - q_{R}^{(1)}$ \\ 
    $Q_{R/L}^{(2)} - {V}^{(0)} - Q_{R/L}^{(0)}$
  & $q_{R/L}^{(1)} - {V}^{(0)} - q_{R/L}^{(1)}$
  & $Q_{L}^{(2)} - {H}^{(0)} - q_{R}^{(0)}$ \\
    $q_{R/L}^{(2)} - {V}^{(0)} - q_{R/L}^{(0)}$
  & $Q_{R/L}^{(2)} - {V}^{(0)} - Q_{R/L}^{(2)}$
  & $Q_{L}^{(0)} - {H}^{(2)} - q_{R}^{(0)}$ \\
    ${V}^{(2)} - {V}^{(0)} - {V}^{(0)}$
  & $q_{R/L}^{(2)} - {V}^{(0)} - q_{R/L}^{(2)}$
  & $Q_{L}^{(0)} - {H}^{(0)} - q_{R}^{(2)}$ \\
   &   &   \\
\hline
 0 & 1 & non-zero  \\
\hline
\end{tabular}
\caption{Classes of different effective (tree level) couplings
(given by equation \ref{eqn:overlap2})
involving the gauge boson ($V$), Higgs ($H$) and the left- and 
right-handed, $SU(2)_W$ doublet ($Q$) and singlet ($q$) quark excitations and their 
relative strengths (shown in the last row)
compared to the {corresponding SM} cases. 
}
\label{tab:overlap}
\end{center}
\end{table}

\subsubsection{Effective couplings {involving the} gauge bosons}
\label{subsubsec:gauge-couplings}
\begin{figure}[t]
\centering
\vspace*{0.1in}
\includegraphics[width=0.4\columnwidth]{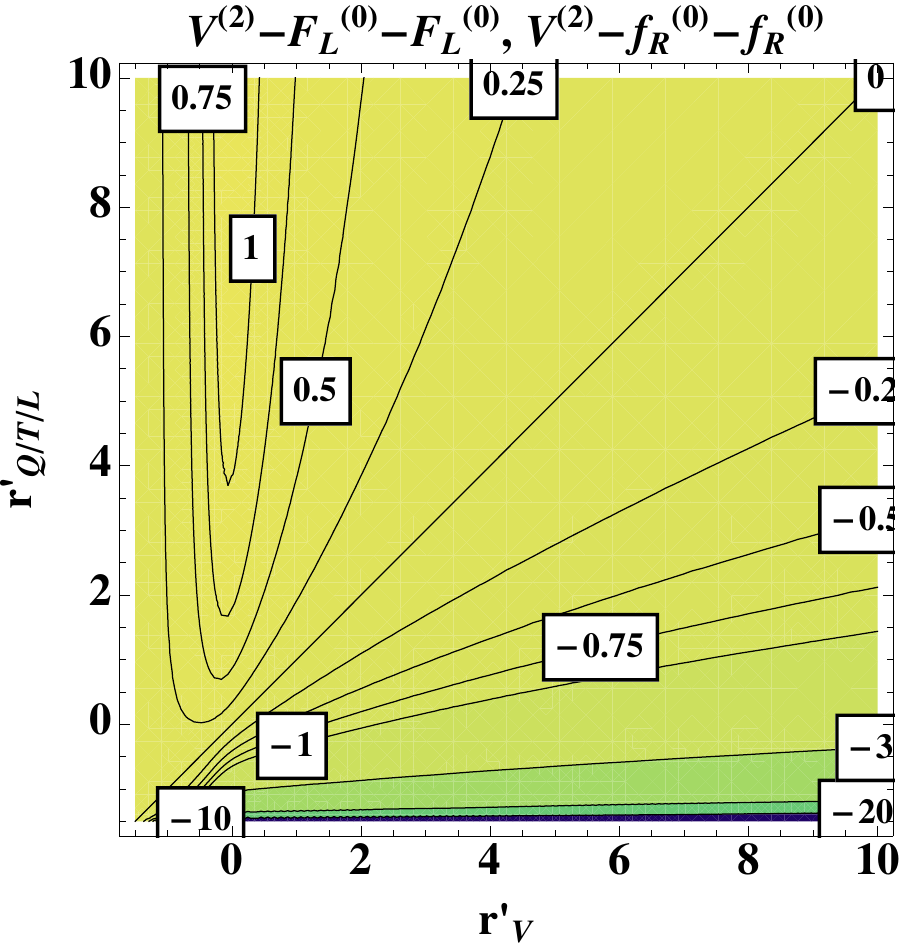}
\hspace*{0.5in}
\includegraphics[width=0.4\columnwidth]{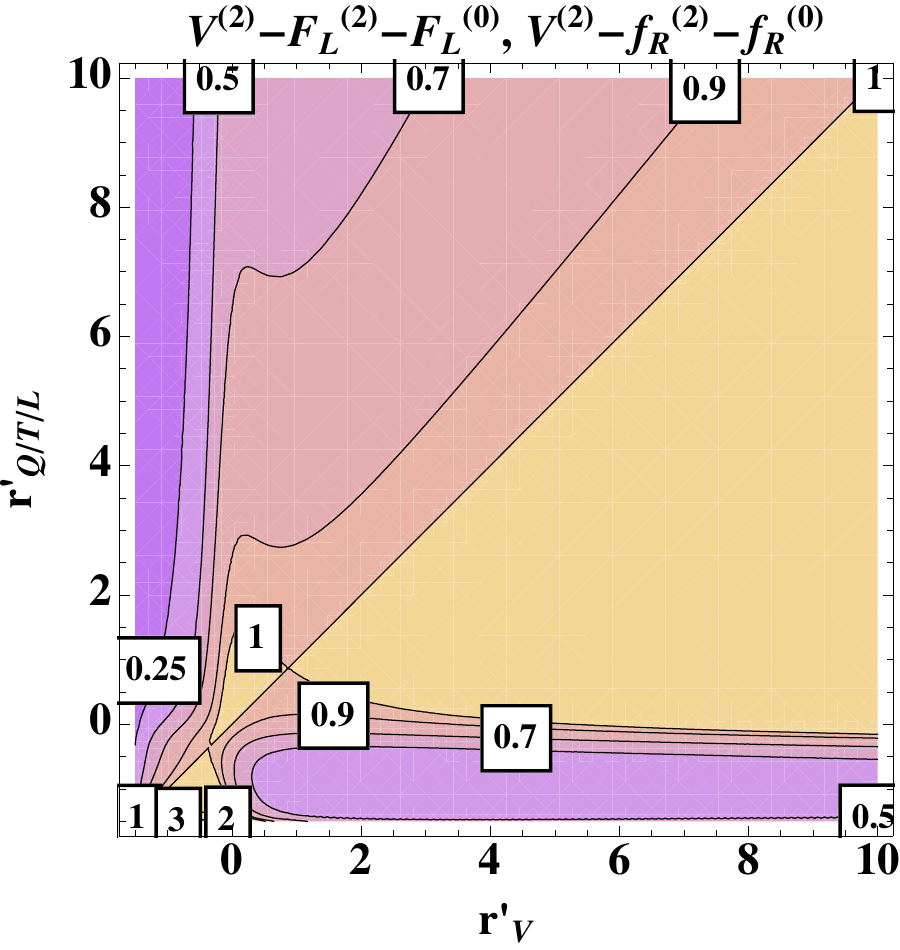}
\\
\caption{Contours of deviation 
for the generic couplings $V^{(2)}$-$F^{(0)}$-$F^{(0)}$ (or
$V^{(2)}$-$f^{(0)}$-$f^{(0)}$) (left) 
and $V^{(2)}$-$F^{(2)}$-$F^{(0)}$ (or $V^{(2)}$-$f^{(2)}$-$f^{(0)}$) (right) from the 
corresponding SM values in the $r^\prime_V-r^\prime_{Q/T/L}$ plane. 
$V$, $F$ and $f$ stand for generic gauge boson, $SU(2)_W$ doublet and 
singlet fermion fields (with corresponding chiralities), respectively.
Note that when $V$ is the (KK) $W$ boson, types of the two 
fermions involved at a given vertex are different.
}
\label{fig:couplings-200-220}
\end{figure}

The set of couplings that we briefly discuss here are those that would 
appear in the production of the KK top quarks at the LHC and their
decays. In figure \ref{fig:couplings-200-220} we illustrate the 
coupling-deviation (a multiplicative factor of the corresponding {SM} 
value at the tree level) 
$g^{(2)}$-$q^{(0)}$-$q^{(0)}$ (left) and $g^{(2)}$-$q^{(2)}$-$q^{(0)}$ 
(right) in the {generic $r'_V-r'_{Q/T/L}$} plane. In both of these plots, the mUED
case is realized along the {diagonals} over which $\rgp=\rqp$.
{In the first case}, the mUED value is known to be vanishing at {the} tree level
since {KK number is violated}. Hence, the diagonal appears with the
contour-value of zero. For vertices involving the top 
quarks, $\rtp$ replaces $\rqp$. 
For a process like 
$pp \to \ttwobarl t$ + h.c., the former kind of coupling appears at 
the parton-fusion 
(initial state) vertex while the latter shows up at the production 
vertex. The combined strength of these two couplings controls the 
production rate for the mentioned process. Further, the situation
is not much different for the level `2' electroweak gauge bosons
except for some modifications due to
mixings present in the electroweak sector. In general, it can be
seen from the first plot of figure \ref{fig:couplings-200-220} that the
coupling $g^{(2)}$-$q^{(0)}$-$q^{(0)}$ picks up a negative sign {for}
$\rgp > \rqp$. This could have nontrivial phenomenological 
implications for processes in which interfering Feynman diagrams 
are present. On the other hand, $g^{(2)}$-$q^{(2)}$-$q^{(0)}$ remains
always positive as is clear from the second plot of figure
\ref{fig:couplings-200-220}.   
Note that the three-point vertex
$V^{(0)}$-$V^{(0)}$-$V^{(2)}$ and the generic
ones of the form $V^{(0)}$-$f^{(0)}$-$f^{(2)}$ are absent because
the corresponding overlap integrals vanish due to orthogonality of 
the involved mode functions. 

%
%
%
%
\begin{figure}[t]
\centering
\vspace*{0.15in}
\includegraphics[width=0.4\columnwidth]{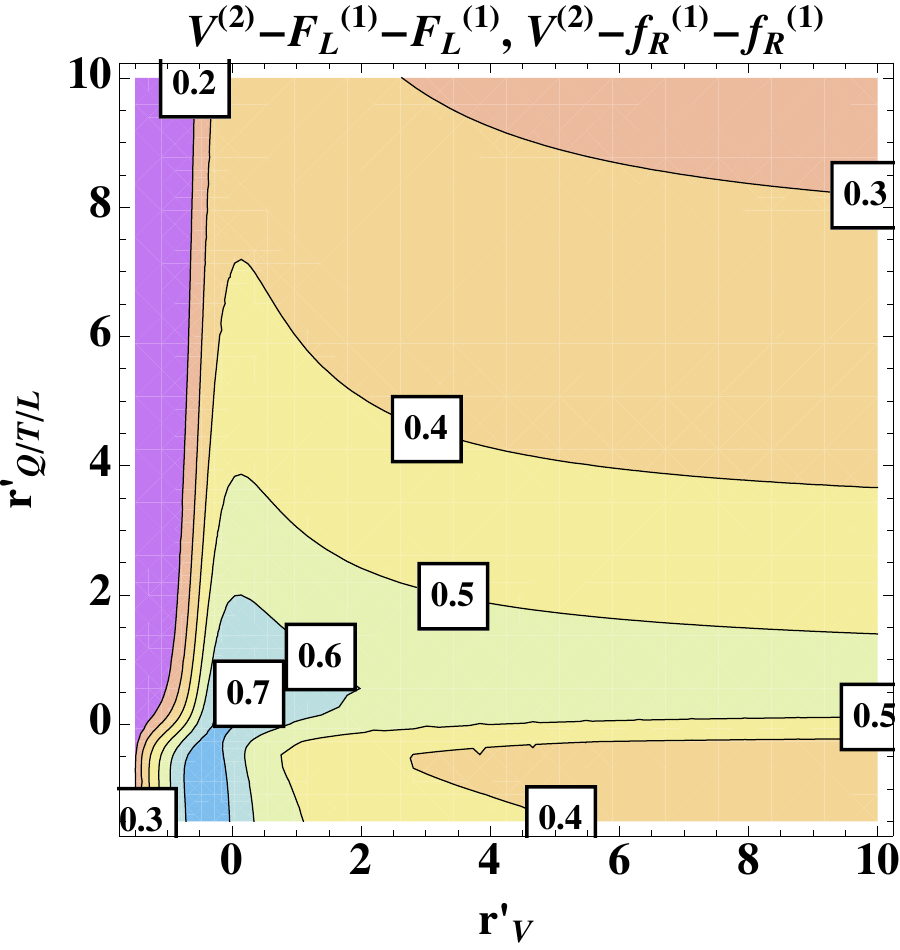}
\hspace*{0.5in}
\includegraphics[width=0.4\columnwidth]{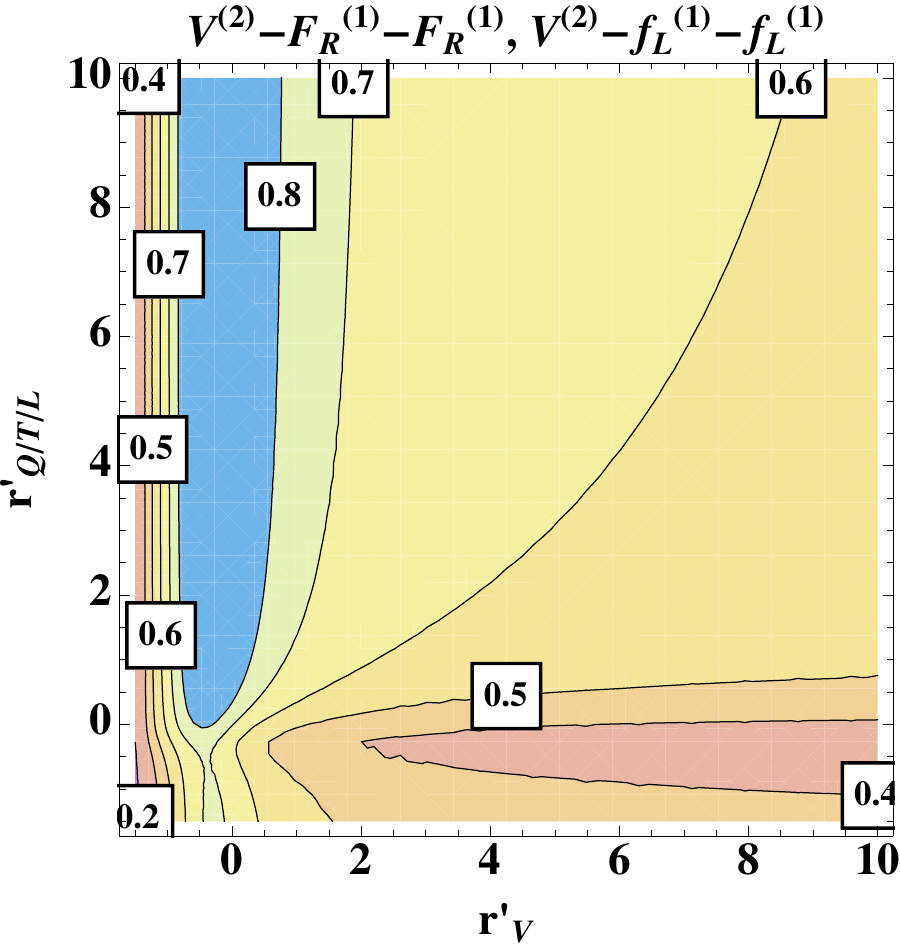}
\caption{Same as in figure \ref{fig:couplings-200-220} but for
the generic couplings
$V^{(2)}$-$F^{(1)}_L$-$F^{(1)}_L$ or
$V^{(2)}$-$f^{(1)}_R$-$f^{(1)}_R$ 
(left) and 
$V^{(2)}$-$f^{(1)}_L$-$f^{(1)}_L$ or
$V^{(2)}$-$F^{(1)}_R$-$F^{(1)}_R$ (right).}
\label{fig:couplings-211}
\end{figure}

In figure \ref{fig:couplings-211} we present the corresponding
contours of {similar deviations} in the couplings involving the level `2' KK 
gauge bosons and the level `1' KK quarks. The plot {on} left shows
the situation for the left- (right-) chiral component of the 
$SU(2)_W$ doublet (singlet) quarks while the plot {on} right illustrates
the case for left- (right-) chiral component of the $SU(2)_W$ singlet 
(doublet) quarks. These are in conformity with the mode functions
for these  individual components of the level `1' KK quarks. 
However, it should be noted that the KK quarks being
vector-like states, each of the $SU(2)_W$ doublet and singlet partners have
both left- and right-chiral components. Thus, the effective
{couplings}
{are} obtained only by {suitably} combining (with appropriate weights) 
the strengths as given by the two
plots. In the case of KK top quarks, the situation would be further 
complicated because of significant mixing between the two gauge
eigenstates. 
For brevity, a list of relevant couplings is presented in 
table \ref{tab:overlap} with mentions of the kind of 
modifications they undergo in the nmUED scenario.
It is clear from
these figures that these (component) couplings involving level `2' KK
states are in general suppressed compared to the relevant SM couplings 
except over a small region with
$r^\prime_{Q/T/L} <0$.
%
%
\subsubsection{Effective couplings {involving the} Higgs bosons}
\label{subsubsec:higgs-couplings}
\begin{figure}[t]
\centering
\vspace*{0.15in}
\includegraphics[width=0.4\columnwidth]{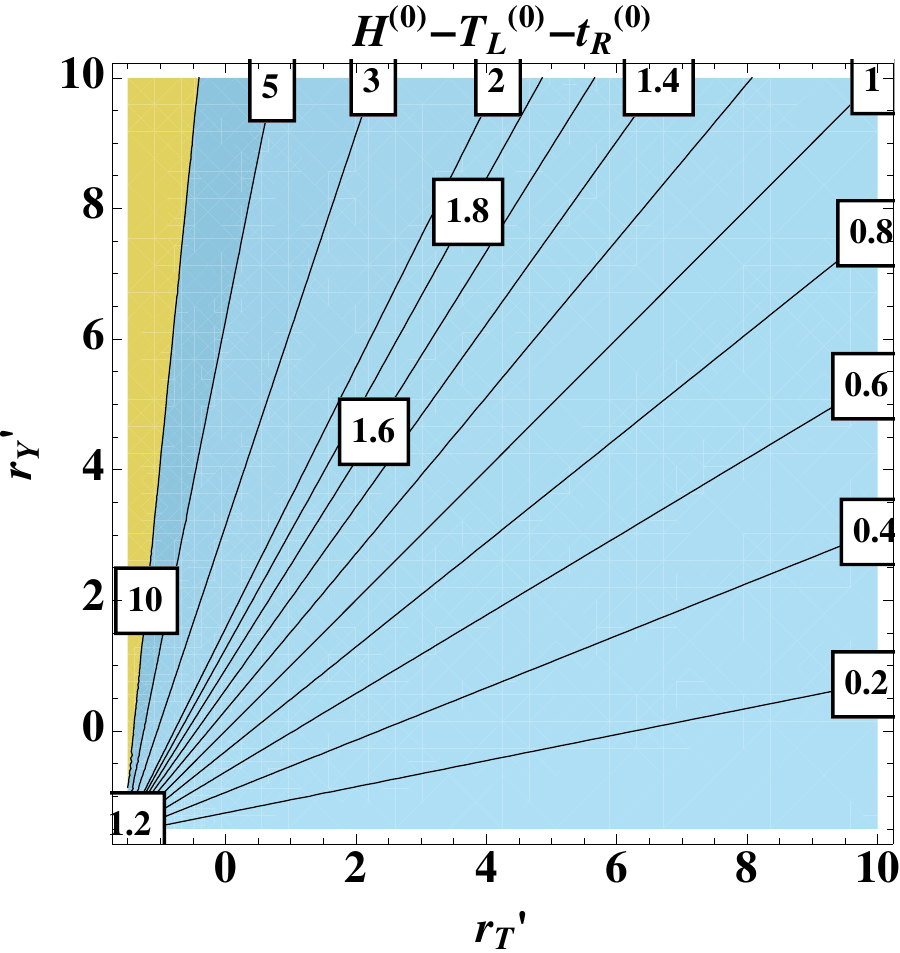}
\hspace*{0.5in}
\includegraphics[width=0.4\columnwidth]{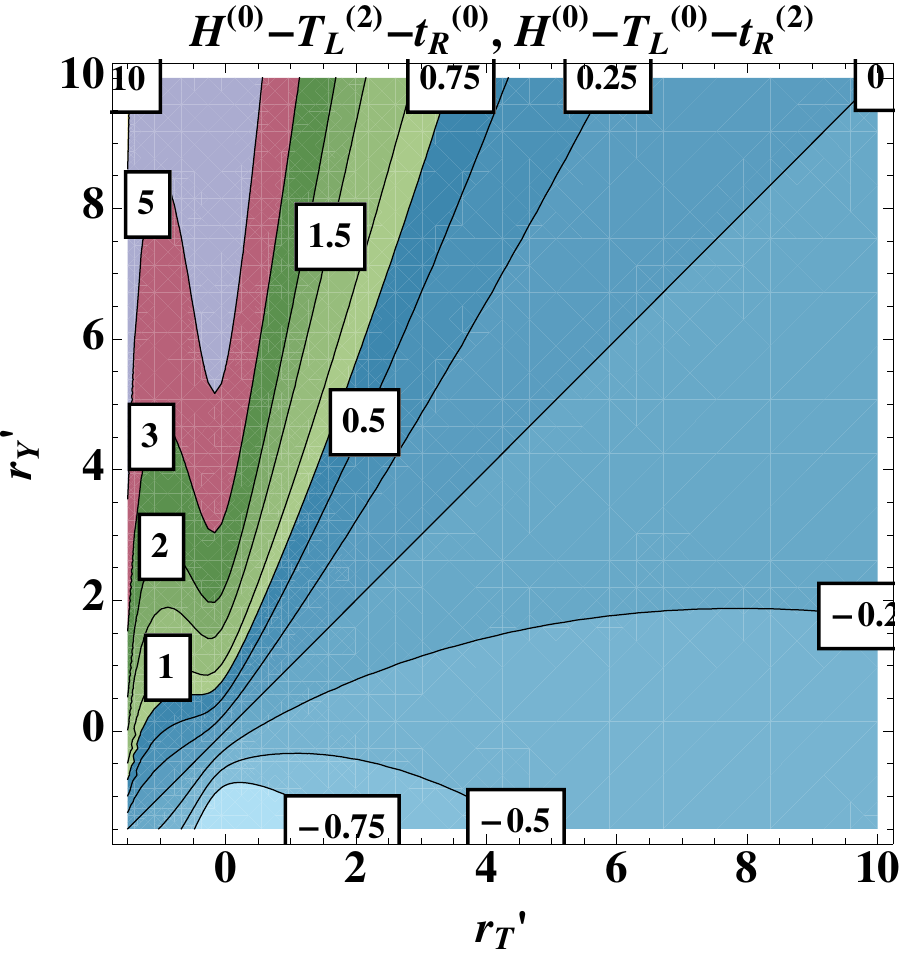}
\caption{
Contours of deviation in the $\rtp-\ryp$ plane for
the generic couplings 
$H^{(0)}$-$T^{(0)}_L$-$t^{(0)}_R$ (left) and
$H^{(0)}$-$T^{(2)}_L$-$t^{(0)}_R$ or $H^{(0)}$-$T^{(0)}_L$-$t^{(2)}_R$
compared to the corresponding SM cases.
}
\label{fig:couplings-higgs}
\end{figure}
The association of the Higgs sector with the third SM family is rather 
intricate and has deep implications which unfold themselves in many
scenarios beyond the SM. SUSY scenarios provide very good examples of
this, some {analyses} have been done in {the} mUED 
\cite{Bandyopadhyay:2009gd} and {the nmUED scenario}  
is also no exception. The couplings among the Higgs bosons 
and the KK top quarks of the nmUED scenario can deviate significantly 
from the corresponding SM Yukawa coupling. {However, the
zero-mode (SM) Higgs Yukawa couplings do not depend upon $\rhp\,(=\rew)$.}
In the left panel of figure
\ref{fig:couplings-higgs} we illustrate the possible deviation in the 
SM Yukawa coupling {itself} in the $\rtp-\ryp$ plane. Along the diagonal of
this figure ({with} $\rtp=\ryp$) the SM value of the Yukawa coupling is 
preserved. Note that the latest LHC data still allows for significant 
deviations in the $H$-$t$-$t$ coupling \cite{Chatrchyan:2013yea,
cms-tth-gamma, atlas-tth-gamma, Nishiwaki:2013cma}. 

In the right panel we
show deviations of the generic $H$-$t^{(2)}$-$t$ which appears at the 
tree level in nmUED. Unlike in the case of the interaction vertex
$V^{(0)}$-$f^{(2)}$-$f^{(0)}$ (where $V^{(0)}$ is a massive SM
gauge boson) where the involved coupling vanishes in the absence of
\emph{level-mixing} between $f^{(2)}$ and $f^{(0)}$, the analogous
Higgs vertex remains non-vanishing even in the absence of
\emph{level-mixing} between the fermions. However, in this case, 
for $\rtp=\ryp$ the coupling vanishes. This implies that {the more 
the} Yukawa coupling involving the level `0' fields appears to agree
with the SM expectation {(from future experimental analyses)}, the weaker 
the coupling $H$-$t^{(2)}$-$t$ in such a scenario {would get to
be}. In both
cases, however, we find that the coupling strengths get enhanced
for smaller values of $\rtp$ with $\rtp < \ryp$.
All these indicate that production of the SM Higgs boson via gluon-fusion
and its decay to di-photon final state can receive non-trivial 
contributions from such couplings and thus might get constrained
from the LHC data.
The issue is currently under study. 
%
%
%
\section{Experimental constraints and benchmark scenarios}
\label{sec:constraints}
Several different experimental observations put constraints of varying
degrees on the parameters (like $\rinv$, $\rtp$, $\rqp$, $\ryp$ and 
the input top quark mass ($\mtopin$)) that control the KK top quark 
sector. First and foremost, $\rinv$ is expected to be constrained
from the searches for level `1' KK quarks and KK gluon at the LHC. In
the absence of {any} such dedicated {search}, a rough estimate of $\rinv>1$ 
TeV has been derived in ref.~\cite{Datta:2012tv} by appropriate recast 
of the LHC constraints obtained for the squarks and {the} gluino
{in SUSY scenarios}.

As discussed in the previous subsection, observed mass of the top 
quark restricts the parameter space in a nontrivial way. Also, 
important 
constraints come from the experimental bounds on flavor changing 
neutral currents (FCNC), electroweak precision bounds in terms of the 
Peskin--Takeuchi parameters ($S, \, T$ and $U$) and bounds on effective
four-fermion interactions. 
In this section we discuss these 
constraints briefly and choose a few benchmark scenarios that satisfy
them and {are} phenomenologically interesting.

\subsection{Constraints from the observed mass of the SM-like top quark}
\label{subsec:topmass}
\begin{figure}[t]
\centering
\includegraphics[width=0.3\columnwidth]{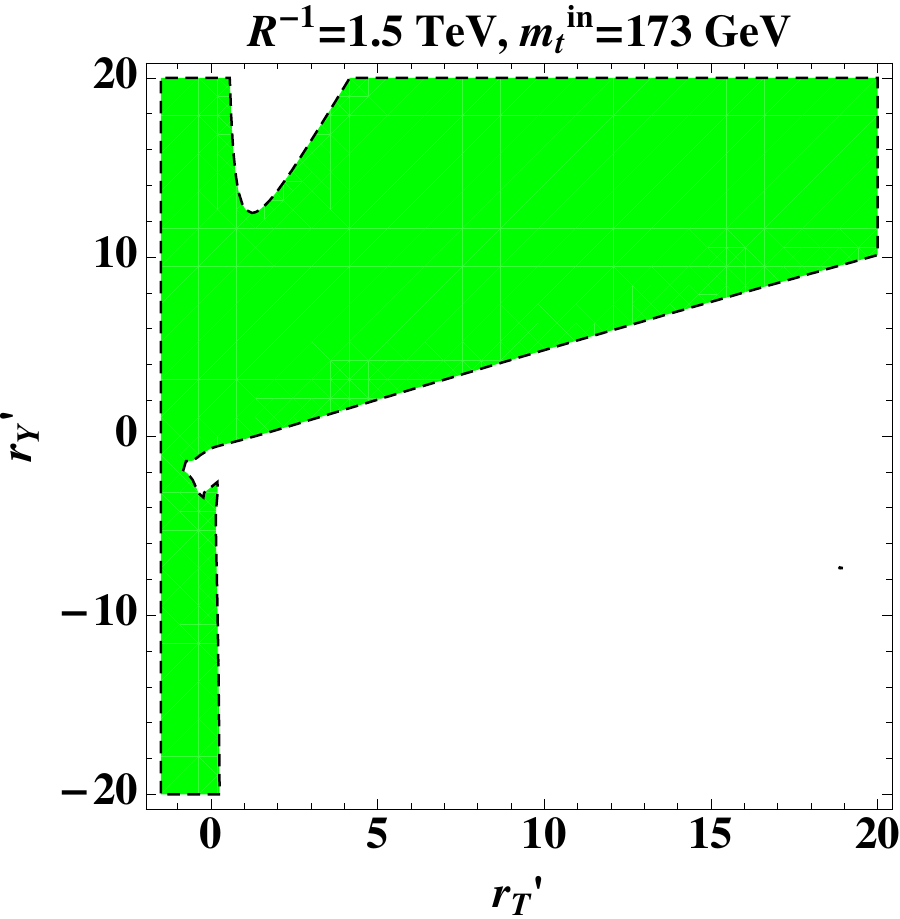}
\hspace*{-0.00in}
\includegraphics[width=0.3\columnwidth]{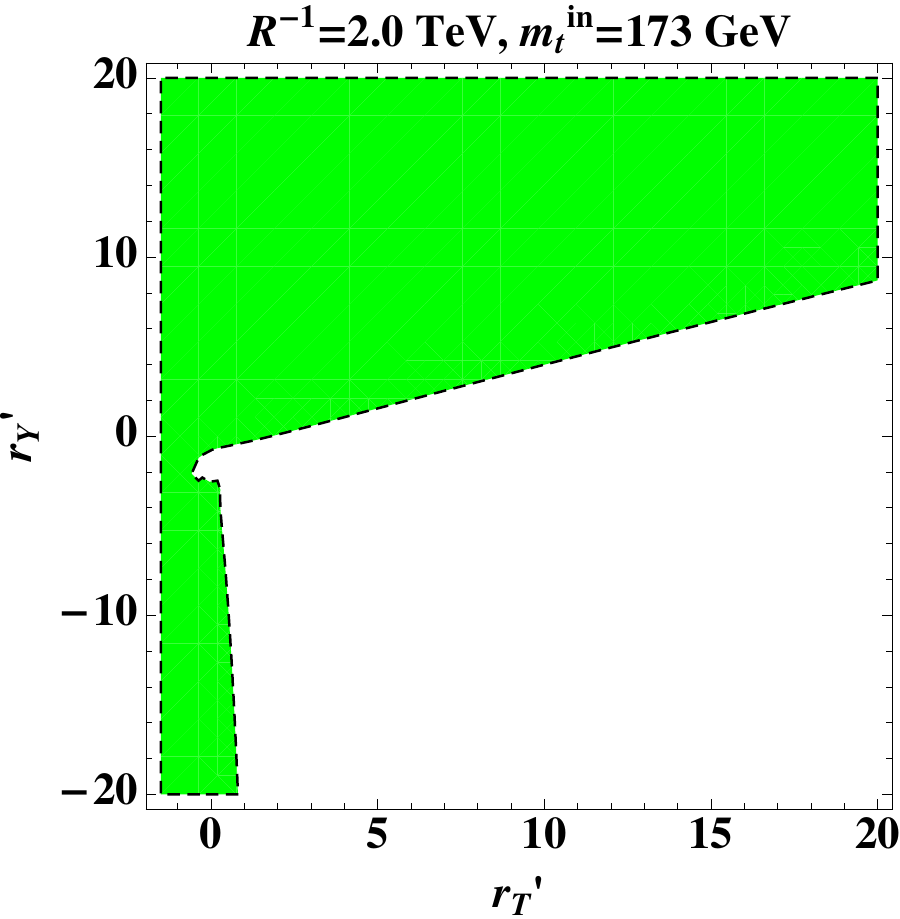}
\hspace*{-0.00in}
\includegraphics[width=0.3\columnwidth]{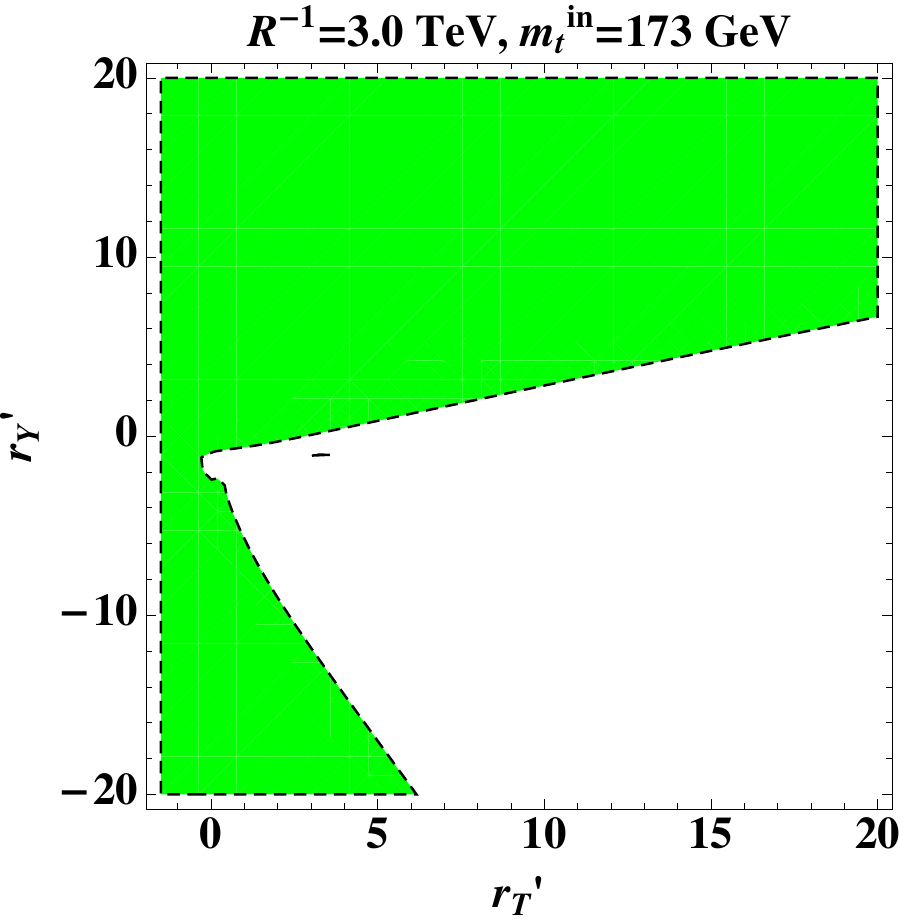}
\\
\includegraphics[width=0.3\columnwidth]{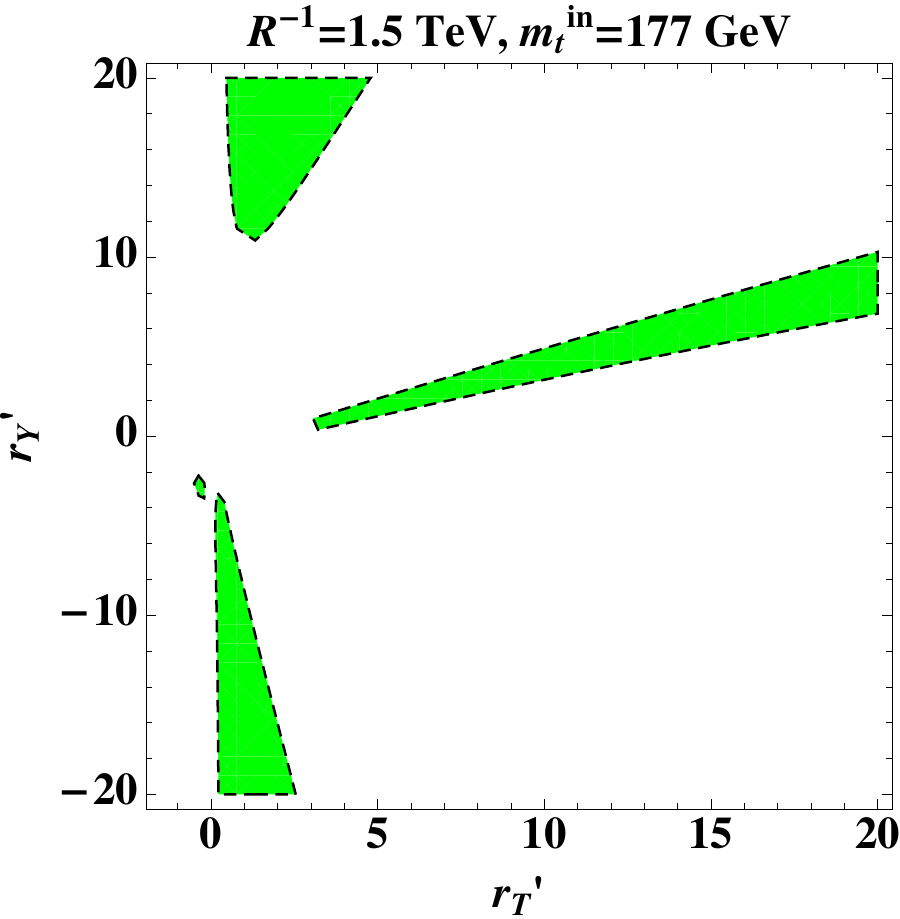}
\hspace*{-0.00in}
\includegraphics[width=0.3\columnwidth]{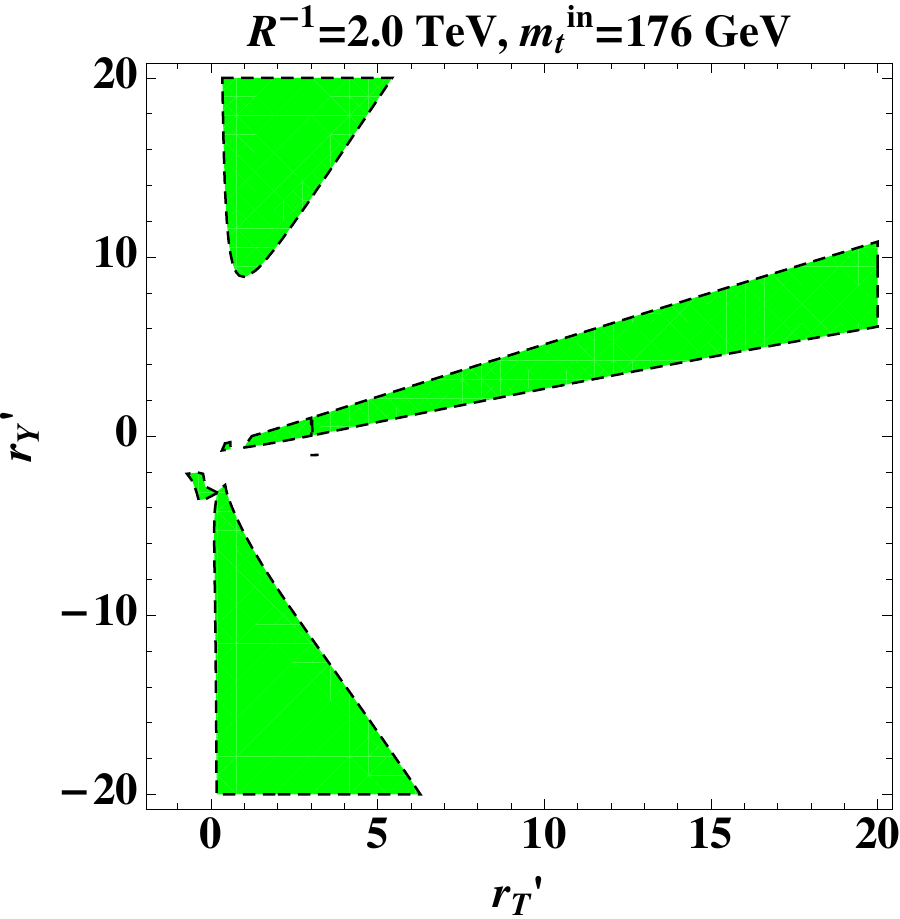}
\hspace*{-0.00in}
\includegraphics[width=0.3\columnwidth]{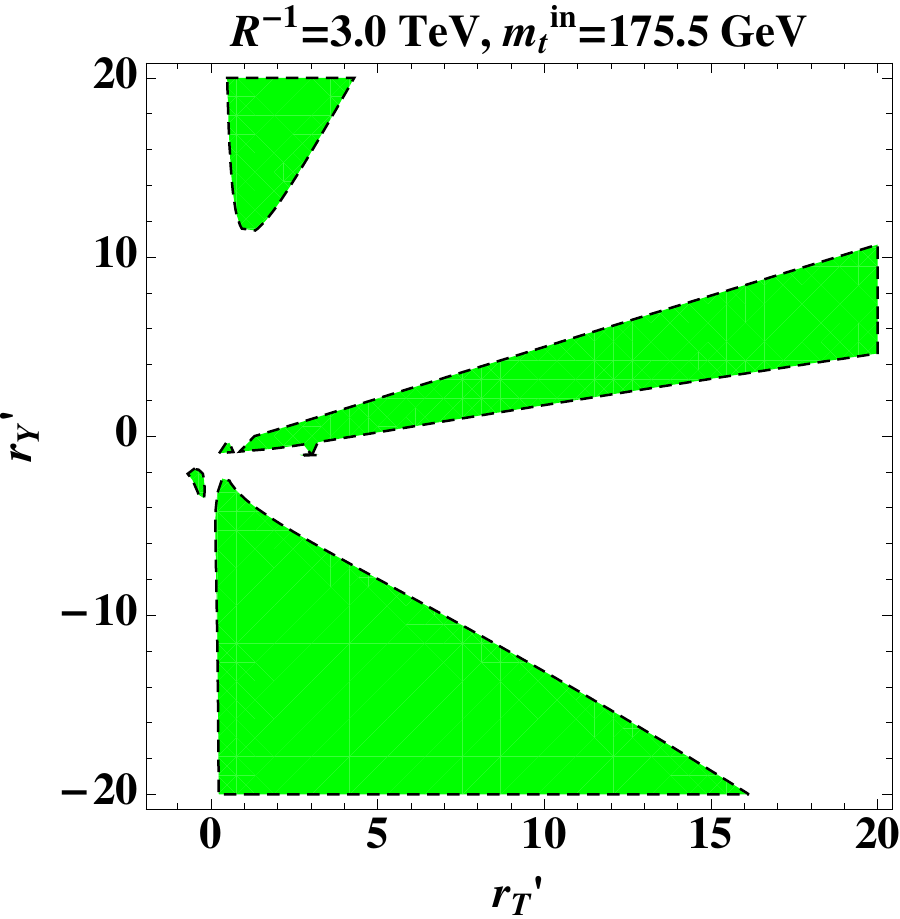}
\caption{Regions {(in green)} in the $\rtp-\ryp$ plane for three $\rinv$ values of
(1.5, 2 and 3 TeV, {varying} along the rows) and for 
{different} suitable values of $\mtopin$ (indicated on 
{top of} each plot) that are consistent with 
physical (SM-like) top quark mass ($\mtopphys$) being within the range 
$m_t^{\text{phys}} = 173 \pm 2$ GeV.
}
\label{constraint-topmass}
\end{figure}

In figure \ref{constraint-topmass} we show the allowed regions in the 
$\rtp-\ryp$ plane that result in top quark pole mass within the range 
171-175 GeV \cite{Alekhin:2012py} (which is argued to be a more
appropriate range than what the 
experiments actually quote \cite{CDF:2013jga}) for given values of
$\rinv$ and input top quark masses.
%
%
\noindent
Some general observations are that the physical top quark mass
($\mtopphys$) rarely becomes larger than the input top quark mass 
($\mtopin$). This means, to have $\mtopphys$ at least of 171 GeV, 
$\mtopin$ has to be larger than 171 GeV. Further, increasing 
$\mtopin$ beyond around 175 GeV, as we go over to the second row of 
figure \ref{constraint-topmass}, opens up disjoint sets of allowed 
islands in the $\rtp-\ryp$ plane with increasing region allowed for
negative $\ryp$ (and extending to larger $\rtp$ values) at the 
expense of the same with positive $\ryp$. Increasing $\mtopin$ further
(beyond say, 180 GeV) results in allowed regions diminishing to 
{an} insignificant level. These features remain more or less unaltered as 
$\rinv$ is increased, as we go from left to right along a
horizontal panel. A palpable direct effect that can be attributed to 
increasing $\rinv$ is in the moderate increase of the region in the 
$\rtp-\ryp$ plane consistent with $\mtopphys$, in particular, for 
negative $\ryp$ values and when $\mtopin$ is not terminally large 
(\emph{i.e.}, below $190$ GeV, say) for the purpose. 

Although a moderate range of input top quark mass 
$171< \mtopin \lesssim 190$ is consistent with $171 < \mtopphys < 175$ 
GeV in the space of $\rinv-\rtp-\ryp$, the allowed region there is 
rather sensitive to the variation in $\mtopin$. Thus, the allowed 
range of the $\mtopphys$ restricts the nmUED parameter space in a 
significant way which, in turn, influences the masses and the 
couplings of the KK top quarks. 
An important point is to be noted here.
The level `1' top quark sector, though does not talk to either level 
`0' or level `2' sector directly (because of conserved KK-parity), 
is influenced by these constraints since $\rtp$, $\ryp$ and $\rinv$ 
also govern the same.
%
%
\subsection{Flavor constraints}
\label{subsec:flavor}
\begin{figure}[t]
\centering
\includegraphics[width=0.3\columnwidth]{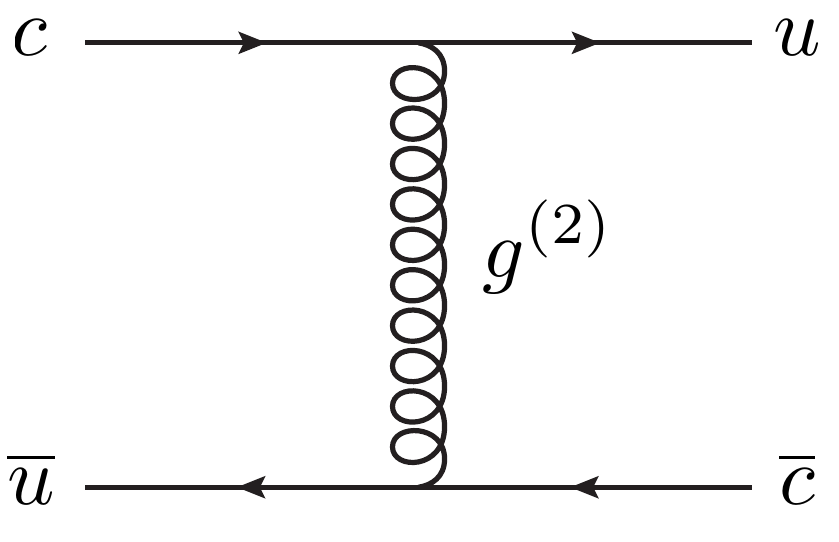}
\caption{Feynman diagram showing the induced FCNC vertex.}
\label{fig:diagram-fcnc}
\end{figure}

The BLKTs ($\rqp$) and the BLYTs ($\ryp$) are matrices in the flavor 
space. Hence, their generic choices may induce large FCNCs at the tree
level. It is possible to choose a basis in which the BLKT matrix is 
diagonal.  This ensures no mixing among fermions of different flavors 
or from different KK levels arising from the gauge kinetic terms. 
However, with the Yukawa sector included, off-diagonal terms (mixings) 
appear in the gauge sector on rotating the gauge kinetic terms into a 
basis where the quark mass matrices are diagonal. These terms could 
induce unacceptable FCNCs at the tree levels and thus, would be 
constrained by experiments. In figure \ref{fig:diagram-fcnc} we 
present the tree level diagram that could {give rise to unwanted} 
FCNC {effects}. 
%
%
%
%

A rather high compactification scale ($\rinv \sim {\cal{O}}(10^5)$ 
TeV; the so-called decoupling mechanism) or a near-perfect 
mass-degeneracy among the KK quarks at a given level 
(${\Delta m\over m^{(1)}} \lesssim 10^{-6}$; across all three 
generations) could suppress the FCNCs to the desired level~\cite{gerstenlauer}. While the 
first option immediately renders all the KK particles rather too 
massive, the second one makes the KK top quarks as heavy as the KK 
quarks from the first two generations thus making them quite
difficult to be accessed at the LHC. A third option in the 
form of ``alignment'' (of the rotation matrices) \cite{gerstenlauer} 
can make way for significant lifting of degeneracy {thus allowing
for light enough quarks from the third generation that are} 
within the reach of the LHC. 
In such a setup, FCNC occurs in the $up$-type doublet sector. 
Hence, the strongest of the bounds in terms of the relevant Wilson
coefficient {($C^1_D$)} comes from the recent observation of $D^0-\overline{D^0}$ 
mixing \cite{Aaij:2012nva} {(and not from the $K$ or the $B$ meson
systems)} and the requirement is 
$|C^1_D| < 7.2 \times 10^{-7}\,\text{TeV}^{-2}$~\cite{Bona:2007vi}, attributed solely to
the gluonic current which is by far the dominant contribution.
The essential contents of the setup are summarized in appendix 
\ref{app:fcnc}. 

\begin{figure}[t]
\centering
\includegraphics[width=0.3\columnwidth]{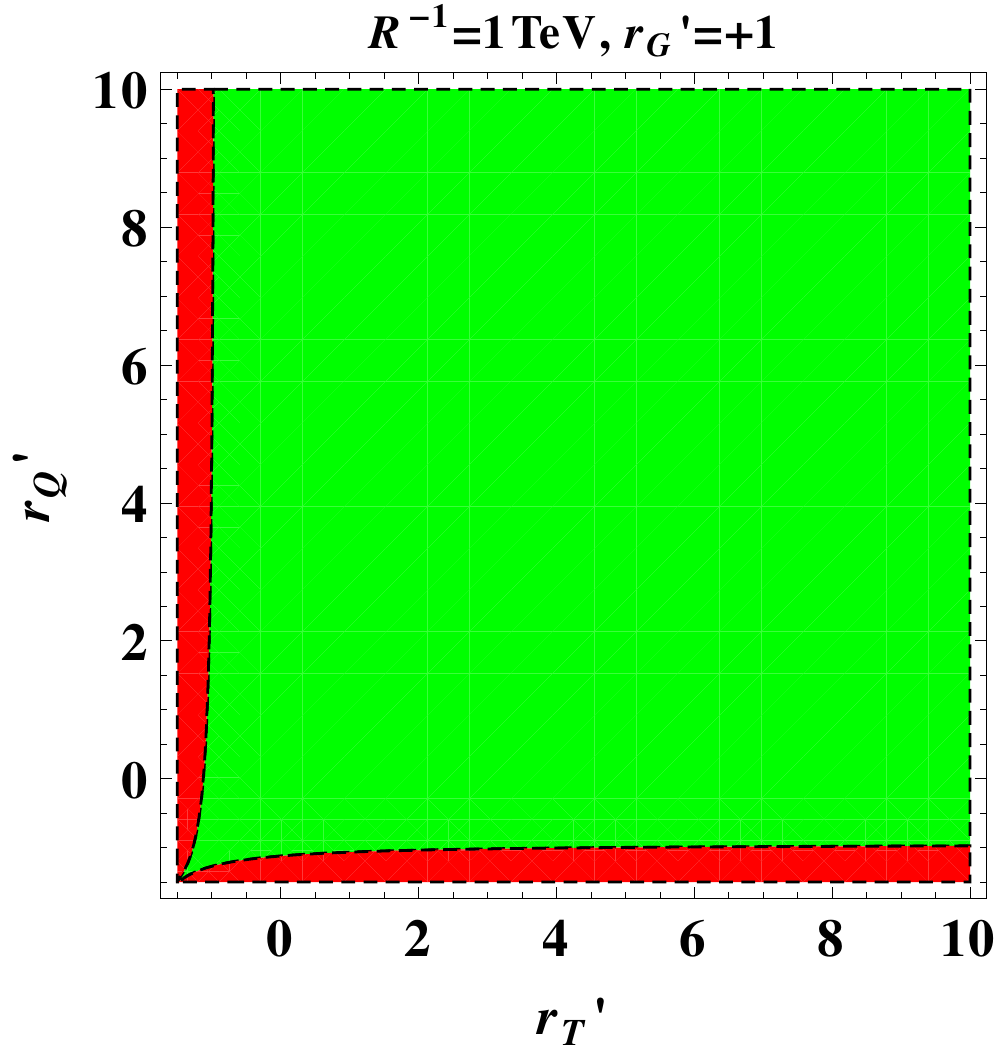}
\hspace*{0.10in}
\includegraphics[width=0.3\columnwidth]{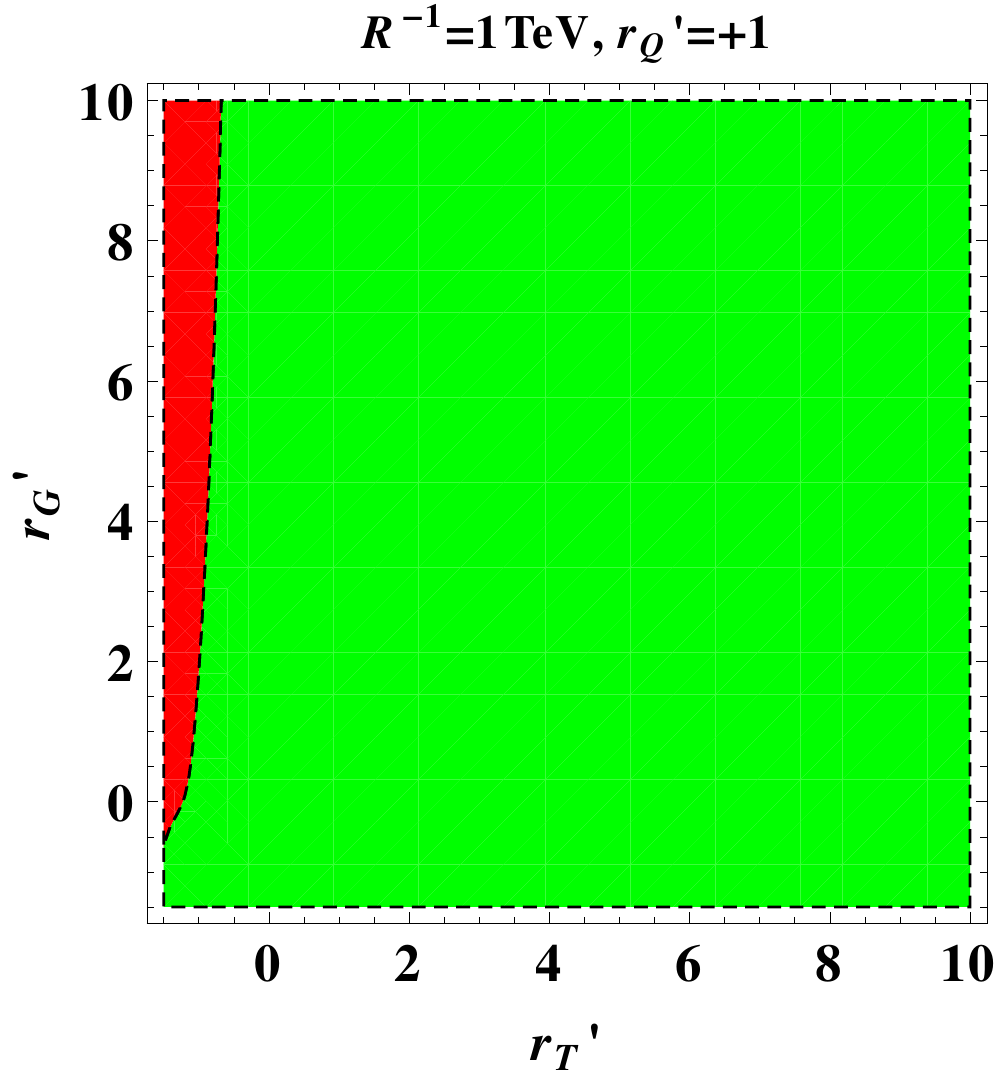}
\hspace*{0.10in}
\includegraphics[width=0.3\columnwidth]{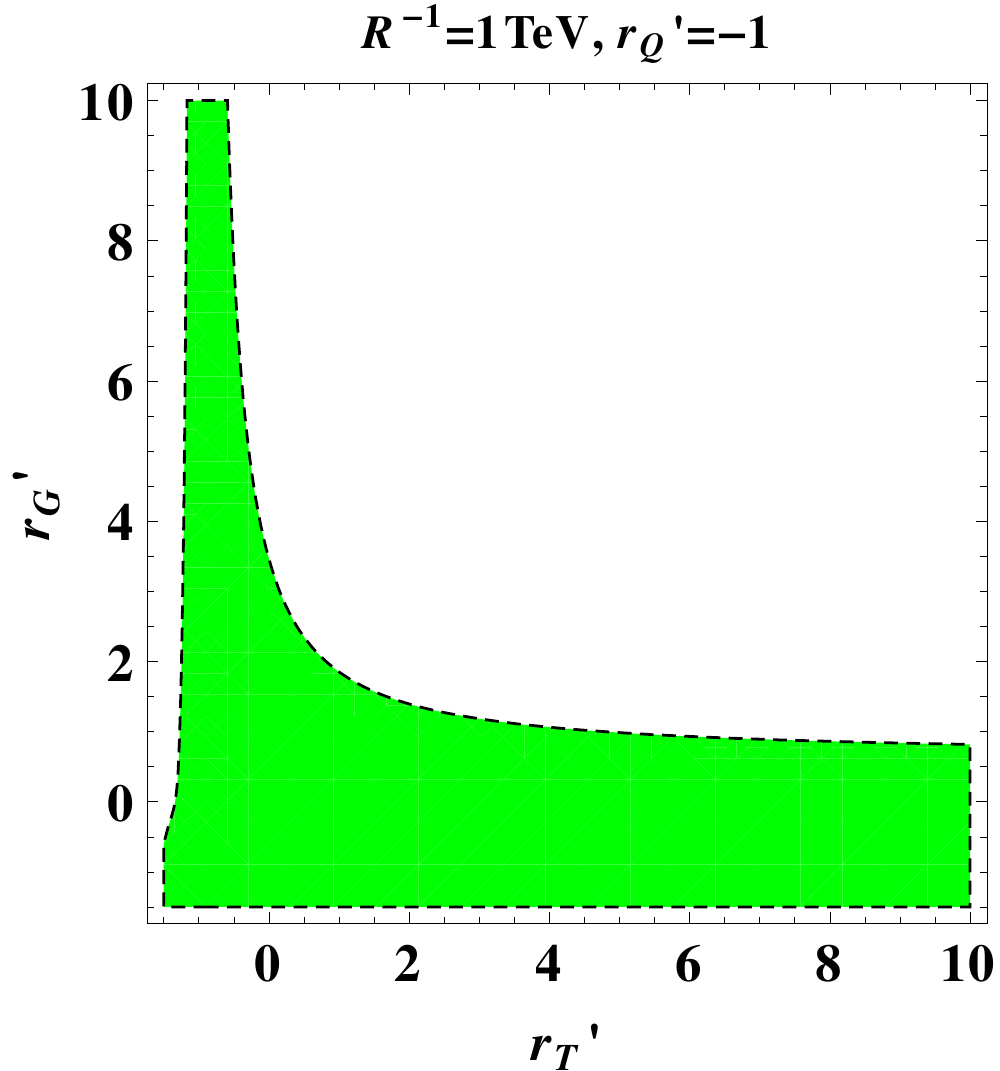}
\caption{Regions in the $\rtp-\rqp$ (for fixed $\rgp$; the left-most
plot) and $\rtp-\rgp$ (for fixed $\rqp$; the middle and the right-most 
plot) planes for $\rinv=1$ TeV 
that are allowed (in green) by FCNC constraints. For the first two
figures, thin strip(s) of the disallowed regions (in red) are 
highlighted for {better} visibility.}
\label{fig:fcnc-plot}
\end{figure}

In the left-most panel of figure \ref{fig:fcnc-plot} we demonstrate the 
allowed/disallowed region in the $\rtp-\rqp$ plane for $\rgp=1$ with $R=1\,\text{TeV}$.
The panel in the middle demonstrates the corresponding regions
in the $\rtp-\rgp$ plane for $\rqp=+1$. It is seen that some region with
$\rtp <0$ is disallowed when $\rgp$ is large, \emph{i.e.},
{when the level `2' KK gluon is relatively light.}
The right-most panel illustrates the {region} allowed in the same plane 
but for $\rqp=-1$. The bearing of the FCNC
{constraint} is most pronounced in this case.
%
{It can be noted that 
the smaller the  value of $\rgp$ is, the heavier is the mass of the level `2' gluon
and hence, the stronger is the suppression of the dangerous FCNC 
contribution.
Such a suppression 
could then allow $\rtp$ to be significantly different from $\rqp$ but 
still satisfying the FCNC bounds.
This feature is apparent from the rightmost panel of figure~\ref{fig:fcnc-plot}.}
Note that a rather minimal value for
$\rinv$ (=1 TeV) is chosen for this demonstration. A larger $\rinv$
results in a more efficient suppression of FCNC effects and hence,
leads to a larger allowed region. In summary, it appears that FCNC
constraints do not seriously restrict the third generation sector as
yet.

%
\vskip 10pt
%
%
\subsection{Precision constraints}
\label{subsec:precision}

It is well known that the Peskin--Takeuchi parameters $S$, $T$ and $U$
that parametrize the so-called oblique corrections to the electroweak
gauge boson propagators \cite{Peskin:1990zt,Peskin:1991sw} put rather 
strong constraints on the mUED scenario. These observables are affected 
by the modification in the
Fermi constant $G_F$ (determined experimentally by studying muon
decay) due to induced effective 4-fermion vertices originating from 
exchange of electroweak gauge bosons from even KK levels.
These were first calculated in refs.~\cite{Kakuda:2013kba,Appelquist:2002wb,Flacke:2005hb,Gogoladze:2006br,Baak:2011ze} assuming mUED
tree-level spectrum while ref.~\cite{Flacke:2013pla} expressed them 
in terms of the
actual (corrected) masses of the KK modes.

As discussed in refs.~\cite{Rizzo:1999br,Davoudiasl:1999tf,
Csaki:2002gy,Carena:2002dz,Flacke:2011nb}, the correction to $G_F$ can 
be incorporated in the electroweak fit via the modifications it induces 
in the Peskin--Takeuchi parameters and contrasting them with the
experimentally determined values of the latter.
Note that in the nmUED scenario we consider, level `2' electroweak gauge
bosons have tree-level couplings to the SM fermions and these modify
the effective 4-fermion couplings. These effects are over and above what 
mUED induces{\footnote{To be precise, in general, the mUED type higher-order 
contributions (usual one-loop-induced oblique corrections) would not 
be exactly the same as that from the actual mUED scenario.
However, as pointed out in ref.~\cite{Flacke:2011nb}, in the 
``minimal'' case of $r_{W} = r_{B} = r_{H}$
along with the requirements on the relations involving $\mu$-s and
$\lambda$-s as given in equations \ref{condition_one_for_minimalcase}
and \ref{eqn:extra-conds}, exact mUED limits for the couplings are 
restored while departures in the KK masses (from the corresponding
mUED values) still remain.}}
where such KK number violating couplings appear only at
higher orders.
It is thus natural to expect that usual oblique corrections to $S$, 
$T$ and $U$ induced at one-loop level would be sub-dominant when 
compared to the above nmUED tree-level contributions. Thus, in our
present analysis, we neglect the one-loop contributions
{but otherwise follow}
{the approach 
originally adopted in ref.~\cite{Flacke:2011nb} and which was 
later used in ref.~\cite{Flacke:2013pla}.}
{The nmUED effects are thus parametrized as:}
\bea
S_{\text{nmUED}} = 0, \qquad \quad
T_{\text{nmUED}} = - \frac{1}{\alpha} \frac{\delta G_F}{G_F}, \qquad \quad
U_{\text{nmUED}} =   \frac{4 \sin^2{\theta_W}}{\alpha} \frac{\delta
G_F}{G_F}
\eea
where $\alpha$ 
is the electromagnetic coupling strength, $\theta_W$ is the 
$\overline{MS}$ Weinberg angle, both given at the scale $M_Z$ and 
$G_F$ is given by 
\al{
G_F = G_F^0 + \delta G_F
}
with $G_F^0$ ($\delta G_F$) originating from the $s$-channel SM (even
KK) $W^\pm$ boson exchange.
The concrete forms of these effects are calculated in our model
following ref.~\cite{Flacke:2011nb}. {Using} our notations, 
{these} are given by:
\al{
G_F^{0} &= \frac{g_2^2}{4\sqrt{2}} \frac{1}{M_W^2},\quad
\delta G_F = \sum_{n \geq 2:\text{even}} \frac{g_2^2}{4\sqrt{2}} \frac{1}{m_{W_{(n)}}^2} \left(g_{_{L_{(0)}W_{(n)}L_{(0)}}}\right)^2, \\
\left.g_{_{L_{(0)}W_{(n)}L_{(0)}}}\right|_{n\text{:even}} &\equiv
	\frac{1}{f_{W^{(0)}}} \int_{-L}^{L} dy
	\left( 1 + r_{\text{EW}} \left[ \delta(y-L) + \delta(y+L) \right] \right)
	f_{L_{(0)}} f_{W_{(n)}} f_{L_{(0)}} \notag \\
	&= \frac{ 2\sqrt{4r_{\text{EW}} + 2\pi R} \left( M_{W_{(n)}}r_L +
	   \tan\left( \frac{M_{W_{(n)}} \pi R}{2} \right) \right) }
	   { M_{W_{(n)}} \left( 2r_L + \pi R \right) \sqrt{4r_{\text{EW}} +
	   \pi R \sec^2\left( \frac{M_{W_{(n)}} \pi R}{2} \right)} +
	   2\tan\left( \frac{M_{W_{(n)}} \pi R}{2} \right)/M_{W_{(n)}}}
	   \label{F0-V2-F0_explicitform}
}
where $M_{W_{(n)}}$ is determined by equation~\ref{masscondition_simplified}.
Even though the KK leptons do not appear in the 
process, the BLKT parameter $r_L$ in the lepton sector (to be precise,
the one for the 5D muon doublet) inevitably influences the 
coupling-strength given in equation \ref{F0-V2-F0_explicitform}. 
We, however, assume a {flavor-}universal BLKT parameter
$r_L$ (just like what we do in the quark sector when we take $r_Q = r_T$)
which {help} trivially
circumvent tree-level contributions to lepton-flavor-violating processes.
%

%
\begin{figure}[t]
\centering
\includegraphics[width=0.3\columnwidth]{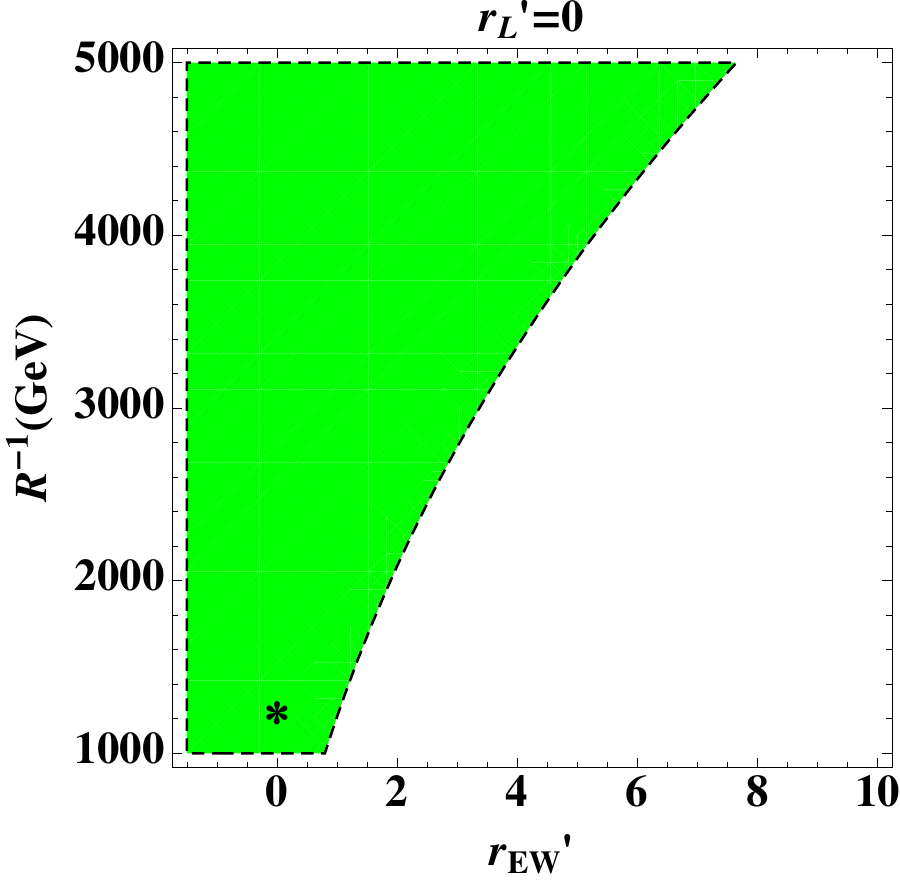}
\includegraphics[width=0.3\columnwidth]{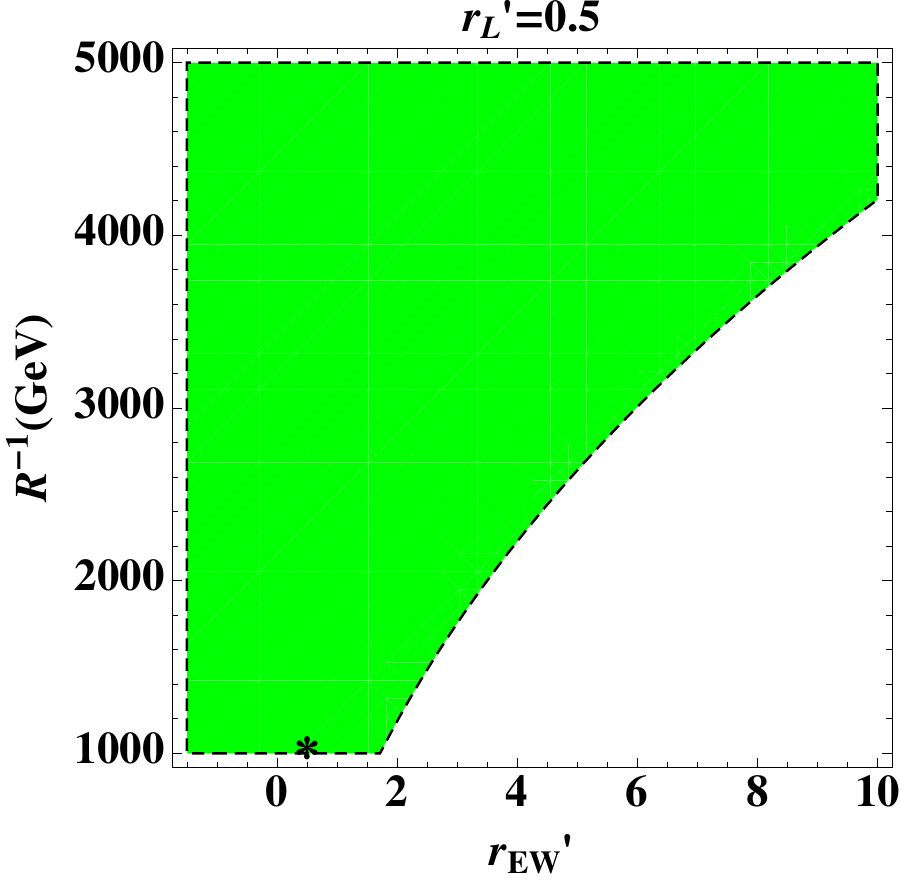}
\includegraphics[width=0.3\columnwidth]{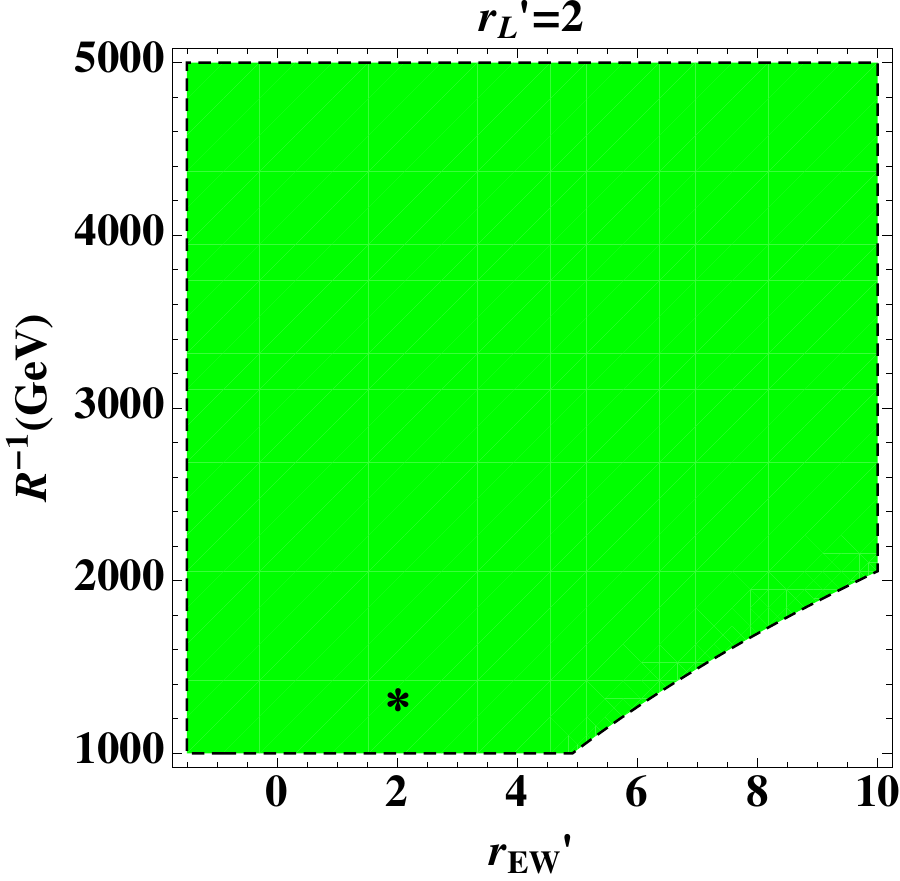}
\caption{Regions (in green) in the $\rew-\rinv$ plane allowed by 
electroweak precision data {at} $95\%$ C.L.
The black asterisks represent the global minimum in each one of them:
$\chi^2_{\text{min}}=8.8 \times 10^{-9}$ at $(\rew, R^{-1})$ = $(6.11 \times 10^{-3}, 1229\,\text{GeV})$ when $\rlp=0$,
$\chi^2_{\text{min}}=3.9 \times 10^{-9}$ at $(\rew, R^{-1})$ = $(0.505, 1029\,\text{GeV})$ when $\rlp=0.5$,
$\chi^2_{\text{min}}=1.5 \times 10^{-8}$ at $(\rew, R^{-1})$ = $(2.02, 1306\,\text{GeV})$ when $\rlp=2$.
}
\label{fig:stu-plot}
\end{figure}

We perform a $\chi^2$ fit of the parameters $S_{\text{nmUED}}$, $T_{\text{nmUED}}$ 
and $U_{\text{nmUED}}$ (with $\delta G_F$ evaluated for $n=2$ only)
for three fixed values of $\rlp$ ($\rlp=r_L \rinv= 0$, $0.5$ and $2$)
to the experimentally fitted values of the 
allowed new physics (NP) components in these respective observables 
as reported by the GFitter group \cite{Baak:2012kk} which are given by
\al{
&S_{\text{NP}} = 0.03 \pm 0.10, & &T_{\text{NP}}= 0.05 \pm 0.12, & &U_{\text{NP}} = 0.03 \pm 0.10, \notag
}
the correlation coefficients being 
\al{
&\rho_{ST} = +0.89, & &\rho_{SU} = -0.54, & &\rho_{TU} = -0.83, \notag
}
and the reference input masses of the SM top quark and the
Higgs boson being $m_t=173$ GeV and $m_H=126$ GeV, respectively.

In figure \ref{fig:stu-plot} we show the $95\%$ C.L. allowed region in the
$\rew-\rinv$ plane as a result of the fit performed.
As can be expected, the bound refers to $\rew$ as the only brane-local 
parameter which, unlike in ref.~\cite{Flacke:2013pla}, can be different 
from the corresponding parameters governing other sectors of the 
theory. Such a constraint is going to restrict the mass-spectrum 
and the couplings in the electroweak sector which is relevant for our
present study. It is not unexpected that for larger values of
$r_{\text{EW}}$ which result in decreasing masses for the electroweak gauge
bosons, only larger values of $\rinv$ (which compensates for the former
effect) remain allowed thus rendering  these excitations (appearing in the
propagators) massive enough to evade the precision bounds. 
Interestingly, it is possible to relax the bounds by introducing 
{a} positive $\rlp$ as shown in figure~\ref{fig:stu-plot}{, a feature
that can be taken advantage of as we explore the nmUED parameter
space further.}
This is since the coupling involved $g_{_{L_{(0)}W_{(n)}L_{(0)}}}$ gets reduced 
in the process (see the left plot in figure \ref{fig:couplings-200-220}).
\subsection{Benchmark scenarios}
\label{subsec:benchmarks}
For our present analysis, we now choose some benchmark scenarios which
satisfy the constraints discussed in the previous subsection. 
The parameter space of these scenarios mainly spans over $\rtp, \, \ryp, \, 
\rinv$ and, as a minimal choice, $\rew=\rhp$\footnote{Departure from
this assumption makes the gauge boson zero modes non-flat and
hence correct values (within experimental errors) of the SM parameters
like $\alpha_{em}, G_f, m_W, m_Z$ can only be reproduced in a constrained
region of $\rew-\rhp$ parameter space \cite{Flacke:2008ne}.}. 
We also include $\rgp$, $\rqp$ and $\rlp$ which are the BLKT parameters
for the KK gluon, the KK quark and the KK lepton sectors, respectively.
$\rgp$ has some non-trivial
implications for the couplings of the KK top quarks to the gluonic
excitations as discussed in section \ref{subsec:couplings}.
The parameter $\rqp$, though enters our discussion primarily
through FCNC considerations (see section \ref{subsec:flavor} and
appendix \ref{app:fcnc}), governs the couplings
$V^{(2)}$-$q^{(0)}$-$q^{(0)}$ (as shown in figure
\ref{fig:couplings-200-220}) that control KK top quark production
processes.
%
%
Both $\rgp$ and $\rqp$ serve as key handles on the masses of the 
KK gluon and the KK quarks from the first two generations,
respectively.
Similar is the status of $\rlp$ which enters through the
oblique parameters and controls the masses and couplings in the
lepton sector.
%
 
In search for suitable benchmark scenarios, we require the
following conditions to be satisfied. We {require} the approximate
lower bound on $\rinv$ to hover around 1 TeV which is obtained by 
recasting the LHC bounds on squarks (from the first two generations) 
and the gluino in terms of level `1' KK quarks and KK gluons in the 
nmUED scenario \cite{Datta:2012tv}. Further, the lighter of the level 
`1' KK top quark ($t^{(1)}_l$) is required to be at least about 500 
GeV. This safely evades current LHC-bounds on similar excitations 
while lower values may still be allowed given that these bounds 
result from model-dependent assumptions.

The above requirements together calls for a non-minimal sector for the 
electroweak gauge bosons ($\rew \neq 0$) such that the lightest KK 
gauge boson, 
the KK photon ($\kkaone$) is the lightest KK particle (LKP, a possible 
dark matter candidate)\footnote{This is a {possible} choice for the dark 
matter candidate in the nmUED scenario. Ref. \cite{Flacke:2008ne} 
{explores other possible candidates in such a
scenario}.}.
Incorporation of a non-minimal gauge sector affects the couplings of 
the gauge bosons which, as we will see, could be phenomenologically 
non-trivial.
The choice $\rew=\rhp$ renders the KK excitations of the gauge and
the Higgs boson very close in mass
thus allowing them to take part in the phenomenology of the KK top 
quarks.
In the present scenario, other BLT parameters in the Higgs sector,
$\mu_b$ and $\lambda_b$, are constrained by equations~\ref{condition_one_for_minimalcase}
and \ref{eqn:extra-conds} in addition to {the measured Higgs mass
as an input.}
Therefore, these are not {independent degrees} of freedom.

In table \ref{tab:bms} we present the spectra for three such benchmark 
scenarios: two of them with $\rinv=1$ TeV and the other with 
$\rinv=1.5$ TeV. 
The BLKT parameters $\rgp$ and $\rqp$ are so chosen such that the 
masses of the level `1' KK gluon are in the range 1.6-1.7 TeV 
(\emph{i.e.}, somewhat above the current LHC lower bounds on similar 
(SUSY) excitations) while the KK quarks from the first two generations
are heavier\footnote{Such a hierarchy of masses {opens up} the 
possibility of level `1' KK top quarks being produced in the 
cascade decays of the KK gluon and the KK quarks.}.
Note that in both cases we are having negative $\rgp$ and $\rqp$. 
In the top quark sector, the BLKT parameter $\rtp$ are fixed at values 
for which both light and heavy level `1' KK top quarks have sub-TeV
masses and hence expected to be within the LHC reach. Also,
{$\ryp$}, the BLT {parameter}
for the Yukawa sector, has been tuned in the process to end up 
with such spectra.  
Note that the choices of values for $\rtp$ and $\ryp$ are consistent 
with the constraints from the physical top quark mass as discussed in 
section \ref{subsec:topmass} and the flavor constraints discussed in 
section \ref{subsec:flavor}. Larger values of $\rinv$ 
would tend to make the level `2' KK top quark a 
little too heavy (${\lesssim} \,1.5$ TeV) to be explored at the LHC while 
{if one requires} the lighter level `1' KK top quark not too light 
($\lesssim 300$ GeV) which can be quickly ruled out by the LHC experiments 
{even in an nmUED scenario which we consider.} 
Nonetheless, the lighter of the level `2' top quark
may anyway be heavy and only the level `1' top quarks remain to be 
relevant at the LHC. In that case, larger values of $\rinv$ also
remain relevant.
Values of $\rew$ are so chosen as to have $\kkaone$ as the LKP 
with masses around half a TeV. This renders the level `2' electroweak 
gauge bosons to have masses around 1.5 TeV thus making them possibly 
sensitive to searches for gauge boson resonances at the LHC 
\cite{Flacke:2012ke, Chatrchyan:2012su}\footnote{The caveats are that 
these level `2' gauge bosons could have very large decay widths 
(exceptionally fat) due to enhanced $V^{(2)}$-$f^{(0)}$-$f^{(0)}$ couplings
as opposed to narrow-width approximation for the 
{resonances} assumed in the experimental analysis 
\cite{Chatrchyan:2012su} and hence need dedicated studies for them at 
the LHC \cite{Kelley:2010ap}. Further, the involved assumption of a 
100\% branching fraction for the resonance decaying to quarks may also 
not hold. These two issues would invariably relax the mentioned bounds 
on level `2' gauge bosons.}. 

In table \ref{tab:bms} we also indicate the masses of the level `2'
KK excitations. It is to be noted that the lighter of the level `2'
KK top quark may not be that heavy ($\lesssim 1.5$ TeV).
Level `2' gluon, for our choices of parameters, is pushed to around 
3 TeV and hence, unless their couplings to quarks (SM ones or
from level `1') are enhanced, LHC may be barely sensitive to their
presence. This is a rather involved issue which {again} warrants 
dedicated studies and is beyond the scope of the present work.

For the first benchmark point (BM1) with $\rinv=1$ TeV, the 
mass-splitting between the two level `1' top quark states is much 
smaller ($\sim 100$ GeV) with a somewhat heavier $\tonel$ when 
compared to the second case (BM2) for which $\rinv=1.5$ TeV. 
We will see  in
section \ref{sec:pheno} that such mass-splittings and the absolute
masses themselves for the KK top quarks have interesting bearing
on their phenomenology at the LHC. Further, the relevant couplings
do change (see figures \ref{fig:couplings-200-220},
\ref{fig:couplings-211} and \ref{fig:couplings-higgs}) in going from one point to the
other. The third benchmark point BM3 is just BM1 but with different
$\ryp$ and $\mtopin$. BM3 demonstrates a situation with enhanced
Higgs-sector couplings and its ramifications at the LHC.
It is found that for all the three benchmark points, the coupling 
$V^{(2)}$-$f^{(0)}$-$f^{(0)}$ get enhanced when level `2' $W$ or $Z$ 
boson is involved.

Note that the KK bottom quark masses are also governed by $\rtp$ and
$\ryp$ for a given $\rinv$. However, since the splitting between the
two physical states at a given KK level is proportional to the SM
bottom quark mass, the KK bottom quarks {at each given level}
are almost degenerate ({just as it is for the} KK quark flavors 
from the first two generations)
in mass unlike their top quark counterparts.
Thus, some of the KK bottom quarks can have
masses comparable to those of the corresponding KK top quark states 
and hence would eventually enter a collider study otherwise dedicated 
for the latter. {A} detailed discussion on the involved issues 
are out of the scope of the present work.

%
\begin{table}[H]
\hspace{-10mm}
\begin{tabular}{|c|c|}
\hline
\hline
BM1 & $\rinv=1$ TeV, $\rgp=-1$, $\rqp=-1.2$, $\rtp=1$, $\ryp=0.5$,
$\rew=1.5$, $\rlp=0.4$, $\mtopin=173$ GeV \\
\hline
\hline
 & \\
Gauge & $\mkkaone=556.9$, $\mkkzone={m_{A^{(1)^{0}}}}=564.4$,
$\mkkwone={m_{H^{(1)^{\pm}}}}=562.7$, $\mkkglone=1653.8$\\
bosons & $\mkkatwo=1301.4$, $\mkkztwo={m_{A^{(2)^{0}}}}=1304.6$,
$\mkkwtwo={m_{H^{(2)^{\pm}}}}=1303.9$, $\mkkgltwo=2780.2$ \\ 
\& Higgs & $m_{H^{(1)^{0}}}=570.8, m_{H^{(2)^{0}}}=1307.4$ \\
 & \\
\hline
 & \\
 & $\mkkqkone=1711.5$, $\mkkqktwo=2816.9$ \\
Quarks & {$\mtopphys=172.6$}, $\mtonel=620.4$, $\mtoneh=714.5$ \\
\& & $\mttwol=1359.6$, $\mttwoh=1471.7$ \\ 
Leptons & $\mbone=638.3$, $\mbtwo=1395.8$ \\ 
 & $m_{l^{(1)}} = 802.3$, $m_{l^{(2)}} = 1631.8$ \\
 & \\
\hline
\hline
BM2 & $\rinv=1.5$ TeV, $\rgp=-0.1$, $\rqp=-1.1$, $\rtp=4$, $\ryp=8$, 
$\rew=5.5$, $\rlp=2$, $\mtopin=173$ GeV\\
\hline
\hline
 & \\
Gauge  & $\mkkaone=487.3$, $\mkkzone={m_{A^{(1)^{0}}}}=495.7$,
$\mkkwone={m_{H^{(1)^{\pm}}}}=493.9$, $\mkkglone=1601.6$\\
bosons & $\mkkatwo=1655.9$, $\mkkztwo={m_{A^{(2)^{0}}}}=1658.4$,
$\mkkwtwo={m_{H^{(2)^{\pm}}}}=1657.8$, $\mkkgltwo=3200.8$ \\
\& Higgs & $m_{H^{(1)^{0}}}=503.0, m_{H^{(2)^{0}}}=1660.6$ \\
 & \\
\hline
 & \\
 & $\mkkqkone=2527.5$, $\mkkqktwo=4200.2$ \\
Quarks & {$\mtopphys=172.4$}, $\mtonel=504.2$, $\mtoneh=813.3$ \\
\& & $\mttwol=1366.3$, $\mttwoh=2220.2$ \\ 
Leptons & $\mbone=561.9$, $\mbtwo=1706.6$ \\ 
 & $m_{l^{(1)}} = 750.0$, $m_{l^{(2)}} = 1865.1$ \\
 & \\
\hline
\hline
BM3 & Input values same as in BM1 except for {$\ryp=5$} and  
$\mtopin=176$ GeV \\
\hline
\hline
 & \\
Gauge &  \\
bosons & Masses same as in BM1 \\
\& Higgs & \\ 
 & \\
\hline
 & \\
Quarks & Masses same as in BM1 except for $\mtopphys=173.4$ and  \\
\&  &     $\mtonel={626.3}$, $\mtoneh={710.5}$ \\
Leptons & $\mttwol={1350.7}$, $\mttwoh={1488.6}$ \\ 
 & \\
\hline
\hline
\end{tabular}
\caption{Masses (in GeV) of different KK excitations in three 
benchmark scenarios. With $\rhp=\rew$, the level `1' Higgs boson 
masses are very much similar to the masses of the level `1' 
electroweak gauge bosons.
Choices of the input parameters satisfy the experimental bounds discussed 
earlier.}
\label{tab:bms}
%
\end{table}

%
%
\section{Phenomenology at the LHC}
\label{sec:pheno}
Given the nontrivial structure of the top quark sector of the nmUED it 
is expected that the same would have a rich phenomenology at the LHC. 
A good understanding of the same requires a thorough study of the
decay patterns of the KK top quarks and their production rates.
In this section we discuss these issues at the lowest order in
perturbation theory.

Towards this we implement the scenario in {\tt MadGraph} 5
\cite{Alwall:2011uj} using {\tt Feynrules} version 1
\cite{Christensen:2008py}
via its {\tt UFO} (Univeral Feynrules Output) 
\cite{Degrande:2011ua, deAquino:2011ub}
{interface}. This now contains the KK
gluons, quarks (including the top and the bottom quarks), leptons\footnote{
The KK leptons would eventually get into the cascades of the KK
gauge bosons.}
and the electroweak gauge bosons up to KK level `2'. Level `1' and level `2' 
KK Higgs bosons
are also incorporated. The mixings in the quark sector, including
`level-mixing' between {KK} level `2' and level `0', have now been 
incorporated in a generic way.
In this section we discuss these with the help of {the} benchmark 
scenarios discussed in section \ref{subsec:benchmarks}. We then 
consolidate the information to summarize the important issues in the 
search for such excitations at the LHC. 

\subsection{Decays of the KK top quarks}
\label{subsec:decays}
\begin{table}[t]
\begin{center}
\begin{tabular}{|c|r|r|r|}
\hline
    &   &   &    \\
BM1 & $\tonel \to b \kkwonep  = {0.597}$ & $\toneh \to b \kkwonep  = {0.615}$ 
    & $\ttwol \to \boneh \kkwonep = {0.351}$ \\
    & $b \kkhonep  = {0.403}$ & $b \kkhonep   = {0.370}$ 
    & $\toneh \kkhpone  = {0.177}$ \\
    &                            & ${\tonel Z = 0.016}$
    & $\boldsymbol{b W^+ = {0.062}}$ \\
    &                            &
    & $\boldsymbol{{t H = 0.062}}$ \\
    &                            &
    & ${\boneh \kkhonep = 0.057}$ \\
    &                            &
    & ${\bonel \kkhonep = 0.055}$ \\ 
    &                            &
    & $\boldsymbol{{t Z = 0.031}}$ \\
\hline
    &   &   &    \\
BM2 & $\tonel \to b \kkhonep  = {0.842}$ 
    & $\toneh \to \boneh W^+  = {0.305}$ 
    & $\ttwol \to \toneh \kkhpone = {0.377}$ \\
    & $b \kkwonep  = {0.158}$ 
    & $\tonel Z = {0.180}$ 
    & $\boneh \kkhonep = {0.208}$ \\
    &                           
    & $\bonel W^+ = {0.141}$
    & $\bonel \kkhonep = {0.200}$ \\
    &                           
    & $t \kkhpone = {0.130}$
    & $\tonel \kkhoneo = {0.109}$ \\
    &                           
    & $\tonel H = {0.126}$ 
    & ${\tonel \kkhpone = 0.055}$ \\
    &                           
    & ${b \kkhonep = 0.069}$ 
    & $\boldsymbol{{t H = 0.014}}$ \\
    &                           
    & ${b \kkwonep = 0.020}$ 
    & $\boldsymbol{{b W^{+} = 0.0022}}$ \\
    &                           
    & ${t \kkhoneo = 0.015}$ 
    & $\boldsymbol{{t Z = 0.00058}}$ \\
\hline
    &   &   &    \\
BM3 & $\tonel \to b \kkhonep  = {0.946}$ 
    & $\toneh \to b \kkhonep  = {0.941}$ 
    & $\boldsymbol{\ttwol \to  t H = {0.448}}$ \\
    & $b \kkwonep = {0.054}$    
    & $b \kkwonep = {0.060}$
    & ${\tonel \kkhpone = 0.102}$ \\ 
    &  
    & 
    &   $\toneh \kkhpone = {0.092}$ \\
    &   
    & 
    & ${\tonel \kkhoneo = 0.082}$ \\
    &
    &
    & ${\toneh \kkhoneo = 0.063}$ \\
    &
    &
    &  $\boldsymbol{{b W^+ = 0.046}}$ \\
    &
    &
    &  $\boldsymbol{{t Z = 0.022}}$ \\
\hline
\end{tabular}
\caption{Decay branching fractions of different KK top quarks for the
three benchmark points presented in table \ref{tab:bms}. Modes having
branching fractions less than about a percent are not presented 
{except for the ones with a pair of SM particles in the final
state}. Tree
level decays of $\ttwol$ to SM states are shown in bold in the
right-most column.}
\label{tab:decays}
\end{center}
\end{table}
Decays of the KK top quarks are mainly governed by the two input
parameters, $\rtp$ and $\rew$, for a given value of 
$\rinv$.\footnote{In the present analysis, the level `1' KK gluon is 
taken to be heavier than all three KK top quark states that are 
relevant for our present work, \emph{i.e.}, 
the two level `1' and the lighter level `2' KK top quarks.} The
dependence is rather involved since these two parameters not only
determine the spectra of the KK top quarks and the KK electroweak
gauge bosons but also {the involved} couplings. The latter, in turn,
are complicated functions of the input parameters as given by equation 
\ref{eqn:overlap2} and as illustrated in figures
\ref{fig:couplings-200-220},
\ref{fig:couplings-211} and \ref{fig:couplings-higgs}.
In the following, we briefly discuss the possible decay modes of the
KK top quarks and the significance of some of them at the LHC. In table
\ref{tab:decays} we list the branching fractions for the three
benchmark points presented earlier in table \ref{tab:bms}.  

For our choices of input parameters, two decay modes are possible for
$\tonel$: $\tonel \to b \kkwonep$ and $\tonel \to b \kkhonep$. Decays
to $t \kkzone / t \kkaone / t \kkhoneo / t \kkhpone$ are 
also possible when the mass-splitting between $\tonel$ and 
$\kkzone / \kkaone / \kkhoneo / \kkhpone$ is larger than the mass of 
the SM-like top quark. In our scenario, its decays to $\bonel$ and 
$\boneh$ are prohibited on kinematic grounds. 
Unlike in some competing scenarios (like the MSSM) where channels 
like, say, $\tilde{t}_1 \to b \chi^+_1$ and 
$\tilde{t}_1 \to t \chi^0_1$) could attain a 100\% branching 
fraction, the 
spectra of the involved KK excitations in our scenario 
would not allow $\tonel$ decaying exclusively to either $b \kkwone$
or $t \kkaone$.
The reason behind this is that $\kkwone$ and $\kkaone$ are rather 
close in mass and hence if decays to $t \kkaone$ is allowed, the same 
to $b \kkwonep$ is {also} kinematically {possible}. 
Further, even the latter mode 
has to compete with $\tonel \to b \kkhonep$ as $\mkkwone \approx
\mkkhonepm$. {Translating} constraints on
such KK top quarks from those obtained in the LHC-studies 
of{, say,} the top 
squarks is not at all straight-forward since the latter explicitly 
assume either
$\tilde{t}_1 \to b \chi^+_1 = 100\%$
\cite{Aad:2013ija,Chatrchyan:2013xna} or
$\tilde{t}_1 \to t \chi^0_1 = 100\%$ \cite{Chatrchyan:2013xna}.
Further, $\kkwone$ (and also $\kkzone$), being among the lighter
most ones of all the level `1' KK excitations, would only undergo three-body decays 
to LKP ($\gamma^{(1)}$) accompanied by leptons or jets that would be
rather soft because of the near-degeneracy of the masses of the level 
`1' KK gauge bosons. This would lead to loss of experimental sensitivity 
for final states with more number of hard leptons and jets 
\cite{Aad:2013ija}.

The situation with $\toneh$ is not qualitatively much different as
long as decay modes similar to $\tonel$ are the dominant ones. This 
is the case with BM1. Under such circumstances, they could 
{turn out to be reasonable} backgrounds to each other 
(if their production rates are comparable)
and dedicated studies would be required to disentangle them. In any case
(even in the absence of good discriminators), simultaneous 
productions of both $\tonel$ and $\toneh$ would enhance the 
{new-physics} signal.
On the other hand, in a situation
like BM2, more decay modes may be available to $\toneh$ 
{although}
decays to level `1' bottom and top quarks along with SM $W^\pm$ and
$Z$ are the dominant ones.
The ensuing cascades of these states would
inevitably make the analysis challenging. However, under favorable 
circumstances, reconstructions of the $W^\pm$ and/or $Z$ bosons
along with $b$- and/or \emph{top-tagging} could help disentangle
the signals. Thus, it appears that search for level `1' KK top quarks 
involves complicated issues (some of which are common to 
top squark searches in SUSY scenarios) and a multi-channel analysis 
could turn out to be very effective.

We now turn to the case of level `2' top KK top quarks. The lighter
of the two states, $\ttwol$ can have substantial rates at the LHC
which is discussed in some detail in section \ref{subsec:prod}.
This motivates us to study the decay patterns of $\ttwol$.
In the last column of table \ref{tab:decays} we present the decay
branching fractions of $\ttwol$.
As can be seen, the decay modes that are usually enhanced involve a pair of
level `1' KK excitations which would cascade to the LKP. 
We, however, strive to understand to what extent $\ttwol$, being an 
even KK-parity
state, could  decay directly to a pair of comparatively light 
(level `0') particles (and hence, boosted) comprising of an SM fermion 
and an SM gauge/Higgs boson\footnote{These may be contrasted with the 
popular SUSY scenarios (sparticles carrying odd $R$-parity) where such
possibilities are absent.}. 
Thus, in the one hand, these decay products are unlikely to be
missed in an experiment while on the other hand, new techniques 
to reconstruct (like the study of jet substructure
\cite{Altheimer:2012mn, Dasgupta:2013ihk} etc.) some of them 
have to be employed.
%

In scenario BM1, the total decay
branching fraction to SM states (shown in bold)
is just about 15\% while in scenario BM2 such decays are practically
absent. Given the large phase space available, such small 
(or non-existent) decay rates to SM particles can only be
justified in terms of rather feeble (effective) couplings among the
involved states.
The couplings of $\ttwol$ to the SM gauge bosons and an SM fermion
would have vanished (due to the orthogonality of 
the mode functions involved) had $\ttwol$ 
been a pure level `2' state. The smallness of these couplings thus
readily follows from the tiny admixture of the SM top quark in the
physical $\ttwol$ state and thus, results in its small branching 
fractions to SM gauge bosons. {The} same argument does not hold 
 for the corresponding coupling $\ttwol$-$t$-$H$ that controls
the other SM decay mode of $\ttwol$, \emph{i.e.}, $\ttwol \to  t H$.
{However, it} is clear from figure \ref{fig:couplings-higgs} that this coupling 
is going to be small for  both the benchmark points BM1 and BM2.

Since direct decays of $\ttwol$ to SM states could provide the
`smoking guns' at the LHC in the form of rather boosted objects (top and
bottom quarks, $Z$, $W^\pm$ and Higgs boson) that could eventually be 
reconstructed to their parent, this motivates us to study if such
decays can ever become appreciable. We find that the coupling 
$\ttwol$-$t$-$H$ gets significantly enhanced with a slight 
modification in the 
{parameters of} BM1 (called BM3 in table \ref{tab:bms}) by setting 
{$\ryp=5$} (see figure \ref{fig:couplings-higgs}) and $\mtopin=176$ 
GeV while keeping other parameters untouched and still satisfying all
the experimental constraints that we discussed. As we can see,
the branching fraction to $tH$ final state could attain a level of
50\% which should be healthy for the purpose. Efficient tagging of 
boosted top quarks \cite{Kaplan:2008ie, CMS:2009lxa, Plehn:2011tg,
Schaetzel:2013vka} 
and boosted Higgs bosons \cite{Butterworth:2008iy} 
would hold the key in such a situation.
Some such techniques have already been proposed in recent
literature \cite{Berger:2012ec}, in particular, in the context of 
vector-like top quarks or
more generally, in the study of `top-partners'.

\begin{figure}[t]
\centering
\includegraphics[width=0.4\columnwidth]{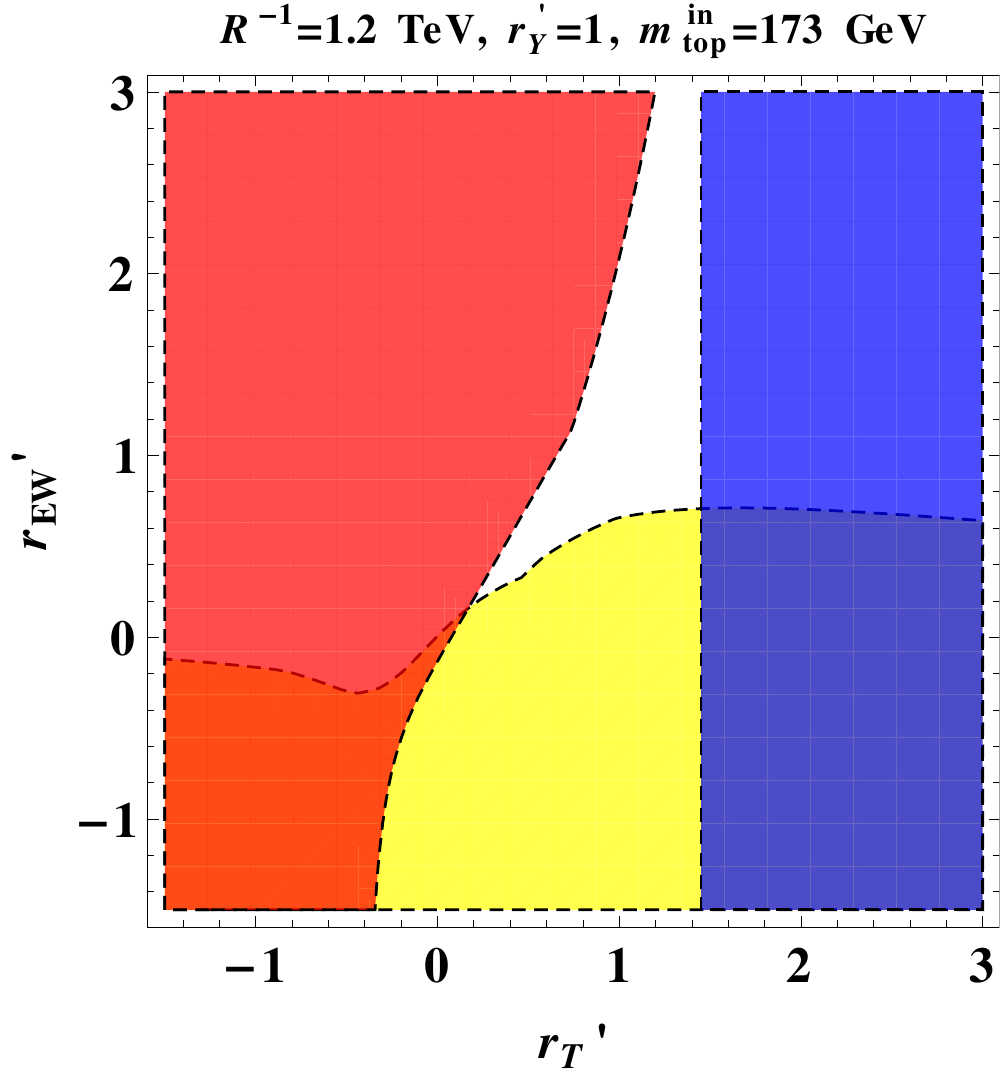}
\vspace*{-0.4cm}
\caption{Region in $\rtp-\rew$ plane 
where the decays $\ttwol \to \tonel \kkaone , \, \tonel \kkzone , 
\, \bonel \kkwonep$
are kinematically prohibited (in yellow),
$\kkaone$ is the LKP with $\mkkaone >
400$ GeV (in red) and $\mttwol < 1.5$ TeV (in blue).
The entire region shown is compatible with the acceptable range of the
mass of the top quark and other precision constraints.}
\label{fig:t2l-massdiff}
\end{figure}

On the other hand, since the $\ttwol$-$t$-$Z$ and $\ttwol$-$b$-$W^\pm$
are dynamically constrained, these could only get enhanced if the
competing modes (decays to a pair of level `1' KK states)
{face closure}. 
As the couplings involved in the latter cases are generically of SM 
strength, these could only be effectively suppressed by having them
kinematically forbidden. From figure \ref{fig:t2l-massdiff} we find that, 
by itself, this is not very difficult to achieve (in yellow shade) 
over the nmUED parameter space. However, rather conspicuously, the 
simultaneous demands for the KK photon to be the LKP with $\mkkaone >
400$ GeV (the red-shaded
region) and that of $\mttwol <1.5$ TeV (in blue shade) 
leave no overlapping region in the nmUED parameter space. 
It may appear that 
one simple way to find some overlap is by moving down in $\rtp$.
However, this implies $\ttwol$ 
{becomes more massive} thus loosing in its production 
cross section in the first place.
Although the interplay of events that leads to this kind of a 
situation is not an easy thing to follow, the issue that is broadly
conspiring is the similarity in the basic evolution-pattern of the 
masses of the KK excitations as functions of the BLKT parameters
(see figure \ref{fig:top-masses} and ref.~\cite{Datta:2012tv}).
\subsection{Production processes}
\label{subsec:prod}
In this section we discuss different production modes of the KK top
quarks at the 14 TeV (the design energy) LHC with reference to the 
nmUED parameter space. These are of following four broad types (in 
line with top squark phenomenology in SUSY scenarios):
\begin{itemize}
\item the generic mode with two top quark excitations in the final 
      state that receives contributions from processes involving both 
      strong and electroweak interactions, 
\item exclusively electroweak processes leading to a single top 
      quark excitation
\item the associated production of a pair of KK top quarks and the (SM)
      Higgs boson and 
\item {production} from the cascades of KK gluons and KK quarks.
\end{itemize}

\subsubsection{Final states with a pair of top quark excitations}
\label{subsubsec:strong-prod}
\begin{figure}[t]
\centering
\includegraphics[width=0.8\columnwidth]{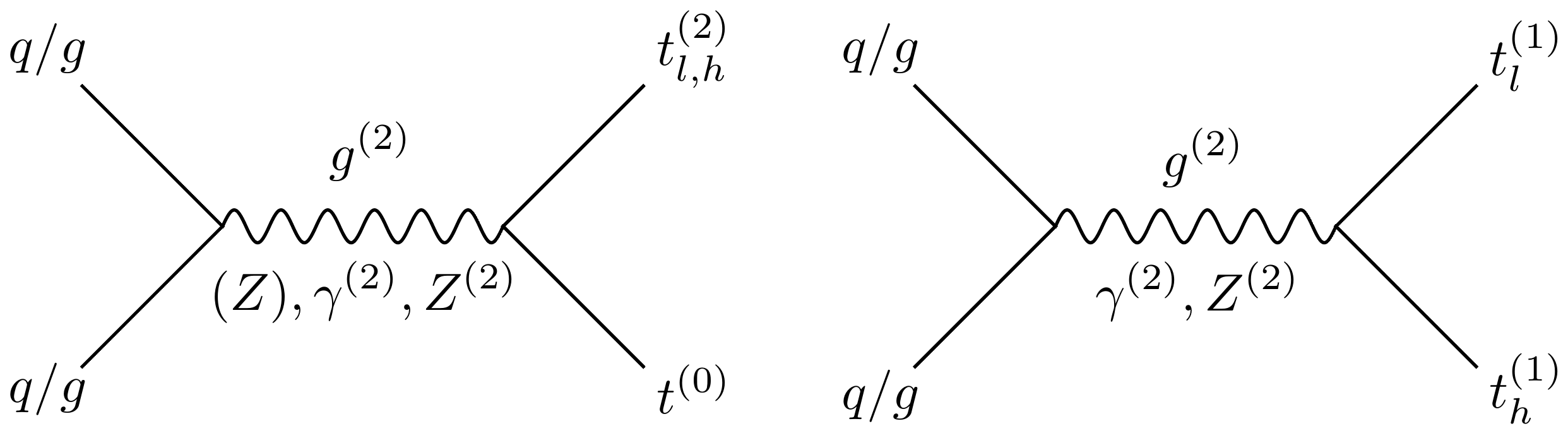}
\caption{Feynman diagrams for the associated $\ttwol-t^{(0)}$ (left) 
and $\tonel-\toneh$ productions at the LHC.
The gluon-initiated
processes are only mediated by $\kkgltwo$ while the quark-initiated
processes are mediated by both $\kkgltwo$ and other electroweak gauge
bosons from level `0' ($Z$) and level `2' ($\kkatwo, \, \kkztwo$).}
\label{fig:feynman1}
\end{figure}

These are the processes where two similar or different kind of top
quark excitations are produced in the final state. The interesting
modes in this category are pair-production of $\tonel$ and $\toneh$
along with the associated productions of $\tonel \toneh$ and 
$\ttwol t$. The latter two processes are possible in an nmUED scenario
and the corresponding Feynman diagrams\footnote{All the
Feynman diagrams in this paper are drawn by use of {\tt Jaxodraw}~\cite{Binosi:2003yf}, based
on {\tt Axodraw}~\cite{Vermaseren:1994je}.} are presented in figure
\ref{fig:feynman1}. Note that the requirement of current conservation
does not allow the massless SM gauge bosons (gluon and photon) to 
mediate these processes while the pair-productions receive
contributions from all possible {mediations}. Also, these two 
associated production modes have no counter-parts in a competing SUSY 
scenario like the MSSM.

\begin{figure}[t]
\centering
\includegraphics[width=0.35\columnwidth]{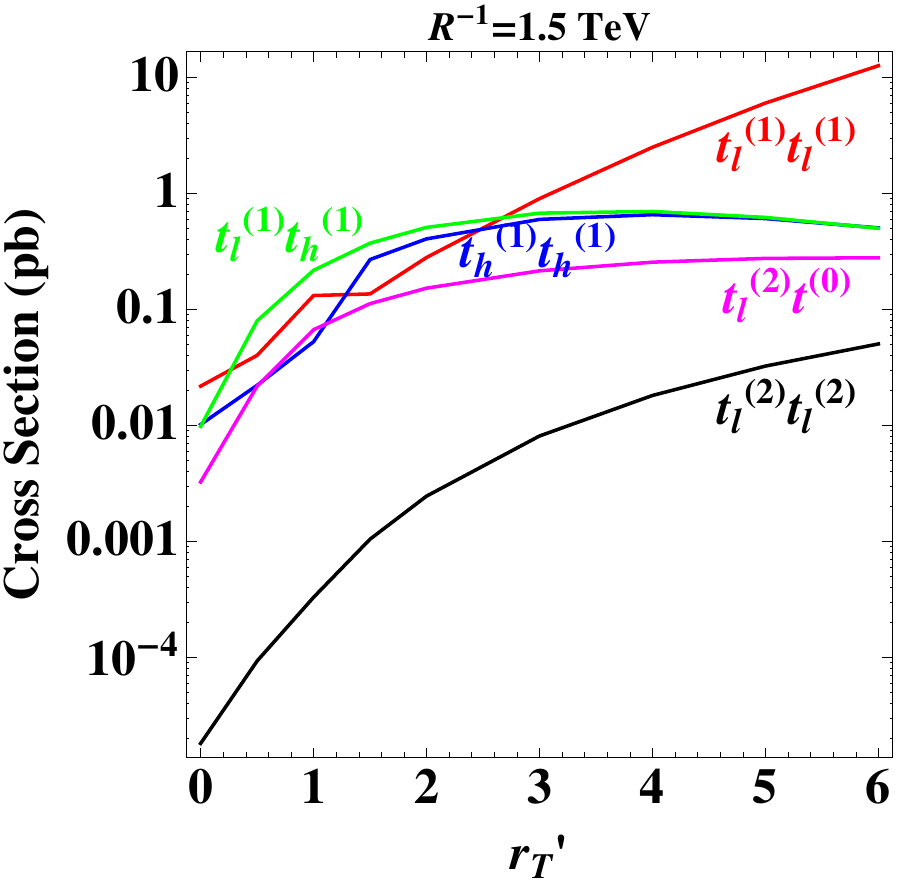}
\hspace*{0.1in}
\includegraphics[width=0.35\columnwidth]{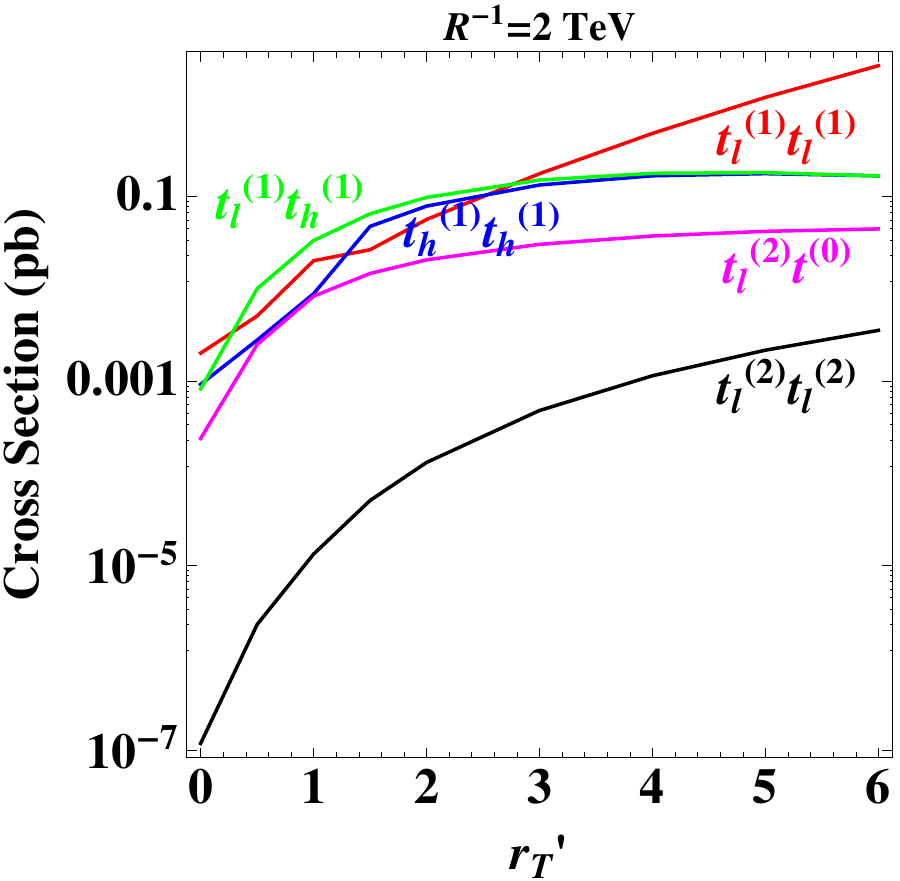}
\vspace*{-0.4cm}
\caption{Cross sections (in picobarns, at tree level) for different production processes
involving the KK top quarks as functions of $\rtp$ at {the} 14 TeV LHC for
$\rinv=1.5$ TeV (left) and $\rinv=2$ TeV (right), $\ryp=3$, $\rgp=0.5$
and the {other parameters} are chosen as in the {BM2}. 
CTEQ6L1 parton distributions~\cite{Nadolsky:2008zw}
are used and the factorization/renormalization scale is set at the sum 
of the masses in the final state.}
\label{fig:xsec-strong}
\end{figure}

In figure \ref{fig:xsec-strong} we illustrate the variations of the 
rates for these processes with $\rtp$ for $\rinv$=1 TeV (left) and 
2 TeV (right). As can be seen, pair production of $\tonel$, has by far
the largest cross section for $\rtp \gtrsim 3$ reaching up to 10 (1)
pb for $\rinv$ = 1.5 (2) TeV. This is not unexpected since $\tonel$ is 
the lightest of the KK top quarks. In this regime, the yields for
$\toneh$-pair and $\tonel \toneh$ associated productions are very
similar touching 1 (0.1) pb for $\rinv$ = 1.5 (2) TeV. The
corresponding rates for $\ttwol t$ associated production do not lag 
much notching 0.5 (0.05) pb, respectively. Further, {the} $\ttwol$-pair
{has} a {trend similar to that of} the $\tonel$-pair in this respect but, 
rate-wise, falls out of the competition.
%
%
%

Note that with increasing $\rtp$ masses of all the KK states decrease. 
Interestingly enough, this effect is reflected in a straight-forward 
manner only in the case of $\tonel$-pair for which the rates increase 
with growing $\rtp$. For other competing processes mentioned above, 
the curves flatten out. This behavior signals non-trivial interplays 
of the intricate couplings involved. These have much to do with when 
all these rates become comparable for $\rtp \lesssim 3$.\footnote{It 
may be noted in this context that an effective $SU(3)$ coupling 
involving a set of KK excitations is not necessarily stronger than the 
effective electroweak coupling among them and these might even have 
relative signs between them (see figures \ref{fig:couplings-200-220}, 
\ref{fig:couplings-211} and \ref{fig:couplings-higgs}). Thus, 
contributions from different mediating processes heavily depend on the 
nmUED parameters.} In the process, the rate for usual $t\bar{t}$ pair production
also gets affected to some extent. However, {our estimates are
all} being at the
tree level, {these do} not pose any immediate concern while
facing {the} measured $t\bar{t}$
cross section which is much larger and agrees with its
{estimation} at
higher orders in perturbation theory.
Also, in table \ref{tab:xsec1} we present the cross sections for
the three benchmark points. 
%

The bottom-line is that the production rates of three
different KK top quark excitations remain moderately healthy over 
favorable region of the nmUED parameter space at a future LHC run.
With the knowledge of their decay patterns 
(see table \ref{tab:decays}) and the associated features discussed
in section \ref{subsec:decays} it is required to chalk out a
strategy to reach out to these excitations.

\begin{table}[t]
\begin{center}
\begin{tabular}{|c|c|c|c|c|}
\hline
Benchmark  & $\tonel \tonebarl$ 
  & $\tonel \tonebarh$ 
  & $\toneh \tonebarh$ 
  & $t \ttwobarl$ \\
  & (pb) & (pb) & (pb) & (pb) \\
\hline
BM1 & 0.63 & 0.10 & 0.35 & 0.07 \\ 
\hline
BM2 & {2.24} & 0.35 & 0.76 & 0.21 \\ 
\hline
BM3 & {0.76} & {0.11} & {0.30} & 0.07 \\ 
\hline
\end{tabular}
\caption{Production cross sections (in picobarns, at tree level) for different 
pairs of KK top quarks for the benchmark points. Contributions from
the Hermitian conjugate processes are taken into account wherever
applicable.
The choices for the parton distribution and the scheme
for determining the factorization/renormalization
scale are the same as in figure~\ref{fig:xsec-strong}.}
\label{tab:xsec1}
\end{center}
\end{table}
%
\begin{figure}[t]
\centering
\includegraphics[width=0.70\columnwidth]{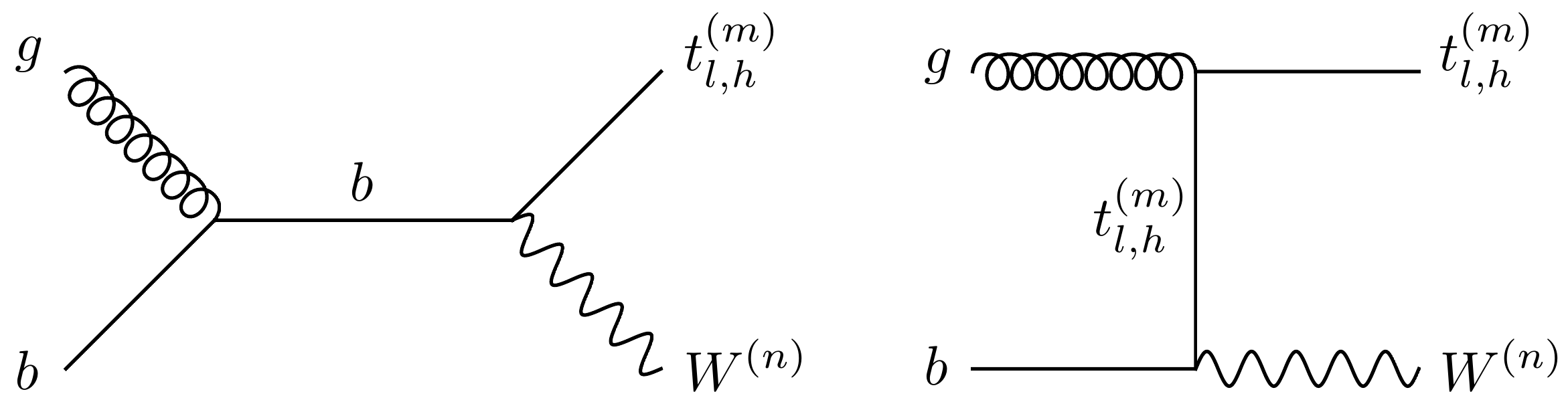}
\vskip 10pt
\includegraphics[width=0.35\columnwidth]{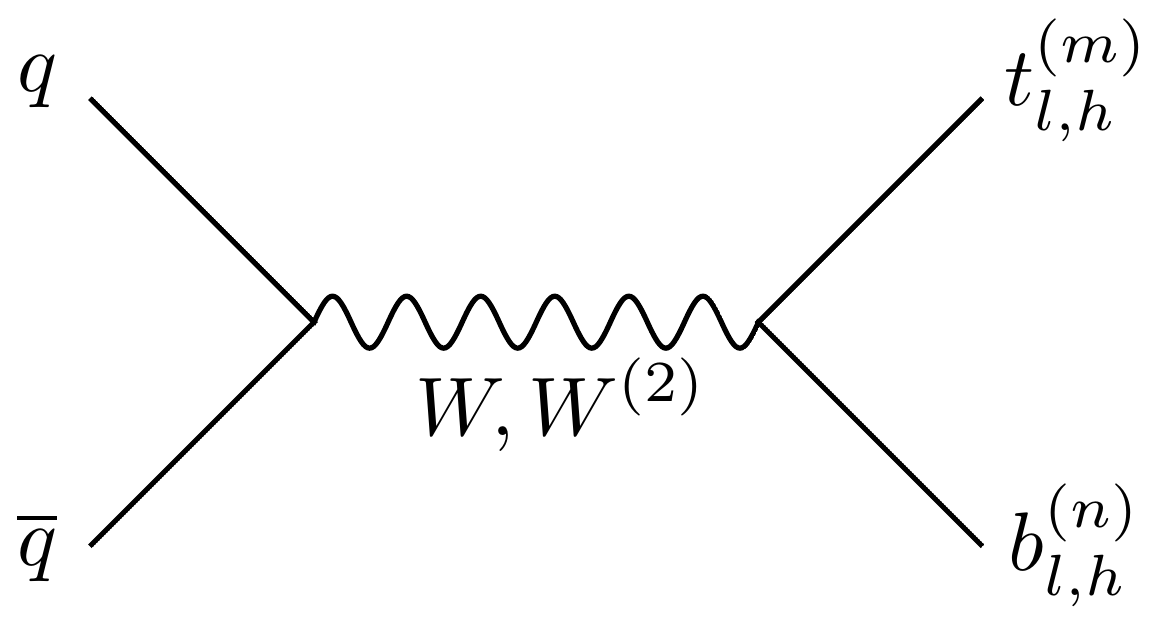}
\caption{Generic Feynman diagrams for the single production of a KK 
top quark along with KK excitations of $W^\pm$ boson (upper panel) and KK 
bottom quark (lower panel) at the LHC. Superscripts $m$ and $n$
standing for the KK levels can be different (like `0' and `2')
but should ensure KK-parity conservation.} 
\label{fig:feynman2}
\end{figure}
%
%
\subsubsection{Single production processes}
\label{subsubsec:single-prod}
{We consider} two broad categories of single production of KK top quarks
which are closely analogous to single top production in the SM once 
the issue of KK-parity conservation is taken into account.
In the first case, a level `1' KK top quark is produced in association 
with level $\kkwone$ or $b^{(1)}$ quark. The second one involves the 
lighter of the level `2' KK top quarks along with an SM $W^\pm$ boson 
or an SM bottom 
quark. The generic, {tree-level} Feynman diagrams that contribute
{to the processes} are
presented in figure \ref{fig:feynman2}. 

\begin{table}[t]
\begin{center}
\begin{tabular}{|c|c|c|c||c|c|c|c|c|}
\hline
Benchmark  & $\tonel \kkwonem$ 
  & $\tonel \bonebarl$  
  & $\ttwol b$  
  & $\tonel \tonebarl H$  
  & $\tonel \tonebarh H$  
  & $\toneh \tonebarh H$ 
  & $t \ttwobarl H$ 
  & $t \bar{t} H$ \\
  & (pb) & (pb) & (pb) & (pb) & (pb) & (pb) & (pb) & (pb) \\
\hline
BM1 & 0.01 & 0.11 & 0.11 & $\sim 10^{-5}$ & $\sim 10^{-4}$ & $\sim 10^{-3}$ & 0.03
& 0.24 \\
\hline
BM2 & 0.04 & 0.21 & 0.13 & {0.73} &  {5.39} & {0.17}
           & {0.11} & {1.25} \\
\hline
BM3 & $\sim 10^{-3}$ & {0.23} & {0.11}  & $\sim 10^{-4}$ 
    & $\sim 10^{-3}$ & 0.01 & 0.04 & {2.21} \\
\hline
\end{tabular}
\caption{Cross sections (in picobarns, at tree level) for single and (SM) 
Higgs-associated KK top quark productions for the benchmark points. 
The mass of the SM Higgs boson is taken to be 125 GeV. Contributions 
from the Hermitian conjugate processes are taken into account wherever
applicable.
The choices for the parton distribution and the scheme for 
determining the factorization/renormalization
scale are the same as in figure~\ref{fig:xsec-strong}.
}
\label{tab:xsec2}
\end{center}
\end{table}

\noindent
\paragraph{\cblack{\underline{Single production of level `1' top
quarks:}}}
Single production of level `1' top quarks along with a level `1' 
$W^\pm$ boson proceeds via $gb$ fusion in $s$-channel and $gb$
scattering in $t$-channel.
The rates are at best a few tens of femtobarns
as can be seen from table \ref{tab:xsec2}. On the other hand, the mode
in which a level `1' bottom quark is produced in association proceeds
through $s$-channel fusion of light quarks and propagated by $W^\pm$
and $\kkwtwo$ bosons. The cross sections are found to be rather healthy
ranging from 110 fb to {230} fb. The observed rates for 
$\tonel \kkwone$ production appear to be consistently lower than that
for $\tonel \bonel$ production. This can be traced back to the
presence of enhanced {$q$-$q'$-$W^{(2)^\pm}$} coupling. 
Moreover, cross sections for other combinations involving heavier
states of $t^{(1)}$ and $b^{(1)}$ in the final state could have 
comparable strengths because of such enhanced couplings.

%

\noindent
\paragraph{\underline{Single production of level `2' top quark}:} 
The associated $\ttwol W^-$ production involves the vertex
$\ttwol$-$W^\pm$-$b$ which, as we discussed earlier (see
sections~\ref{subsubsec:gauge-couplings} and \ref{subsec:decays}), 
vanishes but for a small admixture of level `0' top in the physical 
state $t^{(2)}$. Hence, the rates in this mode {turn out to be}
insignificant. 
Further, the $W^\pm$-mediated diagram in the associated $\ttwol b$ 
production also has the same vertex and thus contributes negligibly.
The only contribution here comes from the diagram mediated by 
$W^{(2)^\pm}$ which is somewhat massive. Thus, the prospect of having 
healthy rates for the single production of $t^{(2)}$ depends entirely
on the coupling strength $\ttwol$-$W^{(2)^\pm}$-$b$ and 
{$W^{(2)^\pm}$-$q$-$q$} (see figure \ref{fig:couplings-200-220}).
Fortunately, this is the case {here} and the cross sections for all three
benchmark points, as can be seen from table \ref{tab:xsec2}, 
are above and around 100 fb.

{We also looked into the production of $\ttwol$ along with 
light quark jets which is analogous to, by far the most dominant, `$t$-channel'
single top production process (the so-called $W$-gluon fusion process) in the
SM. However, in our scenario, such a process with somewhat heavy
$\ttwol$ yields a few tens of a femtobarn for all the three benchmark
points.}

{For both the categories mentioned above, the new-physics
contributions to the corresponding SM processes are systematically
small. This is since these contain the couplings that involve
level-mixing effect in the top-quark sector which is not large.}
%

\subsubsection{Associated production of KK top quarks with the SM
Higgs boson}
\label{subsubsec:asso-prod}
\begin{figure}[t]
\centering
\includegraphics[width=0.95\columnwidth]{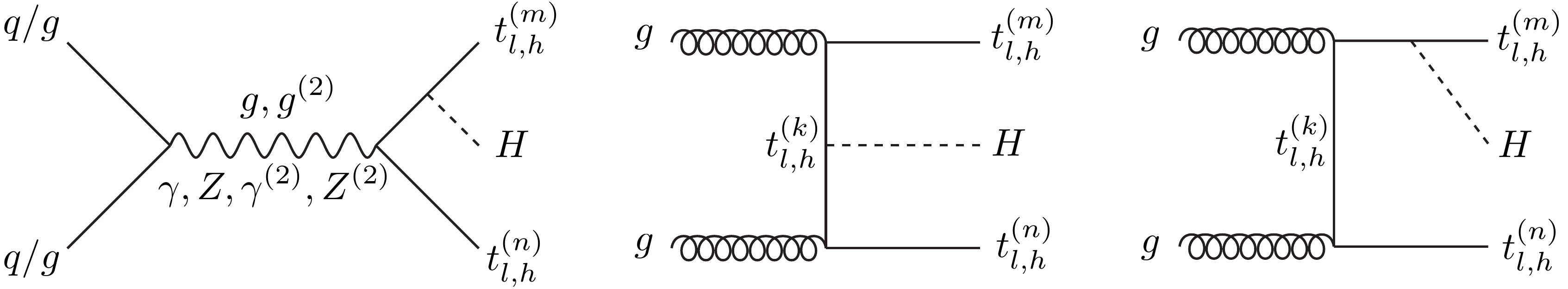}
\caption{Generic Feynman diagrams for the associated (SM) Higgs 
production along with a pair of KK excitations of the top quark. 
Superscripts $k$, $m$ and $n$ can be different (like `0' and `2')
but should ensure KK-parity conservation.}
\label{fig:feynman3}
\end{figure}

The associated Higgs production processes we consider involve both 
light and heavy level `1' top quarks in pairs and the level `2' 
lighter top quark along with the SM top quark. 
The generic {tree level} Feynman diagrams 
are presented in figure \ref{fig:feynman3}.
Given that the study 
of the SM $t\bar{t}H$ production is by itself complicated enough, 
it is only natural to expect that the same with its KK counterparts 
would not be any simpler. 
%
%
%
Cross sections for such processes are listed in table \ref{tab:xsec2}
for the {benchmark points} we consider. To have a feel
about the their phenomenological prospects, these can be compared
with similar processes in the SM and a SUSY scenario like the MSSM.
In the MSSM, the lowest order cross section is around a few tens
of a fb for the process $\tilde{t}_1 \tilde{t}^*_1 H$ with
$m_{\tilde{t}_1} \approx 300$ GeV and for the most
favorable values of the involved couplings 
\cite{Djouadi:1997xx, Djouadi:2005gj} while 
for the SM the corresponding rate is about {430} fb
\cite{Beenakker:2001rj, Djouadi:2005gi}. 
It is encouraging to {find} that the yield for
$t \ttwol H$ is either comparable (for BM1 and BM3) or larger (BM2)
than what can at best be expected in MSSM. 
{Note that 
the level `1' lighter KK top quark is somewhat heavier 
(with mass around or above 500 GeV) for our 
benchmark points when compared to the mass of the top squark
as indicated above}.
For other processes, BM2
consistently leads to larger cross sections. The interplay of
different
Feynman diagrams (see figure \ref{fig:feynman3}) along with the
modified strengths of the participating gauge and Yukawa interactions
play roles in some such enhancements. 

In the last column of table \ref{tab:xsec2} we indicate the lowest 
order cross sections for the SM process $t \bar{t} H$ which now gets
affected in an nmUED scenario. 
Note that for
{BM1 the cross section is smaller than the SM value of
$\approx 430$ fb while for BM2 and BM3 the same is about 3 and 5
times as large, respectively}. 
Such deviations can be expected if we refer back to the left panel of
figure \ref{fig:couplings-higgs} that
illustrates how the $t$-$\bar{t}$-$H$ coupling gets modified over
the nmUED parameter space. Note that, non-observation of such a
process at the LHC, till recently, could only restrict the rate up to 
around five times the SM rate
\cite{Chatrchyan:2013yea,cms-tth-gamma,atlas-tth-gamma}. Thus,
benchmark point BM3, as such, can be considered as a borderline case.
But given that $t \bar{t} H$ cross section depends on other nmUED
parameters like $\rgp$, $\rqp$ etc., one could easily circumvent this
restriction without requiring a compromise with the parameters like
$\rtp$ and $\ryp$ that define the essential feature of BM3,
\emph{i.e.}, the enhanced couplings among the top quark excitations
and the SM Higgs boson. It is interesting to find that in favorable
regions of parameter space, the cross section for Higgs production
in association with a pair of rather heavy KK top quarks could compare with
or even exceed the $t \bar{t} H$ cross section. Note that in the MSSM,
such enhancement only happens for large mixing in the stop sector and
when $m_{\tilde{t}_1} < m_t$ \cite{Djouadi:2005gj}.  

Further, once the level `1' KK Higgs bosons are taken up for studies,
the associated production  of a charged KK Higgs boson (from level 
`1') in the final state $b \tonel H^{{(1)}^\pm}$ would become rather 
relevant and may turn out to be interesting as the total mass involved
in this final state can be comparatively much lower. The prospect 
there depends crucially on the strength of {the involved} 3-point vertex though.

\subsubsection{Production of KK top quarks under cascades}
\label{subsubsec:cascade-prod}
KK gluon(s) and quarks, once produced, can cascade to KK top quarks.
This would result in multiple top quarks (upto four of them) in the
final state at the LHC. In our benchmark scenarios where 
$m_{\kkglone} <  m_{\kkqkone}$, KK 
gluons would directly decay to KK top quarks while KK quarks from the 
first two generations would undergo a two-step decay via KK gluon to 
yield a KK top quark. The latter one has thus suppressed contribution.
We find that the  branching fraction for $\kkglone \to t^{(1)} t$ is
around 50\% for all three benchmark points (the rest 50\% is to
level `1' bottom quark states). With strong production rates for the
$\kkglone$-pair, $\kkglone \kkqkone$ and $\kkqkone$-pair 
ranging between
0.01 pb to 2.6 pb (in increasing order), the yield of a single level 
`1' KK top final state could be anywhere between 10 fb to a few pb.
{These seem quite healthy. However, one has to cope with 
backgrounds which now have enhanced level of jet activity.}
%

\section{Conclusions and outlook}
\label{sec:conclude}
We discuss the structure and the phenomenology of the top
quark sector in a scenario with one flat extra spatial dimension
orbifolded on $S^1/Z_2$ and containing non-vanishing BLTs. The
discussion inevitably draws reference to the gauge and the Higgs
sectors. The scenario, by construct, preserves KK-parity.

The main purpose of the present work is to organize and work out
(following ref.~\cite{Flacke:2008ne}) the
necessary details in the involved sectors and explore the salient
features with their broad phenomenological implications in terms of a
few benchmark scenarios. This lay down the basis for future, detailed
studies of such a top quark sector at the LHC.

The masses and the couplings of the Kaluza-Klein excitations are
estimated at the lowest order in perturbation theory as functions of
$\rinv$ and the BLT parameters. For the KK top quarks, the extended
mixing scheme (originating in the Yukawa sector) is thoroughly worked
out by incorporating \emph{level-mixing}
among the level `0' and the level `2' KK top quark states, a
phenomenon that is not present in the popular mUED scenario.
In addition, unlike in the mUED, tree-level couplings that
violate
KK-number (but conserve KK-parity) are possible. We demonstrate how
all
these new effects, together, attract constraints
from
different precision experiments and shape the phenomenology of such a
scenario.

The nmUED scenario we consider has eight free parameters: $\rinv$ and
the scaled
(by $\rinv$) BLT coefficients $\rqp$, $\rlp$, $\rtp$, $\ryp$, $\rgp$,
$\rew \, (=\rwp=\rbp=\rhp)$ and $\mtopin$. However, in the present study, the
most direct
roles are played by $\rtp$, $\ryp$ and $\rew$ (=$\rhp$) in conjunction
with
$\rinv$. $\rqp$ and $\rgp$ play roles in the production processes by
determining
some relevant gauge-fermion couplings beside controlling the KK quark
and gluon
masses, respectively. On the other hand, $\rlp$ and $\mtopin$ only
play some
indirect roles through their influence on the experimentally measured
effects
that determine the allowed region of the parameter space.

The scenario has been thoroughly implemented in {\tt MadGraph} 5.
Three benchmark scenarios that satisfy all the relevant experimental
constraints are chosen for our study. These give conservatively
light KK spectra with sub-TeV masses for both level `1' electroweak KK
gauge bosons (with $\kkaone$ as the LKP) and the KK top quarks while
having the lighter level `2' top quark  below 1.5 TeV thus making them
all relevant at the LHC. Level `1' KK quarks from the first two
generations and the KK gluon are taken to be heavier than 1.6 TeV.

Near mass-degeneracy of the electroweak KK gauge bosons and the KK
Higgs bosons (at a given KK level) is a feature.
This influences the decays of the KK top quarks.
The lighter of the level `1' KK top quark can never decay 100\% of
the time to a top quark and the LKP photon. This is in sharp contrast
to a similar possibility in a SUSY scenario like the MSSM when a
top squark can decay 100\% of the time to a top quark and the LSP
neutralino,
an assumption that is frequently made by the LHC collaborations.
Instead, such a KK top quark has significant branching fractions to
both charged KK Higgs boson and to KK $W$ bosons at the same time.
Further, split between the KK top quark and the
KK electroweak gauge bosons that is attainable in the nmUED scenario
would generically lead to hard primary jets in the decays of the
former. This is again in clear contrast to the mUED scenario. However,
near mass-degeneracy prevailing in the gauge and the Higgs sector
would still result in rather soft leptons/secondary jets. Limited
mass-splitting among the KK gauge and Higgs bosons is a possibility
that has non-trivial ramifications and hence needs closer scrutiny.

The level `2' KK top quark we consider can decay directly to much
lighter SM particles like the $W$, the $Z$, the Higgs boson
and the top quark. These would then be boosted and hence may serve
as `smoking guns'. Recent studies of the vector-like top partners
\cite{CMS:2012ab, ATLAS:2012qe, atlas:heavytop} are in context.
However, these studies mainly bank on their pair-production and decays
that
comprise only of pairs of SM particles like $bW^\pm$ and/or $tZ$
and/or
$tH$. In the nmUED model that we consider, these are \emph{always}
accompanied
by other modes that may be dominant as well. The level `2' top quark
decaying to a pair of level `1' KK states is one such example.

Thus, phenomenology of the KK top quarks could turn out to be rather
rich (and complex) at the LHC. Clearly, strategies tailor-made for
searches of similar excitations under different scenarios could at
best be of very limited use. Even recasting  the analyses for some of them to the nmUED
scenario is not at all straight-forward. This calls for a dedicated
strategy that incorporates optimal triggers and employs advanced
techniques like analysis of jet-substructures {\it etc.} to tag the boosted
objects in the final states.

In any case, viability of a dedicated hunt depends crucially on
optimal production rates. We study these for the 14 TeV run of the
LHC. For all the possible modes in
which KK top quarks can be produced (like the pair-production, the
single production and the associated production with the SM Higgs
boson), the rates are found to be rather encouraging and may even
exceed
the corresponding MSSM processes, a yard-stick that can
perhaps be used safely (with a broad brush, though) for the purpose.

The LHC experiments are either already sensitive or will be achieving
the same soon in the next run for all the generic processes discussed
in this work. Given that the nmUED provides several top quark
KK excitations with different characteristic decays and production
rates, the sensitivity to them can only be increased if multi-channel
searches are carried out. It is thus possible that the LHC,
running at its design energy of 14 TeV (or even a little less),
finds some of these states. However, concrete studies with rigorous
detector-level simulations are prerequisites to chalking out a robust
strategy.

Last but not the least, the intimate connection between the top quark
and the Higgs sectors raises genuine curiosity in the phenomenology
for the KK Higgs bosons as well. The nmUED Higgs sector holds good promise
for a rather rich phenomenology at the LHC which has become further
relevant after the discovery of the `SM-like' Higgs boson and hence
can turn out to be a fertile area to embark upon.

\vskip 10pt
\noindent
{\bf Acknowledgments}
\vskip 7pt
\noindent
KN and SN are partially supported by funding available from the 
Department of Atomic Energy, Government of India for the Regional 
Centre for Accelerator-based Particle Physics (RECAPP), Harish-Chandra 
Research Institute. The authors like to thank Benjamin Fuks for very 
helpful discussions on issues with FeynRules and 
SN thanks Ujjal Kumar Dey for many helpful discussions.
The authors acknowledge the use of 
computational facility available at RECAPP and thank Joyanto Mitra for 
technical help. 
%
%
\appendix
\section{Gauge and the Higgs sector of the nmUED: some relevant
details}
\label{app:gauge-fixing}
In this appendix we briefly supplement our discussion in section
\ref{subsec:gauge-higgs} with some necessary details pertaining to the
gauge fixing conditions, the inputs that go into the mass-determining
conditions.

\subsection{Gauge fixing conditions}
\label{subsec:fixing}
We introduce the gauge-fixing terms in the bulk and at the boundaries 
{in the following way to obtain} the physical states:

\al{
S_{\text{gf}} &=
	\int d^4x \int_{-L}^{L} dy \Bigg\{
	- \frac{1}{2\xi_A} \left[ \pal_\mu A^{\mu} - \xi_A \pal_y A_y \right]^2
	- \frac{1}{\xi_W} \left| \pal_\mu W^{+\mu} - \xi_W \left( \pal_y W_y^+ +i M_W \phi^{+} \right)\right|^2 \notag \\
&\phantom{=\int d^4x \int_{-L}^{L} dy \Bigg\{ \, \,}
	- \frac{1}{2\xi_Z} \left[ \pal_\mu Z^{\mu} - \xi_Z \left( \pal_y Z_y + M_Z \chi \right) \right]^2 
	- \frac{1}{2\xi_G} \left[ \pal_\mu G^{a\mu} - \xi_G \pal_y G_y^a \right]^2 \notag \\
&- \frac{1}{2 \xi_{A,b}} \Big\{
	\left[ \pal_\mu A^{\mu} + \xi_{A,b} A_y \right]^2 \delta(y-L) +
	\left[ \pal_\mu A^{\mu} - \xi_{A,b} A_y \right]^2 \delta(y+L) \Big\} \notag \\
&- \frac{1}{\xi_{W,b}} \Big\{
	\left| \pal_\mu W^{+\mu} + \xi_{W,b} \left( W_y^+ - i r_H M_W \phi^+ \right) \right|^2 \delta(y-L) + 
	\left| \pal_\mu W^{+\mu} - \xi_{W,b} \left( W_y^+ + i r_H M_W \phi^+ \right) \right|^2 \delta(y+L) \Big\} \notag \\
&- \frac{1}{2 \xi_{Z,b}} \Big\{
	\left[ \pal_\mu Z^{\mu} + \xi_{Z,b} \left( Z_y -  r_H M_Z \chi \right) \right]^2 \delta(y-L) + 
	\left[ \pal_\mu Z^{\mu} - \xi_{Z,b} \left( Z_y +  r_H M_Z \chi \right) \right]^2 \delta(y+L) \Big\} \notag \\
&- \frac{1}{2 \xi_{G,b}} \Big\{
	\left[ \pal_\mu G^{a\mu} + \xi_{G,b} G_y^a \right]^2 \delta(y-L) +
	\left[ \pal_\mu G^{a\mu} - \xi_{G,b} G_y^a \right]^2 \delta(y+L) \Big\} \Bigg\}
}
where the eight gauge-fixing parameters are 
$\xi_A,\, \xi_W,\, \xi_Z,\, \xi_{G}$ {(in the bulk)}, $\xi_{A,b},\, 
\xi_{W,b},\, \xi_{Z,b},\, \xi_{G,b}$ {(at the boundary)}
and $M_W,\, M_Z$ are the masses of the $W$ and 
$Z$ bosons\footnote{This part {of the action} is also symmetric under 
the reflection $y \to -y$.}.

Imposing the unitary gauge in both the bulk and at the boundaries by
setting
\al{
\xi_A,\, \xi_W,\, \xi_Z,\, \xi_{G},\, \xi_{A,b},\, \xi_{W,b},\, \xi_{Z,b},\, \xi_{G,b} \to \infty
}
we obtain the following relations:
\al{
A_y &= 0, &  Z_y \mp r_H M_Z \chi &= 0, \notag \\
W_y^{+} \mp i r_H M_W \phi^{+} &= 0, & G_y^a &= 0, \qquad \text{at }y=\pm L,
\label{boundary_condition}
}
\al{
\pal_y A_y &= 0, &   \pal_y W_{y}^{+} + i M_W \phi^{+} &= 0, \notag \\
\pal_y Z_y + M_Z \chi &=0, &   \pal_y G_y^a &=0, \qquad \text{in the bulk}.
\label{bulk_condition}
}
As we see, $A_y$ and $G_y^a$ are totally gauged away from the 
theory as would-be Nambu-Goldstone bosons.
The two mixed boundary conditions in equation~\ref{boundary_condition}
can be cast into a set containing the individual fields with the help 
of equation~\ref{bulk_condition} as
\al{
\chi \pm r_H \pal_y \chi &= 0,  &   \phi^{+} \pm r_H \pal_y \phi^{+} &= 0, \notag \\
Z_y \pm r_H \pal_y Z_y &= 0,  &   W_y^{+} \pm r_H \pal_y W_y^{+} &= 0, \qquad \text{at }y=\pm L.
}

\vskip 5pt
\subsection{Input parameters for masses of the the KK gauge and Higgs bosons}
\label{subsec:inputs}
Input parameters that determine the masses of the KK gauge and the Higgs bosons of
the nmUED~\cite{Flacke:2008ne} (as solutions for the conditions given in equation (\ref{masscondition}))
are presented in table~\ref{table:massparameters}.
%
%
%

\begin{table}[t]
\begin{center}
\begin{tabular}{|c|c|c|c|}
\hline
Type & $m_F^2$ & $m_{F,b}^2$ & $r_F$ \\ \hline \hline
$W_{\mu}^{+}$ & $M_W^2$ & $r_H M_W^2$ & $r_{\text{EW}}$ \\ \hline 
$Z_{\mu}$ & $M_Z^2$ & $r_H M_Z^2$ & $r_{\text{EW}}$ \\ \hline 
$H$ & $(\sqrt{2} \hat{\mu})^2$ & $(\sqrt{2} \mu_b)^2$ & $r_H$ \\ \hline
$\phi^{+},\, \pal_y W_{y}^{+}$ & $M_W^2$ & $r_H M_W^2$ & $r_H$ \\ \hline 
$\chi,\, \pal_y Z_{y}$ & $M_Z^2$ & $r_H M_Z^2$ & $r_H$ \\ \hline 
\end{tabular}
\caption{Input parameters that determine the masses of the KK gauge and Higgs bosons.
See section~\ref{subsec:gauge-higgs} for notations and conventions.}
\label{table:massparameters}
\end{center}
\end{table}
%
%
\section{Tree-level FCNCs, the ``aligned'' scenario and
constraints from $D^0-\overline{D^0}$ mixing} 
\label{app:fcnc}

It has been demonstrated in ref.~\cite{gerstenlauer} that an 
appropriate short-distance description for a $\Delta F$=2 FCNC process 
like $D^0-\overline{D^0}$ can be found in processes involving only the 
even KK modes 
(starting at level `2') of the gauge bosons and the `0' mode fermions. 
In an effective Hamiltonian approach, such a process would reduce to a 
four-Fermi interaction whose strength is suppressed by the mass of the
exchanged KK gauge boson. The effective FCNC Hamiltonian can be
expressed in terms of suitable fermionic operators and their
associated Wilson coefficients. The latter involve the overlap
matrices in the gauge kinetic terms (by now, suitably rotated to the 
basis where the quark mass matrix is diagonal) which are functions of
the BLKT parameter, $\rqp$ and $\rtp$. Thus, any constraint on the
Wilson coefficients can be translated into constraints in the 
$\rqp$-$\rtp$ plane.

The gauge interactions in the diagonalized basis involving the level
`0' quarks and the KK gluons $g^{(k)}$, with the KK index 
$k$ being even and $k \geq 2$,  are given by:
\al{
&g_s \sum_{i,j,l=1}^{3} \bigg( 
        \overline{q^{(0)}_{iL}} \gamma^{\mu} T^a \Big[ (U_{qL}^\dagger)_{il}
F^{Q,[k]}_{g,ll} (U_{qL})_{lj} \Big] q^{(0)}_{jL} +
        \overline{q^{(0)}_{iR}} \gamma^{\mu} T^a \Big[ (U_{qR}^\dagger)_{il}
F^{q,[k]}_{g,ll} (U_{qR})_{lj} \Big] q^{(0)}_{jR}
\bigg) g_{\mu}^{(k)},
}
where  the 4D and the 5D (the `hatted' one) gauge couplings are 
related by $g_s \,\equiv\, \hat{g}_s/\sqrt{2 r_G + \pi R}$. 
$T^{a}$ represents the $SU(3)$ generators, $a$ being the color index. 
$U_{q(L,R)}$ are the matrices that diagonalize the $q_{L,R}$
fields in the Yukawa sector.
$F^{Q,[k]}_{g,ll}$ and $F^{q,[k]}_{g,ll}$ are the
diagonal overlap matrices (in the original bases)
\al{
F^{Q,[k]}_{g,ll} &= \frac{1}{f_{g^{(0)}}} \int_{-L}^{L} dy
	\left( 1 + r_{Q_l} \left[ \delta(y-L) + \delta(y+L) \right] \right)
	f_{Q^{(0)}_l} f_{g^{(k)}} f_{Q^{(0)}_l}, \\
F^{q,[k]}_{g,ll} &= \frac{1}{f_{g^{(0)}}} \int_{-L}^{L} dy
	\left( 1 + r_{q_l} \left[ \delta(y-L) + \delta(y+L) \right] \right)
	f_{q^{(0)}_l} f_{g^{(k)}} f_{q^{(0)}_l}
}
{while the} explicit form is shown in equation~\ref{F0-V2-F0_explicitform}.
Similar FCNC processes are also induced by the KK photons and the KK
$Z$ bosons.
However, because of weaker couplings their contributions are only
sub-leading and henceforth neglected in the present work.

The so-called ``aligned" scenario in which the rotation matrices for the left- 
and the right-handed quark fields are tuned to avoid as many flavor
constraints as possible can be summarized as
\al{
U_{uR} &= U_{dR} = U_{dL} = {\bf 1}_3, & U_{uL} &= V_{\text{CKM}}^\dagger
\label{eqn:gotomassbases}
}
along with universal BLKT parameters $\rqp$ and $\rtp$, for the first
two and the third quark generations respectively, irrespective of
their chiralities. 
{In such a scenario, by construct, dominant tree-level FCNC is 
induced via KK gluon exchange and only through the doublet up-quark 
sector.} 
Note that no FCNC appears at the up-quark singlet 
part and the down-quark sector. The latter helps evade severe 
bounds from the $K$ and $B$ meson sectors. The forms of the 4D Yukawa 
couplings, before diagonalization, are determined simultaneously as:
\al{
Y^{u}_{ij} = \sum_{l=1}^3 \frac{\left( V_{\text{CKM}}^\dagger \right)_{il}\mathcal{Y}^{u}_{lj}}{F^{d,(0,0)}_{ij}},
\qquad
Y^{d}_{ij} =
\begin{cases}
\displaystyle \frac{\mathcal{Y}^{d}_{ii}}{F^{d,(0,0)}_{ii}} & \text{for}\ i=j, \\
\displaystyle 0 & \text{for}\ i \not= j.
\end{cases}
}
In this configuration, the structure of the vertex
$\overline{u^{(0)}_{iL}} - d^{(0)}_{jL} - W^{+(0)}_{\mu}$ is reduced
to that of the SM. 
The overlap matrices in the gauge 
kinetic sector receive bi-unitary transformations when these terms are 
rotated to a basis where the quark mass matrices in the Yukawa sector 
are diagonal. These rotated overlap matrices are given by  
\al{
\sum_{l=1}^{3}
(U_{uL}^\dagger)_{il} F^{U,[k]}_{g,ll} (U_{uL})_{lj}  &=
\left\{
F^{U,[k]}_{g,11} {\bf 1}_3 +
V_{\text{CKM}}
\begin{pmatrix}
0 & & \\
& 0 & \\
& & \underbrace{F^{U,[k]}_{g,33} - F^{U,[k]}_{g,11}}_{=: \Delta
F^{U,[k]}_{g}}
\end{pmatrix}
V_{\text{CKM}}^{\dagger}  \right\}_{ij} \notag \\
&\simeq
\left\{
F^{U,[k]}_{g,11} {\bf 1}_3 + \Delta F^{U,[k]}_{g}
\begin{pmatrix}
A^2 \lambda^6 & -A^2 \lambda^5 & A \lambda^3 \\
-A^2 \lambda^5 & A^2 \lambda^4 & -A \lambda^2 \\
A \lambda^3 & -A \lambda^2 & 1 
\end{pmatrix}
\right\}_{ij}
\label{eqn:FCNC_matrix}
}
where $A(=0.814)$ and $ \lambda(=0.23)$ are the usual Wolfenstein 
parameters and we use the relation $F^{U,[k]}_{g,11} = F^{U,[k]}_{g,22}$.
Clearly, the difference of the two
overlap matrices in that diagonal term governs the FCNC contribution 
and thus, in turn, relative values of the corresponding BLKT parameters, 
$\rqp$ and $\rtp$ that shape the overlap matrices, get constrained.

To exploit the model independent constraints 
provided by the UTfit collaboration \cite{Bona:2007vi}, the effective 
Hamiltonian for the $t$-channel KK gluon exchange process (that
describes the $D^{0}-\overline{D^0}$ mixing effect) needs to be 
written down in terms of the following quark operators and the 
associated Wilson coefficient:
\al{
\Delta \mathcal{H}_{\text{eff}}^{\Delta C = 2} = C_D^1
(\overline{u}^{a}_L \gamma_{\mu} c^{a}_{L}) (\overline{u}^{b}_L
\gamma^{\mu} c^{b}_{L})
}
where $a$ and $b$ are the color indices and we
use $SU(3)$ algebra and appropriate Fierz transformation to obtain
\al{
C_D^1 = \sum_{k \geq 2:\text{even}} \frac{g_s^2(\mu_D)}{6}
\frac{1}{m_{g^{(2)}}^2} (- A^2 \lambda^5 \Delta F^{U,[k]}_{g})^2
\simeq
\frac{2\pi \alpha_s(\mu_D)}{3m_{g^{(2)}}^2} A^4 \lambda^{10} (\Delta
F^{U,[k]}_{g})^2.
}
As it appears, the value {of $C_D^1$} is highly Cabibbo-suppressed.
Heavier KK gluons (except the one from level `2') effectively decouples.
The QCD coupling at the $D^0$-meson scale $(\mu_D \simeq 2.8\,\text{GeV})$
is estimated by the relation,
\al{
\alpha^{-1}_s(\mu_D) = \alpha^{-1}_s(M_Z) - \frac{1}{6\pi} \left( 23
\ln{\frac{M_Z}{m_b}} + 25 \ln{\frac{m_b}{\mu_D}} \right)
\simeq 1/0.240
}
with $\alpha_s(M_Z) = 0.1184$~\cite{Adachi:2011tn}.
One would now be able to put bounds on the parameter space by use of 
the result by the UTfit collaboration~\cite{Bona:2007vi},
\al{
|C^1_D| < 7.2 \times 10^{-7}\,\text{TeV}^{-2}
}
which, for a given set of values for $\rinv$ and $\rgp$, actually 
exploits the dependence of $\Delta F^{U,[k]}_{g}$ (appearing in 
equation (B.6)) on the BLKT parameters $\rqp$ and $\rtp$.

%
\end{document}